\documentstyle[12pt,axodraw,epsf]{article} 
\catcode`@=11
\newcount\@tempcntc
\def\@citex[#1]#2{\if@filesw\immediate\write\@auxout{\string\citation{#2}}\fi
  \@tempcnta\z@\@tempcntb\m@ne\def\@citea{}\@cite{\@for\@citeb:=#2\do
    {\@ifundefined
       {b@\@citeb}{\@citeo\@tempcntb\m@ne\@citea\def\@citea{,}{\bf ?}\@warning
       {Citation `\@citeb' on page \thepage \space undefined}}%
    {\setbox\z@\hbox{\global\@tempcntc0\csname b@\@citeb\endcsname\relax}%
     \ifnum\@tempcntc=\z@ \@citeo\@tempcntb\m@ne
       \@citea\def\@citea{,}\hbox{\csname b@\@citeb\endcsname}%
     \else
      \advance\@tempcntb\@ne
      \ifnum\@tempcntb=\@tempcntc
      \else\advance\@tempcntb\m@ne\@citeo
      \@tempcnta\@tempcntc\@tempcntb\@tempcntc\fi\fi}}\@citeo}{#1}}
\def\@citeo{\ifnum\@tempcnta>\@tempcntb\else\@citea\def\@citea{,}%
  \ifnum\@tempcnta=\@tempcntb\the\@tempcnta\else
   {\advance\@tempcnta\@ne\ifnum\@tempcnta=\@tempcntb \else \def\@citea{--}\fi
    \advance\@tempcnta\m@ne\the\@tempcnta\@citea\the\@tempcntb}\fi\fi}
\catcode`@=12
\def\theequation{\arabic{section}.\arabic{equation}}

\begin{document}
\vskip -1.5cm
\begin{flushright}
WUE-ITP-2000-024\\[-0.1cm] 
hep-ph/0008268\\[-0.1cm]
August 2000
\end{flushright}

\begin{center}
{\LARGE 
{\bf Higgs Scalars in the Minimal Non-minimal}}\\[0.3cm]
{\LARGE {\bf  Supersymmetric Standard Model}}\\[1.4cm]
{\large C. Panagiotakopoulos$^a$ and A. Pilaftsis$^b$}\\[0.4cm]
$^a${\em Physics Division, School of Technology,
         Aristotle University of Thessaloniki,\\
         54006 Thessaloniki, Greece}\\[0.2cm]
$^b${\em Institut f\"ur Theoretische Physik,
         Universit\"at W\"urzburg,\\
         Am Hubland, 97074 W\"urzburg, Germany}    
\end{center}
\vskip0.7cm  \centerline{\bf ABSTRACT}  
We consider the simplest and  most economic version among the proposed
non-minimal super\-symmetric  models, in which  the $\mu$-parameter is
promoted to a singlet  superfield, whose all self-couplings are absent
from  the renormalizable superpotential.   Such a  particularly simple
form of the renormalizable  superpotential may be enforced by discrete
$R$-symmetries   which    are   extended   to    the   gravity-induced
non-renormalizable operators as well.   We show explicitly that within
the   supergravity-mediated   supersymmetry-breaking   scenario,   the
potentially dangerous divergent  tadpoles associated with the presence
of the  gauge singlet first  appear at loop  levels higher than  5 and
therefore do not destabilize the gauge hierarchy. The model provides a
natural  explanation  for  the   origin  of  the  $\mu$-term,  without
suffering  from  the visible  axion  or  the cosmological  domain-wall
problem.   Focusing on the  Higgs sector  of this  minimal non-minimal
supersymmetric  standard  model,  we  calculate  its  effective  Higgs
potential by integrating out the  dominant quantum effects due to stop
squarks.   We then  discuss the  phenomenological implications  of the
Higgs  scalars predicted  by the  theory  for the  present and  future
high-energy colliders.   In particular, we  find that our  new minimal
non-minimal   supersymmetric  model   can   naturally  accommodate   a
relatively light charged Higgs boson, with a mass close to the present
experimental lower bound.

\newpage

\setcounter{equation}{0}

\section{Introduction}

In  the  well-established  Standard  Model  (SM),  the  generation  of
gauge-invariant,  renormalizable masses  for the  observable fermions,
e.g.\ the electron  and the $t$ quark, and for the  $W$ and $Z$ bosons
is achieved through the so-called Higgs mechanism. Most interestingly,
the  Higgs mechanism  itself predicts  inevitably the  existence  of a
fundamental scalar, known as the Higgs boson. Recently, experiments at
LEP2  have  intensified  their  searches for  directly  observing  the
yet-elusive Higgs boson. Their latest analyses show that its mass must
be  larger   than  113.3~GeV  at   the  95\%  confidence   level  (CL)
\cite{ADLO}.  At the same  time, electroweak  precision data  place an
upper  bound  of  the  order   of  240~GeV  on  the  Higgs-boson  mass
\cite{EWdata}.

So  far, we  have  no much  evidence  to suggest  that the  underlying
structure  of the  Higgs potential  is  indeed that  of the  SM or  it
already  contains components  of a  more fundamental  theory  which is
about to  be unraveled in the next-round  experiments. In particular,
it is known that the SM cannot adequately address the problem of gauge
hierarchy, which is related to the perturbative stability of radiative
effects  between  the  electroweak  scale  and  the  Planck  or  grand
unification  scale.  An appealing  solution  to  this  problem may  be
achieved by means of supersymmetry (SUSY). In order that SUSY theories
avoid  reintroducing the  problem  of gauge  hierarchy,  they must  be
softly broken at a relatively  low scale $M_{{\rm SUSY}}\sim m_{t}$ of
the order of 1~TeV, in agreement with experimental observations.

The  minimal supersymmetric extension   of the   SM, also  called  the
Minimal  Supersymmetric   Standard  Model  (MSSM),  predicts   a  very
constrained  two-Higgs-doublet  potential at    the tree  level, whose
quartic couplings are   determined by the  well-measured SU(2)$_L$ and
U(1)$_Y$ gauge couplings $g_w$ and $g^{\prime}$. As a consequence, the
lightest neutral Higgs boson is  always lighter than  the $Z$ boson at
the tree level. Nevertheless, radiative  corrections to the  effective
Higgs potential are significant and extend  the above mass upper bound
to 110 (130)~GeV for small (large) values of the ratio of Higgs vacuum
expectation    values  (VEV's) $\tan\beta    \approx   2\   (> 15)   $
\cite{SHiggs}. Thus,  a  large portion of  the  parameter space of the
MSSM has   been already excluded  by  the current  LEP2 experiments at
CERN. Moreover, the upgraded Tevatron collider at Fermilab will have a
much higher  reach in discovering   heavier Higgs bosons  with SM-type
couplings and masses up to 140 GeV and therefore will provide a unique
test for the viability of the MSSM.

On the basis  of the above strong experimental  bounds on the lightest
Higgs-boson  mass in  the MSSM  (especially  for low  values of  $\tan
\beta$), it  would be rather  premature to infer that  realizations of
low-energy SUSY  in nature  have a rather  limited range. In  order to
reach  a more  definite conclusion,  it is  very important  to further
analyze the Higgs sectors of minimally extended scenarios of the MSSM.
An  additional reason  for  going beyond  the  MSSM is  the so  called
$\mu$-problem.   The superpotential  of the  MSSM contains  a bilinear
term   $-\mu   \widehat{H}_{1}\widehat{H}_{2}$   involving   the   two
Higgs-doublet  superfields  $\widehat{H}_{1}$  and  $\widehat{H}_{2}$,
known as the $\mu $-term. Although $\mu $ is naturally of the order of
the Planck  scale $M_{{\rm  P}}$, it is  actually required to  be many
orders of magnitude smaller of order $M_{{\rm SUSY}}$ for a successful
Higgs mechanism  at the electroweak  scale.  Many scenarios  have been
proposed in the  existing literature to account for  the origin of the
$\mu$-term, albeit all in extended settings \cite{mu}.

A simple SUSY extension of the  MSSM, which one  might have thought of
considering to address the $\mu$-problem, would be to elevate the $\mu
$-parameter to a dynamical variable by means of a gauge-singlet chiral
superfield $\widehat{S},$  couple the   latter to $\widehat{H}_1$  and
$\widehat{H}_2$  as $\lambda \widehat{S} \widehat{H}_1  \widehat{H}_2$
and arrange that $\widehat{S}$ somehow develops a  VEV of the order of
$M_{{\rm SUSY}} \sim m_{t}$. However, this minimally extended scenario
possesses a global U(1) Peccei-Quinn  (PQ) symmetry, whose spontaneous
breakdown gives rise to a phenomenologically  excluded axion. The most
popular way in the literature of  removing the unwanted PQ symmetry is
to break  the  latter  explicitly by  adding   the cubic self-coupling
$\frac{1}{3}\kappa  \,  \widehat{S}^3$   to  the superpotential.   The
resulting model  has been   termed the Next-to-Minimal  Supersymmetric
Standard Model (NMSSM) \cite{nmssm}.  Unfortunately, the NMSSM is also
plagued by its own problems. The  cubic self-coupling of $\widehat{S}$
leaves invariant a subgroup  of U(1)$_{{\rm PQ}}$, namely the discrete
${\cal Z}_{3}$  symmetry, whose subsequent spontaneous breakdown gives
rise to the formation of cosmologically catastrophic weak-scale domain
walls \cite {KT,ASW}.

Another  well-known  problem  that   a  model  of  low-energy  physics
involving light  gauge-singlets has to face is  the destabilization of
the gauge  hierarchy through the generation of  at least quadratically
divergent  tadpoles for  the singlet  \cite{NIL}.  In  the  context of
$N=1$ supergravity, which  is spontaneously broken by a  set of hidden
sector  superfields,  even  if  one  assumes no  other  scale  between
$M_{{\rm   SUSY}}$  and   $M_{{\rm  P}}$,   the  simple   presence  of
gravity-induced non-renormalizable operators in the superpotential and
the K\"{a}hler potential is able to generate such tadpoles \cite{Bag}.
Using the Planck mass $M_{{\rm P}}$ as a physical cut-off energy, such
divergences contribute  tadpole terms of  order $(1/16\pi^2)^n M_{{\rm
    P}}M_{{\rm  SUSY}}^2  S$ to  the  effective  potential, where  $n$
indicates the loop  level at which the tadpole  divergence appears. It
is obvious that for small values of $n$, e.g.\ $n\le 4$, the generated
tadpole terms  lead generically to  unacceptably large values  for the
VEV  of   $S$  (the   scalar  component  of   $\widehat{S}$),  thereby
destabilizing the gauge hierarchy.

In the case of the aforementioned  extensions of the MSSM, the problem
of destabilization does not occur as long as  the U(1)$_{{\rm PQ}}$ or
${\cal Z}_{3}$ symmetries   are  imposed  on   the  complete set    of
non-renormalizable operators as  well.  However, any attempt  to break
these  unwanted  symmetries   through a subset   of non-renormalizable
operators   would, as an   immediate consequence, destabilize the weak
scale.  This  aspect  has been  emphasized in  Ref.\ \cite{ASW,SA}, in
connection with the ${\cal Z}_{3}$ symmetry of the NMSSM.

Recently, it  has been realized  that the unwanted ${\cal  Z}_{3}$ and
U(1)$_{{\rm   PQ}}$   symmetries    present   in   the   corresponding
supersymmetric extensions of the  MSSM could be effectively broken not
by  the non-renormalizable  operators  themselves, but  rather by  the
tadpoles   generated   by  them   \cite{PT1,PT2}.    For  the   ${\cal
  Z}_{3}$-symmetric extension  of the MSSM, harmless  tadpole terms of
order $(1/16\pi^2)^n  M_{{\rm SUSY}}^{3} S$, with $2\le  n\le 4$, were
sufficient for the  breaking of the ${\cal Z}_{3}$  symmetry.  For the
PQ-symmetric  extension instead,  it  was necessary  that the  harmful
tadpoles  of order  $(1/16\pi^{2})^{n}M_{{\rm  P}}M_{{\rm SUSY}}^{2}S$
(using  $M_{  {\rm  P}}$  as  a  cut-off  scale)  be  generated  at  a
sufficiently high  loop level  $n$, with  $5 \leq n  \leq 8$.   In the
${\cal Z}_3$ case, the harmful  tadpoles were forbidden by imposing on
the operators of  the non-renormalizable superpotential and K\"{a}hler
potential the ${\cal Z}_{2}$ $R$-symmetry of the cubic superpotential,
under which all  superfields as well as the  superpotential flip sign. 
However, the  desirable form of the  renormalizable superpotential was
enforced by imposing a larger  group, namely the product of the ${\cal
  Z}_2$ matter parity with a ${\cal Z}_4$ $R$-symmetry \cite{PT1}.  In
the  U(1)$_{{\rm  PQ}}$ case,  a  ${\cal  Z}_{5}$ $R$-symmetry  proved
sufficient to enforce  the desirable renormalizable superpotential and
postpone the  appearance of the  harmful divergent tadpoles  until the
sixth loop order \cite{PT2}.  Thus,  in both cases the breaking of the
unwanted symmetries was  successfully implemented without jeopartizing
the  stability of  the electroweak  scale and  without  generating new
cosmological problems.

In  the present  paper,  we shall   study in  detail  the new  minimal
supersymmetric extension  of the MSSM, in  which the linear, quadratic
and cubic terms involving  the singlet superfield $\widehat{S}$ itself
are absent  from   the  renormalizable part  of the   superpotential.  
Hereafter, we shall call such  a supersymmetric extension the  Minimal
Non-minimal Supersymmetric Standard Model  (MNSSM).  In particular, we
shall explicitly show that with the imposition  of the discrete ${\cal
  Z}_5$ and ${\cal Z}_7$ $R$-symmetries on the complete superpotential
and on  the  K\"ahler potential   of the  corresponding   supergravity
models, the potentially  dangerous tadpole divergences first appear at
the six- and seven- loop levels, respectively, and hence are naturally
suppressed to the order  of $M^3_{ {\rm  SUSY}}\, S$.  Evidently,  the
resulting  model constitutes the   simplest and most  economic version
among  the non-minimal     supersymmetric models   proposed in     the
literature.  In order  to properly study  the properties of  the Higgs
bosons predicted by the theory, we will  calculate the effective Higgs
potential by  taking into  account  the  dominant stop-loop  effects.  
Finally,  we shall  analyze the  phenomenological  implications of the
MNSSM for  direct  Higgs-boson searches at the  LEP2  and the upgraded
Tevatron colliders.

The organization of the paper is as follows: in  Section 2 we describe
the    Higgs  sector  of the   MNSSM    and show  that harmful tadpole
divergences first appear at the six- and seven- loop levels, after the
aforementioned   discrete   ${\cal   Z}_{5}$  and   ${\cal     Z}_{7}$
$R$-symmetries   are respectively imposed   on  the theory.  Technical
details of the argument are relegated to Appendix  A.  In Section 3 we
compute the effective Higgs  potential by integrating out the dominant
radiative effects due   to stop  squarks,  from  which  we derive  the
CP-even  and  CP-odd  Higgs-boson mass matrices.      In Section 4  we
investigate  the  theoretical  differences  of  the  Higgs-boson  mass
spectrum    between   the     MNSSM   under consideration     and  the
frequently-discussed  NMSSM.   In Section   5  we  present   numerical
estimates of the   Higgs-boson masses and  their couplings  associated
with  the   $Z$  boson   in   these  two   models, and   discuss   the
phenomenological implications of the MNSSM Higgs sector for the direct
Higgs-boson searches at  LEP2 and  for  the upcoming  searches at  the
upgraded Tevatron collider.  Section 6 contains our conclusions.

\setcounter{equation}{0}

\section{MNSSM: Symmetries and stability of \\ the electroweak scale}

In this section we shall consider  the simplest extension of the MSSM,
the MNSSM, within    the context of  $N=1$ supergravity  spontaneously
broken by a set of hidden sector fields at  an intermediate scale.  In
the   MNSSM,  the $\mu$-parameter is  promoted  to  a dynamical chiral
superfield,\footnote[1]{An earlier  suggestion along  these  lines was
  discussed in \cite{Pomarol}.}  with  the linear, quadratic and cubic
terms involving only the singlet superfield $\widehat{S}$ being absent
from  the  renormalizable superpotential.  Such  a particularly simple
form of the superpotential may be enforced by discrete $R$-symmetries,
e.g.\ ${\cal Z}_{5}^{R}$ and  ${\cal Z} _{7}^{R}$, which are  extended
to the  non-renormalizable   parts  of  the  superpotential  and   the
K\"{a}hler potential as   well.  Adopting the standard power  counting
rules  \cite{Bag,SA}, we shall show   that in such $N=1$  supergravity
scenarios,  the     potentially dangerous   tadpole    divergences are
suppressed  by loop factors $1/(16\pi   ^{2})^{n}$ of order $n=6$  and
higher,  and therefore do   not   destabilize  the gauge  hierarchy.   
Technical details are given in Appendix A.

The renormalizable superpotential  of  the MNSSM under  discussion  is
given by
\begin{equation}
W_{{\rm ren}}\ =\ h_{l}\,\widehat{H}_{1}^{T}i\tau
_{2}\widehat{L}\widehat{E} \:+\:h_{d}\,\widehat{H}_{1}^{T}i\tau
_{2}\widehat{Q}\widehat{D}\:+\:h_{u}\, \widehat{Q}^{T}i\tau
_{2}\widehat{H}_{2}\widehat{U}\:+\:\lambda \,\widehat{S}\,
\widehat{H}_{1}^{T}i\tau _{2}\widehat{H}_{2}\,, \label{Wren}
\end{equation}
where $\tau  _{2}$ is  the usual  $2\times 2$  Pauli matrix.   In Eq.\ 
(\ref{Wren}),   the    Higgs      superfields,   $\widehat{H}_{1}$ and
$\widehat{H}_{2}$, as well as the  quark and lepton chiral multiplets,
$\widehat{Q}$  and  $\widehat{L}$,  are SU(2)$_L$-doublets, while  the
remaining superfields  $\widehat{S}$, $\widehat{U}$, $\widehat{D}$ and
$\widehat{E}$  are  singlets under  SU(2)$_L$.   The chiral multiplets
also carry the following hypercharges:
\begin{equation}
{\rm U(1)}_{Y}:\quad \widehat{H}_{1}\,(-1),\ \widehat{H}_{2}\,(1),\
\widehat{ S}\,(0),\ \widehat{Q}\,(1/3),\ \widehat{U}\,(-4/3),\
\widehat{D}\,(2/3),\ \widehat{L}\,(-1),\ \widehat{E}\,(2)\,,
\end{equation}
where  the hypercharge  of  each superfield  is  indicated within  the
parentheses. In addition to the baryon (B) and lepton (L) numbers, the
renormalizable superpotential $W_{{\rm ren}}$ respects the global U(1)
PQ and $R$ symmetries:
\begin{eqnarray}
{\rm U(1)}_{{\rm PQ}} &:&\widehat{H}_{1}\,(1),\ \widehat{H}_{2}\,(1),\
\widehat{S}\,(-2),\ \widehat{Q}\,(-1),\ \widehat{U}\,(0),\ \widehat{D}
\,(0),\ \widehat{L}\,(-1),\ \widehat{E}\,(0)\,; \nonumber \label{U1}
\\ {\rm U(1)}_{R} &:&\widehat{H}_{1}\,(0),\ \widehat{H}_{2}\,(0),\
\widehat{S} \,(2),\ \widehat{Q}\,(1),\ \widehat{U}\,(1),\
\widehat{D}\,(1),\ \widehat{L} \,(1),\ \widehat{E}\,(1),\ W_{{\rm
ren}}\,(2)\,.\qquad
\end{eqnarray}
Note that  $W_{{\rm ren}}$ has  charge  2 under ${\rm U(1)}_{R}$.  The
symmetry  group ${\rm U(1)}_{R}$ is non-anomalous  with respect to QCD
interactions,  but gets broken  by   the soft SUSY-breaking  trilinear
couplings  down to its maximal non-$R$  ${\cal  Z}_{2}$ subgroup which
becomes  the known  matter-parity.   Instead,  the  anomalous symmetry
${\rm  U(1)}_{{\rm PQ }}$  remains unbroken  by the soft SUSY-breaking
terms. Neglecting QCD-instanton effects, ${\rm U(1)}_{{\rm PQ }}$ will
remain unbroken, unless  a gravity-induced tadpole  operator linear in
$S$ gets generated from  the non-renormalizable sector  of the theory. 
The  tadpole   operator  generically  contributes   to  the  effective
potential a term
\begin{equation}
V_{{\rm tad}}\ \sim \ \frac{1}{(16\pi ^{2})^{n}}\,M_{{\rm P}}M_{\rm
  SUSY}^2 S\quad +\quad {\rm h.c.}\,,  \label{Vtad}
\end{equation}
where $n$  is the loop level at  which the  tadpole divergence occurs,
using the Planck mass $M_{{\rm P}}$ as  an energy cut-off. The tadpole
term $V_{  {\rm tad}}$ together with the  soft SUSY-breaking mass term
$M_{{\rm SUSY} }^{2}S^{*}S$ lead to a VEV for the singlet field $S$ of
order  $(1/16\pi  ^{2})^{n}M_{{\rm  P}}$.  To  avoid destabilizing the
gauge  hierarchy, one  must require \cite{PT2}  that $\langle S\rangle
\sim M_{{\rm SUSY}}\sim (1/16\pi  ^{2})^{n}M_{{\rm P}}$, with $M_{{\rm
    SUSY}}\sim 1$~TeV.   This  requirement can only be   fulfilled for
sufficiently high  values  of $n$,  i.e.\ for $n\ge  5$.   Finally, we
should remark that  the full renormalizable Lagrangian,  including the
tadpole term, preserves the B    and L numbers. However, the   quantum
numbers   B  and  L may  be   violated   by certain non-renormalizable
operators, which are hopefully of sufficiently high order in order not
to upset the laboratory limits on proton instability. We can therefore
conclude that the renormalizable superpotential $W_{{\rm ren}}$ of Eq.
(\ref{Wren}) supplemented with  a sufficiently suppressed  tadpole for
the singlet $S$ leads to  a model without any obvious phenomenological
or cosmological problem.

One may now wonder whether there exists a  symmetry giving rise to the
above-described model that  includes a  tadpole  term for $S$  of  the
desirable order.  To address this question, let us consider the global
symmetry defined as a linear combination  $R^{\prime }=3R+{\rm PQ}$ of
U(1)$_{R}$ and U(1)$_{{\rm PQ}}$, with
\begin{equation}
{\rm U(1)}_{R^{\prime }}:\quad \widehat{H}_{1}\,(1),\
\widehat{H}_{2}\,(1),\ \widehat{S}\,(4),\ \widehat{Q}\,(2),\
\widehat{U}\,(3),\ \widehat{D}\,(3),\ \widehat{L}\,(2),\
\widehat{E}\,(3),\ W_{{\rm ren}}\,(6)\, .
\end{equation}
Observe that   the   imposition  of ${\rm U(1)}_{R^{\prime      }}$ is
sufficient to ensure the  form (\ref{Wren}) for $W_{{\rm  ren} }$.  We
should now examine whether ${\rm U(1)}_{R^{\prime  }}$ also allows the
generation of     a tadpole    term.    The symmetry      group  ${\rm
  U(1)}_{R^{\prime  }}$  is explicitly  broken  by the  trilinear soft
SUSY-breaking interactions down to its maximal non-$R$ subgroup ${\cal
  Z}_{6}$ which is isomorph (equivalent) to the  product group $ {\cal
  Z}_{2}\times  {\cal  Z}_{3}$.    The   symmetry ${\cal Z}_{2}$    is
essentially  the  ordinary matter-parity,   under which  the   tadpole
remains invariant.  Instead, the symmetry ${\cal  Z}_{3}$ is broken by
the tadpole   of $S$.   Consequently,  a  tadpole   term can only   be
generated if  the whole symmetry group  U(1)$_{R^{\prime }}$ or one of
its subgroups  that  includes  ${\cal Z}  _{3}$  is  violated by   the
higher-order non-renormalizable operators.

%******************************************************************
%%% Figure 1 
%******************************************************************
\begin{figure}[t]
\begin{center}
\begin{picture}(400,200)(0,0)
\SetWidth{0.8}
 
\ArrowLine(100,20)(100,50)\Vertex(100,50){2.5}
\ArrowArc(100,70)(20,90,-90)\ArrowArcn(100,70)(20,90,-90)\Vertex(100,90){2.5}
\ArrowArcn(100,120)(30,90,-20)\ArrowArc(100,120)(30,-90,-20)
\ArrowArcn(100,120)(30,20,90)\ArrowArc(100,120)(30,20,-90)
\ArrowArc(100,170)(20,-90,90)\ArrowArcn(100,170)(20,-90,90)
\ArrowLine(100,150)(100,190)\Vertex(100,190){2.5}\Vertex(100,150){2.5}
\ArrowLine(100,150)(72,110)\ArrowLine(100,150)(128,110)
\Vertex(72,110){2.5}\Vertex(128,110){2.5}
\Text(94,30)[r]{$\widehat{S}$}
\Text(76,70)[r]{$\widehat{H}_1$}\Text(126,70)[l]{$\widehat{H}_2$}
\Text(78,93)[r]{$\widehat{S}$}\Text(122,93)[l]{$\widehat{S}$}
\Text(68,130)[r]{$\widehat{H}_1$}\Text(134,130)[l]{$\widehat{H}_2$}
\Text(98,120)[r]{$\widehat{H}_2$}\Text(104,120)[l]{$\widehat{H}_1$}
\Text(76,170)[r]{$\widehat{H}_1$}\Text(126,170)[l]{$\widehat{H}_2$}
\Text(94,170)[r]{$\widehat{S}$}

\Text(100,0)[]{\bf (a)}

\ArrowArc(220,120)(20,180,0)\ArrowArcn(220,120)(20,180,0)\Vertex(200,120){2.5}
\ArrowLine(200,120)(240,120)\Vertex(240,120){2.5}
\ArrowArcn(270,120)(30,0,60)\ArrowArc(270,120)(30,180,75)
\ArrowArcn(270,120)(30,120,0)\ArrowArc(270,120)(30,120,180)
\ArrowLine(260,147)(300,120)\ArrowLine(260,93)(300,120)\Vertex(260,147){2.5}
\Vertex(260,93){2.5}\Vertex(300,120){2.5}\Vertex(340,120){2.5}
\ArrowArcn(320,120)(20,0,180)\ArrowArc(320,120)(20,0,180)
\ArrowLine(340,120)(300,120)\ArrowLine(300,40)(300,120)
\Text(220,145)[b]{$\widehat{H}_1$}\Text(220,125)[b]{$\widehat{S}$}
\Text(220,95)[t]{$\widehat{H}_2$}\Text(245,143)[b]{$\widehat{S}$}
\Text(245,97)[t]{$\widehat{S}$}\Text(287,152)[b]{$\widehat{H}_1$}
\Text(287,90)[t]{$\widehat{H}_2$}\Text(270,131)[]{$\widehat{H}_2$}
\Text(270,112)[]{$\widehat{H}_1$}\Text(320,145)[b]{$\widehat{H}_1$}
\Text(320,125)[b]{$\widehat{S}$}\Text(320,95)[t]{$\widehat{H}_2$}
\Text(305,60)[l]{$\widehat{S}$}

\Text(300,0)[]{\bf (b)}

\end{picture}
\end{center}
\caption{Typical harmful tadpole divergences at the (a) six- and (b) 
seven- loop  levels.}
\label{f1}
\end{figure}
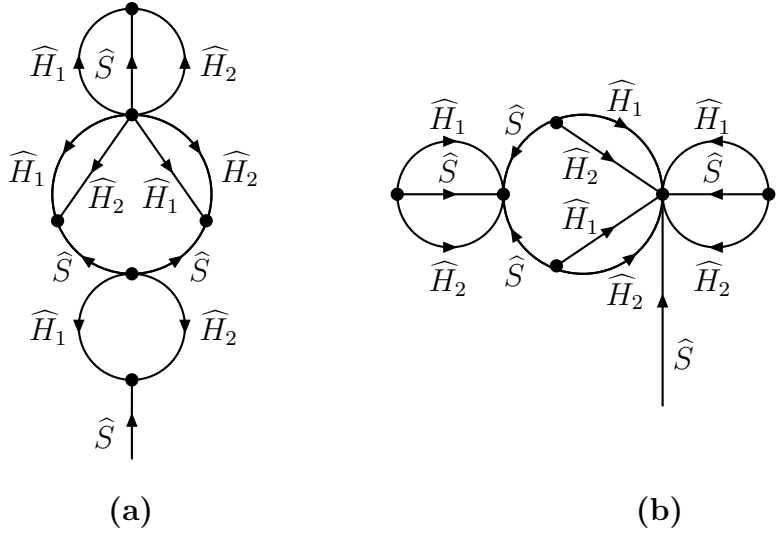

The above arguments  seem to suggest that  the symmetry we are looking
for  is likely to   be  a subgroup   of U(1)$_{R^{\prime }}$  which is
sufficiently large  to enforce the form  of  $W_{{\rm ren}}$  given by
Eq.\ (\ref{Wren}), but does not contain the ${\cal Z}_{3}$ subgroup of
U(1)$_{R^{\prime }}$. Subgroups  of  U(1)$_{R^{\prime }}$  obeying the
above criteria are the  discrete  $R$-symmetries ${\cal Z}   _{5}^{R}$
\cite{PT2} and ${\cal Z}_{7}^{R}$.

Let  us  first consider  the  ${\cal  Z}_{5}^{R}$  case. Under  ${\cal
Z}_{5}^{R}$,  the  chiral multiplets  as  well  as the  superpotential
$W_{{\rm ren}}$ transform as follows:
\begin{eqnarray}
{\cal Z}_{5}^{R} &:&(\widehat{H}_{1},\ \widehat{H}_{2})\to \omega \,(
\widehat{H}_{1},\ \widehat{H}_{2})\,,\quad (\widehat{Q},\ \widehat{L})\to
\omega ^{2}\,(\widehat{Q},\ \widehat{L})\,,\quad (\widehat{U},\ \widehat{D}
,\ \widehat{E})\to \omega ^{3}\,(\widehat{U},\ \widehat{D},\ \widehat{E})\,,
\nonumber  \label{Z5R} \\
&&\widehat{S}\to \omega ^{4}\widehat{S}\,,\quad W_{{\rm ren}}\to \omega W_{
{\rm ren}}\,,
\end{eqnarray}
with $\omega =\exp (2\pi i/5)$  and $\omega ^{5}=1$.  The discrete $R$
-symmetry ${\cal Z}_{5}^{R}$ is imposed on the complete superpotential
and K\"{a}hler  potential. By means of  standard power  counting rules
governing  the  harmful tadpole divergences,  it   can  be shown  that
harmful tadpoles first appear  at the six-loop  level.  As can be seen
from Fig.\ \ref{f1}(a), a typical harmful six-loop tadpole diagram can
be induced by appropriately combining the non-renormalizable operators
of the K\"{a}hler potential
\begin{equation}
K_{2}\ =\ \kappa _{2}\,\frac{\widehat{S}^{2}\,(\widehat{H}_{1}^{T}i\tau _{2}
\widehat{H}_{2})}{M_{{\rm P}}^{2}}\ +\ {\rm h.c.}\,,\qquad K_{5}\ =\ \kappa
_{5}\,\frac{\widehat{S}\,(\widehat{H}_{1}^{T}i\tau _{2}\widehat{H}_{2})^{3}}{
M_{{\rm P}}^{5}}\ +\ {\rm h.c.}\,,
\end{equation}
and         four  times      the        renormalizable  term  $\lambda
\,\widehat{S}\widehat{H}  _{1}^{T}i\tau _{2}\widehat{H}_{2}$   of  the
superpotential  (\ref{Wren}).  The analytic steps  of the argument are
presented in Appendix  A. Thus, the  induced harmful divergent tadpole
term has the form
\begin{equation}
V_{{\rm tad}}\ \sim \ \frac{\kappa _{2}\kappa _{5}\lambda ^{4}}{(16\pi
^{2})^{6}}\,M_{{\rm P}}M_{{\rm SUSY}}^{2}S\ +\ {\rm h.c.}  \label{Vtad6}
\end{equation}
{}From Eq.\ (\ref{Vtad6}), it is easy  to see that the tadpole term is
of  order $(1\  {\rm  TeV})^{3}$, e.g.\  for  $\kappa _{2}\sim  \kappa
_{5}\sim  0.1$, $\lambda  \sim 0.6$  and $M_{{\rm  SUSY}}\sim  1$ TeV,
and does not destabilize the gauge hierarchy.

In the ${\cal Z}_{7}^{R}$  case, the harmful tadpole divergence occurs
at  one loop-order  higher, namely at   the  seven-loop level,  so the
generated  tadpole terms   can   naturally be   as  low as  $(100~{\rm
  GeV})^3$.  In  detail, under  this new  discrete $R$  -symmetry, the
superfields and $W_{{\rm ren}}$ transform in the following way:
\begin{eqnarray}
{\cal Z}_{7}^{R} &:&\!\!(\widehat{H}_{1},\ \widehat{H}_{2})\to \omega \,(
\widehat{H}_{1},\ \widehat{H}_{2})\,,\quad (\widehat{Q},\ \widehat{L})\to
\omega ^{2}\,(\widehat{Q},\ \widehat{L})\,,\quad (\widehat{U},\ \widehat{D}
,\ \widehat{E})\to \omega ^{3}\,(\widehat{U},\ \widehat{D},\ \widehat{E})\,,
\nonumber  \label{Z7R} \\
&&\!\!\widehat{S}\to \omega ^{4}\widehat{S}\,,\quad W_{{\rm ren}}\to \omega
^{6}W_{{\rm ren}}\,,
\end{eqnarray}
with $\omega =\exp (2\pi i/7)$ and $\omega ^{7}=1$. Following the same
line of steps as above,   we impose the discrete $R$-symmetry   ${\cal
  Z}_{7}^{R}$ on the complete superpotential and K\"{a}hler potential.
Based on standard power counting rules, we show in Appendix A that the
potentially harmful tadpole divergences first appear at the seven-loop
level.  A typical harmful tadpole diagram at  seven loops is displayed
in  Fig.\   \ref{f1}(b), and   can  be  obtained  by    combining  the
non-renormalizable operators of the K\"{a}hler potential
\begin{equation}
K_{3}^{(1)}\ =\ \kappa _{3}^{(1)}\,\frac{\widehat{S}^{3}\,(\widehat{H}
_{1}^{T}i\tau _{2}\widehat{H}_{2})}{M_{{\rm P}}^{3}}\ +\ {\rm h.c.}\,,\qquad
K_{6}\ =\ \kappa _{6}\,\frac{\widehat{S}^{2}\,(\widehat{H}_{1}^{T}i\tau _{2}
\widehat{H}_{2})^{3}}{M_{{\rm P}}^{6}}\ +\ {\rm h.c.}\,,
\end{equation}
and four times the renormalizable term $\lambda \,\widehat{S}\widehat{H}
_{1}^{T}i\tau _{2}\widehat{H}_{2}$ of $W_{{\rm ren}}$. The size of the
so-generated tadpole term may be estimated as
\begin{equation}
V_{{\rm tad}}\ \sim \ \frac{\kappa _{3}^{(1)}\kappa _{6}\lambda ^{4}}{(16\pi
^{2})^{7}}\,M_{{\rm P}}M_{{\rm SUSY}}^{2}S\ +\ {\rm h.c.}\ \sim \ (1\ {\rm 
TeV})\times M_{{\rm SUSY}}^{2}\,S\ +\ {\rm h.c.}\,,  \label{Vtad7}
\end{equation}
for $\kappa _{3}^{(1)}\sim \kappa _{6}\sim  1$ and $\lambda \sim 0.6$. 
If $ \kappa    _{3}^{(1)}\sim   \kappa _{6}\sim 0.1$   and    $M_{{\rm
    SUSY}}\sim 1$ TeV,  the size of $V_{{\rm tad}}$  can be as low  as
$(0.2\ {\rm TeV})^{3}$.

We  conclude this  section   by noticing that  although the   discrete
$R$-symmetries ${\cal Z}^R_5$  and ${\cal Z}^R_7$  do not contain  the
usual ${\cal Z}_2$ matter parity,  they still prohibit the presence of
all  dimension $d=4$ B-  and  L-  violating  operators as well  as the
dangerous B-   and  L- violating  operators  $\widehat{Q}  \widehat{Q}
\widehat{Q} \widehat{L}$  and $\widehat{U}    \widehat{U}  \widehat{D}
\widehat{E}$ of dimension 5.  However,  the symmetries ${\cal  Z}^R_5$
and ${\cal  Z}^R_7$    allow the  L-violating   operator  $\widehat{L}
\widehat{L} \widehat{H}_2 \widehat{H}_2$  of  $d=5$, which is  able to
generate Majorana   masses   for the  light   left-handed   neutrinos. 
Moreover,  ${\cal  Z}^R_5$   allows the  $d=5$  L-violating  operators
$\widehat{S}  \widehat{S} \widehat{L} \widehat{H}_2$,     $\widehat{S}
\widehat{L}    \widehat{L} \widehat{E}$  and $\widehat{S}  \widehat{L}
\widehat{Q} \widehat{D}$,  whereas  ${\cal  Z}^R_7$ allows  the  $d=5$
B-violating     operator    $\widehat{S}     \widehat{U}   \widehat{D}
\widehat{D}$.  Although these last operators   are unable to lead   by
themselves  to   an observable proton  decay,   they  still render the
lightest  supersymmetric particle  (LSP) unstable.  However, estimates
based on  naive dimensional analysis  show  that the LSP  is very long
lived  with   a lifetime larger   than  the age   of  the Universe and
therefore safely qualifies to be a  dark-matter candidate.  Of course,
the LSP can be made absolutely stable  by the additional imposition of
the ${\cal Z}_2$ matter parity.

\setcounter{equation}{0}

\section{The Higgs sector of the MNSSM}

In this section  we  shall study  the  low-energy Higgs sector of  the
MNSSM.  After discussing    its  tree-level structure, we  will   then
calculate the  one-loop  effective Higgs potential  by integrating out
the dominant loop effects due to stop/top quarks, from which we derive
analytic  expressions for the  Higgs-boson masses and their respective
mixing  angles. We shall then  focus on the gaugino-Higgsino sector of
the  MNSSM, and  briefly  discuss  possible laboratory  limits  on the
would-be $\mu$-parameter due to the presence  of a light quasi-singlet
neutralino   state.  Finally,   for  our  forthcoming phenomenological
discussion in Section 5,   we shall present the effective  Higgs-boson
couplings to the $W$ and $Z$ bosons.

\subsection{Higgs-boson masses at the tree level}

In addition to   terms proportional  to  $S$,  another  effect of  the
tadpole   supergraphs of  Fig.\  \ref{f1}  is  the generation of terms
proportional to $F_{S}$, namely  to the auxiliary scalar  component of
$\widehat{S}$.  As a  consequence, the  effective renormalizable Higgs
superpotential of the MNSSM reads
\begin{equation}
W_{{\rm Higgs}}^{{\rm eff}}\ =\ \lambda \,\widehat{S}\widehat{H}
_{1}^{T}i\tau _{2}\widehat{H}_{2}\:+\:\xi _{F}M_{{\rm SUSY}}^{2}\widehat{S}\,,
\label{WHiggs}
\end{equation}
where $\xi _{F}$ is a model-dependent constant. Moreover, the Lagrangian
describing the soft SUSY-breaking Higgs sector is given by 
\begin{eqnarray}
-\,{\cal L}_{{\rm soft}} &=&\Big( \xi _{S}M_{{\rm SUSY}}^{3}S\ +\ {\rm h.c.}
\Big)\:+\:m_{1}^{2}\,\widetilde{\Phi }_{1}^{\dagger }\widetilde{\Phi }
_{1}\:+\:m_{2}^{2}\,\Phi _{2}^{\dagger }\Phi _{2}\:+\:m_{S}^{2}\,S^{*}S 
\nonumber  \label{softH} \\
&&+\,\Big( \lambda A_{\lambda }\,S\widetilde{\Phi }_{1}^{\dagger }i\tau
_{2}\Phi _{2}\ +\ {\rm h.c.}\Big)\,,
\end{eqnarray}
where $\widetilde{\Phi  }_{1}=i\tau _{2}\Phi _{1}^{*}$ and $\Phi _{2}$
are  the physical bosonic degrees of  freedom of $\widehat{H}_{1}$ and
$\widehat{H}  _{2}$,  respectively. After including the  relevant $F$-
and  $D$-term   contributions in addition  to  the  soft SUSY-breaking
terms, we obtain the   complete renormalizable Higgs potential of  the
model of interest
\begin{eqnarray}
-\,{\cal L}_{V}^{0} &=&\Big( t_{S}\,S\:+\:{\rm h.c.}\Big)\:+\:m_{1}^{2}\,\Phi
_{1}^{\dagger }\Phi _{1}\:+\:m_{2}^{2}\,\Phi _{2}^{\dagger }\Phi
_{2}\:+\:m_{S}^{2}\,S^{*}S\:+\:\Big( m_{12}^{2}\,\Phi _{1}^{\dagger }\Phi _{2}\:+\:
{\rm h.c.}\Big)  \nonumber  \label{LV} \\
&&+\,\Big( \lambda A_{\lambda }\,S\Phi _{1}^{\dagger }\Phi _{2}\:+\:{\rm h.c.}
\Big)\:-\:\lambda _{1}(\Phi _{1}^{\dagger }\Phi _{1})^{2}\:-\:\lambda
_{2}\,(\Phi _{2}^{\dagger }\Phi _{2})^{2}\:-\:\lambda _{3}\,(\Phi
_{1}^{\dagger }\Phi _{1})(\Phi _{2}^{\dagger }\Phi _{2})  \nonumber \\
&&-\,(\lambda _{4}-\lambda ^{2})\,(\Phi _{1}^{\dagger }\Phi _{2})(\Phi
_{2}^{\dagger }\Phi _{1})\:+\:\lambda ^{2}\,S^{*}S\Big( \Phi _{1}^{\dagger
}\Phi _{1}\:+\:\Phi _{2}^{\dagger }\Phi _{2}\Big)\,,
\end{eqnarray}
with 
\begin{eqnarray}
  \label{Lpar} 
t_{S}\! &=&\!\xi _{S}\,M_{{\rm SUSY}}^{3}\,,\qquad m_{12}^{2}\ =\ \lambda
\xi _{F}M_{{\rm SUSY}}^{2}\,,  \nonumber\\
\lambda _{1}\! &=&\!\lambda _{2}\ =\ -\,\frac{g_{w}^{2}+g^{\prime }{}^{2}}{8}
\,,\quad \lambda _{3}\ =\ -\,\frac{g_{w}^{2}-g^{\prime }{}^{2}}{4}\,,\quad
\lambda _{4}\ =\ \frac{g_{w}^{2}}{2}\ .
\end{eqnarray}
Here, $g_{w}$ ($g^{\prime }$) is  the coupling  constant of the  gauge
group  SU(2)$_{L}$ (U(1)$_{Y}$).  As   was discussed in the   previous
section,  the tadpole prefactor $\xi_{S}$  in Eq.\  (\ref{Lpar}) is of
order unity.  However, the size of  $\xi_{F}$ crucially depends on the
VEVs of the scalar   components of the hidden-sector superfields  that
break SUSY   \cite{Bag}.    The VEVs    remain unconstrained   by  the
requirement that the breaking of SUSY takes place at some intermediate
scale in  the hidden sector, in  which the $F$-terms of the respective
hidden-sector superfields  are involved.   In  case that  some  of the
hidden-sector fields acquire VEVs of order $M_{ {\rm P}}$, the tadpole
prefactors $\left| \xi_{F}\right|$  and $\left| \xi_{S} \right|$ could
be comparable.  Otherwise,  it is $\left|  \xi  _{F}\right| \ll \left|
\xi  _{S}\right|$. In the following, we  shall treat the ratio $\left|
\xi _{F}\right| /\left| \xi _{S}\right|$ as  a free parameter which is
always less than unity.

We shall now derive the minimization conditions of the Higgs potential
in Eq.\ (\ref{LV}). Throughout the paper, we shall assume that CP is a
good symmetry of the theory. Under this assumption, we can perform the
following linear expansions of the Higgs fields about their VEV's:
\begin{eqnarray}
\Phi _{1}\! &=&\!\left( 
\begin{array}{c}
\phi _{1}^{+} \\ 
\frac{1}{\sqrt{2}}\,(v_{1}\,+\,\phi _{1}\,+\,ia_{1})
\end{array}
\right) \,,\qquad \Phi _{2}\ =\ \left( 
\begin{array}{c}
\phi _{2}^{+} \\ 
\frac{1}{\sqrt{2}}\,(v_{2}\,+\,\phi _{2}\,+\,ia_{2})
\end{array}
\right) \,,  \nonumber  \label{Phi123} \\
S\! &=&\!\frac{1}{\sqrt{2}}\,\Big( v_{S}\,+\,\phi _{S}\,+\,ia_{S}\Big)\,.
\end{eqnarray}
The minimization conditions are then determined by the vanishing of the
tadpole parameters 
\begin{eqnarray}
  \label{Tphi1}
T_{\phi _{1}} &\equiv &\bigg<\,\frac{\partial {\cal L}_{V}}{\partial \phi
_{1}}\,\bigg>\ =\ -\,v_{1}\,\bigg[ \,m_{1}^{2}\ \ +\ 
\bigg(\,\frac{1}{\sqrt{2}}\,\lambda A_\lambda v_{S}\: +\: m_{12}^{2}\, 
\bigg)\,t_\beta \ -\ \lambda _{1}v_{1}^{2}  \nonumber\\
&&-\,\frac{1}{2}\,(\lambda _{3}+\lambda _{4}-\lambda ^{2})\,v_{2}^{2}\ +\ 
\frac{1}{2}\,\lambda ^{2}v_{S}^{2}\, \bigg]\,, \\
  \label{Tphi2}
T_{\phi _{2}} &\equiv &\bigg<\,\frac{\partial {\cal L}_{V}}{\partial \phi
_{2}}\,\bigg>\ =\ -\,v_{2}\,\bigg[\,m_{2}^{2}\ +\ 
\bigg(\, \frac{1}{\sqrt{2}}\,\lambda A_\lambda v_{S}\: +\: m_{12}^{2}\,\bigg)\,
t_{\beta }^{-1}\ -\ \lambda _{2}v_{2}^{2} 
\nonumber\\
&&-\,\frac{1}{2}\,(\lambda _{3}+\lambda _{4}-\lambda ^{2})\,v_{1}^{2}\ +\ 
\frac{1}{2}\,\lambda ^{2}v_{S}^{2}\, \bigg]\,,\\
  \label{Tphi3}
T_{\phi _{S}} &\equiv &\bigg<\,\frac{\partial {\cal L}_{V}}{\partial \phi
_{S}}\,\bigg>\ =\ -\,v_{S}\,\bigg(\, m_{S}^{2}\ +\ \lambda A_{\lambda }\,
\frac{v_{1}v_{2}}{\sqrt{2}\,v_{S}}\ +\ \frac{1}{2}\,\lambda ^{2}\,v^{2}\,\ +
\frac{\sqrt{2}\,t_{S}}{v_{S}}\,\bigg)\,,
\end{eqnarray}
with $v=\sqrt{v_1^2 + v_2^2} = 2 M_W / g_w$ and $t_\beta = v_2 / v_1$.
Our  earlier assumption of  CP  invariance entails  that all kinematic
parameters  involved, e.g.\  $\lambda$ and  $A_{\lambda  }$, are real,
namely there are no  explicit sources of  CP violation in the  theory. 
Also, it  is important to remark  that based on  Romao's no-go theorem
\cite{JCR}, CP  invariance cannot be broken  spontaneously at the tree
level in the  MNSSM.\footnote[2]{We find that this property  persists,
  even  if CP-conserving   radiative effects mediated   by large  stop
  mixing are included in our model.}

It proves now convenient to perform a change of the weak basis for the
charged and CP-odd scalars:
\begin{equation}  \label{rot}
\left( 
\begin{array}{c}
\phi^+_1 \\ 
\phi^+_2
\end{array}
\right)\ =\ \left( 
\begin{array}{cc}
c_\beta & -s_\beta \\ 
s_\beta & c_\beta
\end{array}
\right)\, \left( 
\begin{array}{c}
G^+ \\ 
H^+
\end{array}
\right)\, ,\qquad \left( 
\begin{array}{c}
a_1 \\ 
a_2
\end{array}
\right)\ =\ \left( 
\begin{array}{cc}
c_\beta & -s_\beta \\ 
s_\beta & c_\beta
\end{array}
\right)\, \left( 
\begin{array}{c}
G^0 \\ 
a
\end{array}
\right)\, ,
\end{equation}
where $s_\beta = v_2/v$ and $c_\beta = v_1/v$, such that $H^+$ becomes
the mass  eigenstate of the charged  Higgs boson, and  $G^+$ and $G^0$
are the  would-be Goldstone  bosons which constitute  the longitudinal
degrees of freedom of the $W^+$ and $Z$ bosons, respectively.

Let us first  consider the charged Higgs sector.  In the newly defined
weak basis  of Eq.\  (\ref{rot}), the tree-level  mass of  the charged
Higgs boson may easily be computed by
\begin{equation}
  \label{MH0plus}
M_{H^+}^{2(0)}\ =\ \frac{1}{s_\beta c_\beta }\,\Big(\,\mu A_\lambda
\: - \: m_{12}^{2}\,\Big)\ +\ M_W^2\ -\ \frac{1}{2}\,
\lambda^{2}\, v^2 \,,
\end{equation}
where 
\begin{equation}
\mu \ =\ -\,\frac{1}{\sqrt{2}}\,\lambda \,v_{S}  \label{mu}
\end{equation}
is the  would-be  $\mu$-parameter of  the  MSSM.    Here and  in   the
following,  we  adhere the  superscript $(0)$  to a specific kinematic
quantity    in   order to  emphasize    its  tree-level  origin, e.g.\ 
$M_{H^{+}}^{2(0)}$.

Since the  would-be Goldstone  boson $G^{0}$ does  not mix  with other
fields, the tree-level CP-odd mass  matrix takes on the simple form in
the reduced weak basis $\{a,\ a_{S}\}$:
\begin{equation}
  \label{MCPodd0}
M_{P}^{2(0)}\ =\ \left( 
\begin{array}{cc}
M_{a}^{2(0)} & \frac{\displaystyle v}{\displaystyle v_{S}}\Big(\,
s_\beta c_\beta\,M_{a}^{2(0)}\:+\:\,m_{12}^{2}\Big) \\[0.1cm] 
\frac{\displaystyle v}{\displaystyle v_{S}}\Big(\,s_\beta c_\beta\,
M_{a}^{2(0)}\:+\:\,m_{12}^{2}\Big) & \quad \frac{\displaystyle v^{2}}{
\displaystyle v_{S}^{2}}\,\,s_{\beta }c_{\beta }\Big(s_{\beta }c_{\beta
}\,M_{a}^{2(0)}\:+\: m_{12}^{2}\Big)\: +
\:\frac{\displaystyle \lambda\,t_S}{\displaystyle \mu}
\end{array}
\right) \,,  
\end{equation}
with 
\begin{equation}
  \label{Ma}
M_{a}^{2(0)}\ =\ M_{H^{+}}^{2(0)}-M_{W}^{2}+\frac{1}{2}\,\lambda ^{2}v^{2}\,.
\end{equation}
In deriving the above form of $M_{P}^{2(0)}$,  we have also considered
the  tadpole constraints given  by Eqs.\ (\ref{Tphi1})--(\ref{Tphi3}). 
In  the   MSSM  limit,     which   is obtained    for    $v_S  \approx
-\sqrt{2}\,t_S/m^2_S \gg v$   with the would-be  $\mu$-parameter being
kept fixed ($\lambda \to 0$), the mass eigenvalues  of the CP-odd mass
matrix can easily be approximated by
\begin{equation}
M_{A_{1}}^{2(0)}\ \approx \ M_{a}^{2(0)}\,,\qquad M_{A_{2}}^{2(0)}\ \approx
\ \frac{\lambda t_{S}}{\mu}\ ,  \label{MA012}
\end{equation}
with  $\lambda t_S/\mu > M^{2(0)}_a$.  Furthermore,  in  the limit, in
which the tadpole parameters $m_{12}^{2}$ and $t_S$ vanish, the CP-odd
mass matrix contains a massless  state, i.e.\ a PQ  axion, as a result
of the spontaneous breakdown of the symmetry group $U(1)_{ {\rm PQ}}$.

Taking    into   account    the   tadpole    constraints    of   Eqs.\
(\ref{Tphi1})--(\ref {Tphi3}), the  tree-level CP-even mass matrix may
be expressed in the weak basis $\{\phi _{1},\ \phi _{2},\ \phi _{S}\}$
as follows:
\begin{eqnarray}
  \label{CPeven0}
(M_S^{2(0)})_{11}\! &=&\!c_{\beta }^{2}\,M_{Z}^{2}\ +\ s_{\beta
}^{2}\,M_{a}^{2(0)}\,,  \nonumber\\
(M_S^{2(0)})_{12}\! &=&\!(M_{S}^{2(0)})_{21}\ =\ 
-\,s_{\beta }c_{\beta }\Big( M_{a}^{2(0)}\ +\ M_{Z}^{2}\ 
    -\ \lambda ^{2}v^{2}\,\Big)\,,  \nonumber \\
(M_S^{2(0)})_{13}\! &=&\!(M_{S}^{2(0)})_{31}\ =\ -\,\frac{v}{v_S}\, \Big( 
s_\beta^2 c_\beta\,M_{a}^{2(0)}\ -\ 2c_{\beta }\mu ^{2}\ +\ s_\beta 
m_{12}^2\,\Big)\,,  \nonumber \\
(M_S^{2(0)})_{22}\! &=&\!s_{\beta }^{2}\,M_{Z}^{2}\ +\ c_\beta^{2}\,
              M_{a}^{2(0)}\,,  \nonumber \\
(M_S^{2(0)})_{23}\! &=&\!(M_{S}^{2(0)})_{32}\ =\ -\,\frac{v}{v_{S}}\,\Big( 
s_{\beta }c_{\beta }^{2}\,M_{a}^{2(0)}\ -\ 2s_\beta \mu ^{2}\ +\
c_\beta m_{12}^2\,\Big)\,,  \nonumber \\
(M_S^{2(0)})_{33}\! &=&\!\frac{v^{2}}{v_{S}^{2}}s_{\beta }c_{\beta }\Big( 
\,s_{\beta }c_{\beta }\,M_{a}^{2(0)}\ +\ \,m_{12}^{2}\,\Big)\ 
+\ \frac{\lambda\,t_S}{\mu}\ ,
\end{eqnarray}
with $M_Z = \sqrt{g_w^2 +  g^\prime{}^2}\,v/2$.  In the MSSM limit, in
which $v_S\approx  -\sqrt{2}\,t_S/m^2_S \gg  v$ with $\mu$  fixed, the
Higgs-singlet  components decouple  from the  tree-level  CP-even mass
matrix $M_S^{2(0)}$.  In this case,  the heaviest Higgs boson $H_3$ is
predominantly singlet  and has a squared  mass $M_{H_3}^{2(0)} \approx
\lambda  t_S/\mu$; $H_{3}$  becomes mass  degenerate with  $A_2$ (cf.\ 
Eq.\ (\ref{MA012})).

Apart from  the MSSM  limit  mentioned  above, there exists  a   novel
non-trivial decoupling limit for the heavy Higgs sector  in the MNSSM. 
This  decoupling limit is obtained  for  large values  of the  tadpole
parameter  $|t_S|$,   where all other    kinematic parameters are kept
fixed.  In this  case, the Higgs  states $A_2$ and $H_3$ are singlets,
i.e.\ $A_2\equiv a_S$ and $H_3\equiv \phi_S$, and so decouple from the
remaining    Higgs  sector while   being  degenerate  in   mass, i.e.\ 
$M_{A_2}^{2(0)} \approx M_{H_3}^{2(0)}  \approx \lambda t_S/\mu$.   An
immediate consequence of this is the relation
\begin{equation}
  \label{tadec}
M^{2(0)}_{A_1}\  \approx\ M_{a}^{2(0)}\, ,
\end{equation}
where $M_{a}^{2(0)}$ is defined in Eq.\ (\ref{Ma}).  Most importantly,
in this limit  the structure of the low-energy  Higgs sector, although
reminiscent  of, is  {\em not}  identical to  that of  the  MSSM.  For
example,  as opposed  to the  MSSM  limit, the  terms proportional  to
$\lambda^2 v^2$, which  occur in the CP-odd and  CP-even mass matrices
of  Eqs.\  (\ref{MCPodd0})  and  (\ref{CPeven0}), do  not  necessarily
vanish in the decoupling limit due to a large tadpole.  Thus, contrary
to  the  MSSM, Eq.~(\ref{tadec})  implies  that  for  large values  of
$\lambda$,  e.g.\ $\lambda  \sim  g_w$, the  charged Higgs-boson  mass
$M^{(0)}_{H^+}$ can become even  smaller than the mass $M^{(0)}_{A_1}$
of the non-decoupled CP-odd scalar.  As we will see in Section 5, this
last fact  plays a very important  r\^ole in lowering the  mass of the
MNSSM charged Higgs boson up to its experimental lower bound, i.e.\ up
to $M_{H^+}  \sim 80$~GeV~\cite{ADLO,PDG}.  Moreover, in  Section 4 we
shall see  that this new non-trivial  decoupling limit due  to a large
tadpole parameter  $\lambda t_S/\mu$ is only attainable  in the MNSSM,
and no analogue of this exists in the NMSSM.

As in the  MSSM \cite{GT}, an upper bound  on the mass of the lightest
CP-even Higgs  boson  in the MNSSM  with  large $|t_S|$ may easily  be
derived in the decoupling limit of a heavy  charged Higgs boson, i.e.\ 
for
\begin{equation}  
  \label{Hpldec}
 \frac{\lambda\,t_S}{\mu}\ \gg\ M^{2(0)}_{H^+}\ \gg\ M^2_Z\, .
\end{equation}
In this limit, in addition to $A_2$ and $H_3$, the Higgs scalars $A_1$
and $H_2$  decouple from the lightest  Higgs  sector as  well, and are
almost  mass degenerate  with   the charged Higgs boson  $H^+$.  After
taking  into consideration the   heavy $H^+$-decoupling limit of  Eq.\ 
(\ref{Hpldec}), the mass of the lightest  CP-even Higgs state $H_1$ is
found to satisfy the inequality
\begin{equation}
  \label{M0H1}
M_{H_1}^{2(0)}\ \le \ M_Z^2\ \bigg(\, \cos^2 2\beta \: +\:
\frac{2\lambda^2}{g_w^2 + g^\prime{}^2}\,\sin^2 2\beta \,\bigg)\,.  
\end{equation}
Note that as  opposed to the  MSSM where $\lambda =0$, $M_{H_1}^{(0)}$
can be larger than $M_Z$ in the MNSSM, especially  for small values of
$\tan\beta$.  This  prediction is very similar to  the one obtained in
the frequently-discussed  NMSSM \cite{nmssm}.  However, as $\tan\beta$
increases,    e.g.\  for  $\tan\beta  \stackrel{>}{{}_\sim}   5$,  the
$\lambda$-dependent   term in   Eq.~(\ref{M0H1}) becomes   negligible. 
Thus,  in  the  large-$\tan\beta$  case,    the upper bound   on   the
$H_1$-boson mass is almost identical to the one obtained  in the MSSM. 
Finally,  an important  property of the  MNSSM is  that the tree-level
neutral Higgs-boson masses satisfy the equality
\begin{equation}
  \label{sumrule}
M_{H_{1}}^{2(0)}\ +\ M_{H_{2}}^{2(0)}\ +\ M_{H_{3}}^{2(0)}\ =\ M_{Z}^{2}\ +\
M_{A_{1}}^{2(0)}\ +\ M_{A_{2}}^{2(0)}\,.
\end{equation}
It is interesting to notice  the striking similarity of the above mass
sum rule  with the corresponding one  in the MSSM  \cite{GT}, in which
case  the mass  terms  $M_{H_3}^{2(0)}$ and  $M_{A_2}^{2(0)}$ are  not
present in  Eq.\ (\ref{sumrule}).  The above observation  allows us to
advocate that the  structure of the MNSSM Higgs  sector departs indeed
minimally from that of the  MSSM.  Nevertheless, exactly as happens in
the MSSM \cite{SHiggs}, the  tree-level Higgs sector receives sizeable
quantum corrections due to stop squarks, leading to a violation of the
mass sum rule (\ref{sumrule}).

\subsection{One-loop effective potential}

We  shall  now calculate  the  dominant  one-loop  corrections to  the
effective  potential due  to  top ($t$)  and scalar-top  ($\tilde{t}$)
quarks.   As a good  approximation, we  neglect the  one-loop $D$-term
contributions as well as  bottom ($b$) and scalar-bottom ($\tilde{b}$)
quark  effects  by assuming  a  vanishing  $b$-quark Yukawa  coupling,
i.e.~$h_b  =   0$.   The  above  approximations   are  reasonable  for
relatively  small values  of  $\tan\beta$, e.g.\  $\tan\beta \le  10$,
where the MNSSM predictions for the lightest Higgs sector are expected
to deviate considerably from the ones obtained in the MSSM.

The  interaction  Lagrangians  relevant  for the  computation  of  the
one-loop effective potential are given by
\begin{eqnarray}  \label{Lint}
-{\cal L}_{{\rm fermion}} &=& h_t \bar{Q}_L i\tau_2 \Phi^*_2 t_R\quad +
\quad {\rm h.c.},  \nonumber \\
-{\cal L}_F &=& h^2_t\, |\Phi^T_2 i\tau_2 \widetilde{Q}_L |^2\ +\ \Big(\,
\lambda h_t\, S \widetilde{Q}^\dagger_L i\tau_2 \Phi_1^\ast \widetilde{t}_R\
+\ {\rm h.c.}\,\Big)\ +\ h^2_t \tilde{t}_R \Phi^\dagger_2\Phi_2 \widetilde{t}
^*_R\,,  \nonumber \\
-{\cal L}_{{\rm soft}} &=& \widetilde{M}^2_Q\, \widetilde{Q}^\dagger_L 
\widetilde{Q}_L\ +\ 
\widetilde{M}^2_t\, \widetilde{t}^\ast_R \widetilde{t}_R\ +\ 
\Big( h_t A_t\, \widetilde{Q}^\dagger_L i\tau_2 \Phi^*_2\widetilde{t}_R\ 
+ \ {\rm h.c.}\Big)\,,\quad
\end{eqnarray}
where $\widetilde{Q}_L   = (\widetilde{t}_L,\  \widetilde{b}_L)^T$ and
$Q_L = (t_L,\ b_L)^T$ are the bosonic and fermionic degrees of freedom
of the third-generation left-handed quark superfield.

Equipped with the  Lagrangians in Eq.\ (\ref{Lint}), we  can now derive
the  Higgs-dependent $t$  and  $\tilde{t}$ masses.  Thus, the  squared
$t$-quark mass in the Higgs background is given by
\begin{equation}  \label{mtH}
\bar{m}^2_t\ =\ h^2_t\, \Phi^\dagger_2 \Phi_2\ .
\end{equation}
The corresponding  background-dependent stop masses  may be determined
from the  $3\times 3$  squark mass matrix,  which is expressed  in the
weak basis $\{ \widetilde{Q}_L,\ \widetilde{t}_R \}$ as:
\begin{equation}  
  \label{squark}
\widetilde{{\cal M}}^2\ =\ \left( 
\begin{array}{cc}
\widetilde{M}^2_Q\, {\bf 1}_2\: +\: h^2_t\, ( \Phi^\dagger_2\Phi_2 {\bf 1}_2
- \Phi_2 \Phi^\dagger_2 ) & \quad h_t A_t\, i\tau_2 \Phi^*_2\: +\: \lambda
h_t\, S i \tau_2 \Phi^*_1 \\ 
-h_t A_t\, \Phi^T_2 i\tau_2\: -\: \lambda h_t\, S^\ast \Phi^T_1 i \tau_2 & 
\widetilde{M}^2_t\: +\: h^2_t\, \Phi^\dagger_2 \Phi_2
\end{array}
\right)\, .
\end{equation}
The squared squark-mass matrix $\widetilde{{\cal M}}^2$ has three mass
eigenvalues.  For  $\phi^\pm_{1,2} =  0$, these are  given by  the two
squared Higgs-dependent $\tilde{t}$-quark masses
\begin{equation}  
  \label{mstop}
\widetilde{m}^2_{t_1\,(t_2)}\ =\ \frac{1}{2}\, \Big( \widetilde{M}^2_Q\, +\, 
\widetilde{M}^2_t\, +\, 2h^2_t\, |\phi^0_2|^2\ +(-)\ \sqrt{ 
(\widetilde{M}^2_Q\, -\, \widetilde{M}^2_t)^2\, +\, 4 h^2_t |A_t\,
\phi^0_2\, +\, 
\lambda\, S^\ast \phi^0_1 |^2 }\ \Big)\, \quad
\end{equation}
and  by  the squared  left-handed  sbottom  mass $m^2_{\tilde{b}_L}  =
\widetilde{M}^2_Q$,  where  $\phi^0_{1,2} =  (v_{1,2}  + \phi_{1,2}  +
ia_{1,2} )/\sqrt{2}$ are the neutral parts of $\Phi_{1,2}$.

In  the  $\overline{\rm MS}$  scheme,  the one-loop  Coleman--Weinberg
effective  potential  \cite{CW}  may  be  expressed in  terms  of  the
relevant     squared     Higgs-dependent     masses     $\bar{m}^2_t$,
$\widetilde{m}^2_{t_1}$,          $\widetilde{m}^2_{t_2}$          and
$\widetilde{m}^2_{b_L}$ as follows:
\begin{equation}  \label{LVeff}
-{\cal L}_V\ =\ -{\cal L}^0_V\, +\, \frac{3}{32\pi^2}\, \bigg[
\,\sum\limits_{k = t_1,t_2,b_L}\!\!\!\! \widetilde{m}^4_k\, 
\bigg( \ln \frac{\widetilde{m}^2_k}{Q^2}\, 
-\, \frac{3}{2}\,\bigg)\, -\, 2\bar{m}^4_t\, \bigg( \ln 
\frac{\bar{m}^2_t}{Q^2}\, -\, \frac{3}{2}\,\bigg)\, \bigg]\, ,
\end{equation}
where  $-{\cal L}^0_V$  is  the  bare Higgs  potential  given by  Eq.\ 
(\ref{LV}).  With  the help  of ${\cal L}_V$,  we can now  compute the
radiatively corrected mass of the  charged Higgs boson by means of the
relation
\begin{eqnarray}
M^2_{H^+} \!&=&\! \frac{1}{s_\beta c_\beta}\, \bigg< \,\frac{\partial^2 
{\cal L}_V}{\partial \phi^+_1\, \partial \phi^-_2 }\, \bigg>\ = \
M^{2(0)}_{H^+}\ +\ \Delta M^2_{H^+}  \nonumber \\
\!&=&\! M^{2(0)}_{H^+}\ -\, \frac{3}{16\pi^2\, s_\beta c_\beta}\,
\bigg[\,\sum\limits_{k = t_1,t_2,b_L}\!\!\!
 \bigg< \,\frac{\partial^2 \widetilde{m}^2_k }{
\partial \phi^+_1\, \partial \phi^-_2 }\, \bigg>\, m^2_{\tilde{k}}\, 
\bigg( \ln \frac{m^2_{\tilde{k}}}{Q^2}\, -\, 1\,\bigg)\, \bigg]\,,
\end{eqnarray}
where  $M^{2(0)}_{H^+}$ is  the tree-level  contribution  and $\langle
\widetilde{m}^2_k\rangle = m^2_{\tilde{k}}$ .  Following the procedure
outlined  in  \cite{CEPW},\footnote[3]{A  similar procedure  was  also
  followed in \cite{EKW}.} we find
\begin{eqnarray}  
  \label{derHplus}
\bigg< \,\frac{\partial^2 \widetilde{m}^2_{t_1 (t_2)} }{\partial \phi_1^+\,
\partial \phi^-_2 }\, \bigg> &=& -(+) \, 
\frac{h^2_t \mu A_t}{ m^2_{\tilde{t}_1} - m^2_{\tilde{t}_2} }\
+(-)\ 
\frac{h^4_t \mu^2 v_1 v_2}{ 2\,\big(m^2_{\tilde{t}_1(\tilde{t}_2)} 
- m^2_{\tilde{b}_L}\big)\,
\big( m^2_{\tilde{t}_1} - m^2_{\tilde{t}_2}\big) }\ ,\nonumber\\
\bigg< \,\frac{\partial^2 \widetilde{m}^2_{b_L} }{\partial \phi_1^+\,
\partial \phi^-_2 }\, \bigg> &=& 
\frac{h^4_t \mu^2 v_1 v_2}{ 2\,\big(m^2_{\tilde{t}_1} 
- m^2_{\tilde{b}_L}\big)\,
\big( m^2_{\tilde{t}_2} - m^2_{\tilde{b}_L}\big) }\ .
\end{eqnarray}
Then, the one-loop correction to $M^2_{H^+}$, $\Delta M^2_{H^+}$, is
given by
\begin{equation}  \label{DMHplus}
\Delta M^2_{H^+}\ =\ \frac{3 h^2_t \mu A_t}{32\pi^2\, s_\beta c_\beta}\, 
\bigg[\, \ln \bigg(\frac{m^2_{\tilde{t}_1} m^2_{\tilde{t}_2}}{Q^4}\bigg)\:
+\: g(m^2_{\tilde{t}_1},m^2_{\tilde{t}_2}) \,\bigg]\ + \ \delta_{\rm rem}\, ,
\end{equation}
with
\begin{equation}
  \label{g}
g(m^2_1,m^2_2)\ =\ \frac{m^2_1\, +\, m^2_2}{m^2_1\, -\, m^2_2}\, \ln \bigg(
\frac{m^2_1}{m^2_2}\bigg)\ -\ 2\, .
\end{equation}
In Eq.\  (\ref{DMHplus}), the quantity  $\delta_{\rm rem}$ summarizes
the remaining $Q^2$-independent corrections:
\begin{eqnarray}
  \label{drem}
\delta_{\rm rem} &=& \frac{3h^4_t \mu^2 v^2}{64\pi^2}\, \bigg[\,
\frac{m^2_{\tilde{b}_L}}{ \big(m^2_{\tilde{t}_1} 
- m^2_{\tilde{b}_L}\big)\,
\big( m^2_{\tilde{t}_2} - m^2_{\tilde{b}_L}\big) }\ 
\ln\bigg( \frac{m^2_{\tilde{t}_1}\,m^2_{\tilde{t}_2}}{
m^4_{\tilde{b}_L}}\bigg)\nonumber\\
&& -\ \frac{1}{m^2_{\tilde{t}_1} - m^2_{\tilde{t}_2}}\, \bigg(\,
\frac{m^2_{\tilde{t}_1}}{m^2_{\tilde{t}_1} 
- m^2_{\tilde{b}_L}}\ +\ \frac{m^2_{\tilde{t}_2}}{m^2_{\tilde{t}_2} 
- m^2_{\tilde{b}_L}}\, \bigg)\, \ln\bigg(
\frac{m^2_{\tilde{t}_1}}{m^2_{\tilde{t}_2}} \bigg)\, \bigg]\nonumber\\
&\approx& -\, \frac{3h^4_t}{32\pi^2}\, \frac{\mu^2 v^2}{
m^2_{\tilde{t}_1} + m^2_{\tilde{t}_2}}\ .
\end{eqnarray}
As we will see below, the presence of $\delta_{\rm rem}$ gives rise to
a  modification of  the  tree-level relation  between $M^2_{H^+}$  and
$M^2_a$ in  Eq.\ (\ref{Ma}).  Nevertheless,  it can be  estimated from
Eq.~(\ref{drem})  that this  modification, which  scales quadratically
with  the $\mu$-parameter,  is insignificant  for almost  all relevant
values of  $\mu$ of interest  to us, i.e.\  for $|\mu/m_{\tilde{t}_1}|
\stackrel{<}{{}_\sim} 2$.

We  now calculate the one-loop radiative  shift $\Delta  M^2_P$ to the
CP-odd Higgs-boson mass matrix. The analytic  result may be completely
expressed in terms of  $\Delta M^2_a =  \Delta M^2_{H^+} - \delta_{\rm
rem}$ as
\begin{eqnarray}  
  \label{DMP}
\Delta M^2_P\ =\ \Delta M^2_a\, \left( 
\begin{array}{cc}
1 & \quad \frac{\displaystyle v}{\displaystyle v_S}\, s_\beta c_\beta\\[0.1cm] 
\frac{\displaystyle v}{\displaystyle v_S}\, s_\beta c_\beta & \quad \frac{
\displaystyle v^2}{\displaystyle v^2_S}\, s^2_\beta c^2_\beta
\end{array}
\right)\, .
\end{eqnarray}
It is  easy to see that  the one-loop radiative shift  may be entirely
absorbed  into the  tree-level  mass matrix  in Eq.\  (\ref{MCPodd0}),
after performing  an one-loop  re-definition of $M^{2  (0)}_a$, namely
$M^2_a =  M^{2 (0)}_a +  \Delta M^2_a$. After this  re-definition, the
tree-level mass relation  in Eq.~(\ref{Ma}) gets radiatively corrected
as follows:
\begin{equation}
M^2_a\ =\ M^2_{H^+}\: -\: M^2_W\: +\: \frac{1}{2}\,\lambda^2 v^2\: -\:
\delta_{\rm rem}\, .
\end{equation}

The one-loop  Born-improved CP-odd mass matrix, $M^2_P  = M^{2(0)}_P +
\Delta   M^2_P$,   may   be   diagonalized   through   an   orthogonal
transformation of the weak fields
\begin{equation}  
 \label{OA}
\left( 
\begin{array}{c}
a \\ 
a_S
\end{array}
\right)\ =\ O^A\, \left( 
\begin{array}{c}
A_1 \\ 
A_2
\end{array}
\right)\,,\qquad {\rm with}\quad O^A\ =\ \left( 
\begin{array}{cc}
\cos\theta_A & \sin\theta_A \\ 
-\, \sin\theta_A & \cos\theta_A
\end{array}
\right)\, .
\end{equation}
The CP-odd fields $A_1$ and $A_2$ are the mass eigenstates of $M^2_P$, with
squared masses 
\begin{equation}  
  \label{MA12}
M^2_{A_1(A_2)}\ =\ \frac{1}{2}\, \Big(\, {\rm Tr}\, M^2_P\ - (+)\ \sqrt{{\rm 
Tr}^2\, M^2_P\: -\: 4\,{\rm det}\, M^2_P}\, \Big)\, .
\end{equation}
The mixing angle $\theta_A$ relating the weak to the mass eigenstates is
uniquely determined by 
\begin{equation}
\cos\theta_A\ =\ \frac{ |(M^2_P)_{12}|}{\sqrt{ (M^2_P)^2_{12}\: +\:
[(M^2_P)_{11} - M^2_{A_1} ]^2} }\ ,\quad \sin\theta_A\ =\ \frac{
|(M^2_P)_{11}\: -\: M^2_{A_1} |}{\sqrt{ (M^2_P)^2_{12}\: +\: [(M^2_P)_{11}
- M^2_{A_1} ]^2} }\ .\quad
\end{equation}

Finally, we calculate the radiative  corrections $\Delta M^2_S$ to the
  CP-even Higgs-boson mass matrix.   The individual matrix elements of
  $\Delta M^2_S$ are given by
\begin{eqnarray}  \label{DMS}
(\Delta M^2_S)_{11} \!&= &\! s^2_\beta\, \Delta M^2_a\ -\ \frac{3h^4_t
v^2_2}{16\pi^2}\, \frac{ \mu^2\, X^2_t}{(m^2_{\tilde{t}_1} - m^2_{\tilde{t}
_2})^2}\ g(m^2_{\tilde{t}_1},m^2_{\tilde{t}_2})\, ,  \nonumber \\
(\Delta M^2_S)_{12} \!&=&\! (\Delta M^2_S)_{21}\ =\ - s_\beta c_\beta\,
\Delta M^2_a\ -\ \frac{3 h^4_t v^2_2}{16\pi^2}\, \bigg[ \, \frac{\mu X_t
}{m^2_{\tilde{t}_1} - m^2_{\tilde{t}_2}}\, \ln \bigg(\frac{m^2_{\tilde{t}_1} 
}{m^2_{\tilde{t}_2}}\bigg)  \nonumber \\
&&-\, \frac{\mu A_t X^2_t}{(m^2_{\tilde{t}_1} - m^2_{\tilde{t}_2})^2}\
g(m^2_{\tilde{t}_1},m^2_{\tilde{t}_2})\, \bigg]\,,  \nonumber \\
(\Delta M^2_S)_{13} \!&=&\! (\Delta M^2_S)_{31}\ =\ -\,\frac{v}{v_S}\,
s^2_\beta c_\beta\, \bigg[\, \Delta M^2_a\ +\, \frac{3 h^4_t v^2}{16\pi^2
}\, \frac{\mu^2\, X^2_t}{(m^2_{\tilde{t}_1} - m^2_{\tilde{t}_2})^2}\ g(m^2_{
\tilde{t}_1},m^2_{\tilde{t}_2})\, \bigg]  \nonumber \\
&&+\, \frac{3 h^2_t}{16\pi^2}\, \bigg(\frac{v c_\beta}{v_S}\bigg)\, \mu^2\, 
\bigg[\ln \bigg(\frac{m^2_{\tilde{t}_1} m^2_{\tilde{t}_2}}{Q^4}\bigg)\: +\:
g(m^2_{\tilde{t}_1},m^2_{\tilde{t}_2}) \, \bigg]\, ,  \nonumber \\
(\Delta M^2_S)_{22} \!&=&\! c^2_\beta\, \Delta M^2_a\ +\ \frac{3 h^4_t
v^2_2}{16\pi^2}\, \bigg[ \, \ln \bigg(\frac{m^2_{\tilde{t}_1} m^2_{\tilde{t}
_2}}{m^4_t}\bigg)\ +\ \frac{2 A_t\, X_t}{m^2_{\tilde{t}_1} - m^2_{\tilde{t}
_2}}\, \ln \bigg(\frac{m^2_{\tilde{t}_1} }{m^2_{\tilde{t}_2}}\bigg) 
\nonumber \\
&&-\, \frac{A^2_t X^2_t}{(m^2_{\tilde{t}_1} - m^2_{\tilde{t}_2})^2}\ g(m^2_{
\tilde{t}_1},m^2_{\tilde{t}_2})\, \bigg]\, ,  \nonumber \\
(\Delta M^2_S)_{23} \!&=&\! (\Delta M^2_S)_{32}\ =\ -\,\frac{v}{v_S}\,
s_\beta c^2_\beta\, \bigg[\, \Delta M^2_a\ +\ \frac{3 h^4_t v^2}{16\pi^2}
\, \frac{t_\beta \mu\, X_t}{m^2_{\tilde{t}_1} - m^2_{\tilde{t}_2}}\ \ln 
\bigg(\frac{m^2_{\tilde{t}_1} }{m^2_{\tilde{t}_2}}\bigg)  \nonumber \\
&&-\, \frac{3 h^4_t v^2}{16\pi^2}\, \frac{t_\beta \mu A_t \, X^2_t}{(m^2_{
\tilde{t}_1} - m^2_{\tilde{t}_2})^2}\ g(m^2_{\tilde{t}_1},m^2_{\tilde{t}
_2})\, \bigg]\,,  \nonumber \\
(\Delta M^2_S)_{33} \!&=&\! \frac{v^2}{v^2_S}\, s^2_\beta c^2_\beta\, \bigg[
\, \Delta M^2_a\ -\ \frac{3 h^4_t v^2}{16\pi^2}\, \frac{\mu^2\, X^2_t}{
(m^2_{\tilde{t}_1} - m^2_{\tilde{t}_2})^2}\ g(m^2_{\tilde{t}_1},m^2_{\tilde{t
}_2})\, \bigg]\, ,
\end{eqnarray}
where $X_t  = A_t -  \mu / t_\beta$.   Again, we find that  almost the
entire  $Q^2$-dependence  of  the radiatively-corrected  CP-even  mass
matrix given  in Eq.\ (\ref{DMS}) can  be absorbed into  $M^2_a$ by an
one-loop re-definition  of $M^{2(0)}_a$. An  exception to this  is the
mass-matrix  element $(\Delta  M^2_S)_{13}$.  The  $Q^2$-dependence of
the $\{ 13\}$ element  can be eliminated by the $\Phi_2$-wave-function
counter term (CT) which is contained in the $\lambda$-parameter.

To   make   this   last   point   explicit,   we   shall   apply   the
non-renormalization  theorem  of the  superpotential  to the  coupling
$\lambda\, \widehat{S} \widehat{H}^T_1  i\tau_2 \widehat{H}_2$ in Eq.\ 
(\ref{WHiggs}). Since this operator  does not receive any ultra-violet
(UV) infinite radiative corrections  to all orders, the wave-functions
of  $\widehat{S}$,  $\widehat{H}_1$  and $\widehat{H}_2$,  denoted  as
$Z_{\widehat{S}}$,  $Z_{\widehat{H}_1}$  and $Z_{\widehat{H}_2}$  must
cancel against the CT of $\lambda$, $\delta \lambda$, that is
\begin{equation}  \label{dlambda}
\delta \lambda\ =\ \Big(\,Z^{-1/2}_{\widehat{S}} Z^{-1/2}_{\widehat{H}_1}
Z^{-1/2}_{\widehat{H}_2}\: -\: 1\,\Big)\, \lambda\ =\ -\frac{1}{2}\, \Big(\,
\delta Z_{\widehat{S}}\: +\: \delta Z_{\widehat{H}_1}\: +\: \delta Z_{
\widehat{H}_2}\, \Big)\, \lambda\,
\end{equation}
where $Z^{1/2}_z = 1 + \frac{1}{2}\delta Z_z$, with $z = \widehat{S},\
\widehat{H}_1,$ and $\widehat{H}_2$. Since only the wave-function of $
\widehat{H}_2$ receives  quantum corrections  due to top  quarks, Eq.\
(\ref {dlambda}) becomes
\begin{equation}  \label{dl}
\delta \lambda\ =\ -\frac{1}{2}\, \delta Z_{\widehat{H}_2}\, \lambda\ =\ -\,
\frac{3 \lambda h^2_t}{32\pi^2}\, \ln \bigg(\frac{ m^2_t }{Q^2}\bigg)\ .
\end{equation}
Here, we   have  implicitly assumed  that  the  coupling $\lambda$  is
renormalized at the  scale $Q^2 = m^2_t$.  Returning  now  to the bare
Higgs   potential in  Eq.\   (\ref{LV}),   we  see that  the  operator
$\lambda^2\, (S^* S) (\Phi^\dagger_1 \Phi_1)$ induces the CT $2\lambda
\delta  \lambda$,   which gives  rise  to a   corresponding CT  in the
tree-level mass-matrix element $(M^2_S)_{13}$,
\begin{equation}  
  \label{dMS13}
(\delta M^2_S)_{13}\ =\ 4\, \bigg(\frac{v c_\beta}{v_S}\bigg)\, 
\bigg( \frac{\delta \lambda}{\lambda}\bigg)\, \mu^2\ =\ 
-\,\frac{3 h^2_t}{8\pi^2}\, \bigg( \frac{v c_\beta}{v_S}\bigg)\, 
\mu^2\, \ln \bigg(\frac{ m^2_t }{Q^2}\bigg)\, .
\end{equation}
Adding  the CT $(\delta M^2_S)_{13}$   to the one-loop result $(\Delta
M^2_S)_{13}$, we readily see  that $Q^2$ gets  substituted by $m^2_t$. 
Finally, it is not difficult  to convince ourselves  that there are no
analogous   $\delta   \lambda$-dependent   CTs  for    the   operators
$\lambda^2\,  (S^*  S)   (\Phi^\dagger_2  \Phi_2)$  and   $\lambda^2\,
(\Phi^\dagger_1 \Phi_1) (\Phi^\dagger_2  \Phi_2)$, as they are exactly
canceled by the wave-function renormalization of $\Phi_2$.

The  one-loop  radiatively corrected  CP-even  mass  matrix, $M^2_S  =
M^{2(0)}_S + \Delta M^2_S$, is  diagonalized by means of a $3\times 3$
orthogonal matrix $O^H$, i.e.\
\begin{equation}  
  \label{MH123}
(O^H)^T\, M^2_S\, O^H\ =\ {\rm diag}\, \Big(\, M^2_{H_1},\ M^2_{H_2},\
M^2_{H_3}\,\Big)\, ,
\end{equation}
with $M^2_{H_1}  \le M^2_{H_2}  \le M^2_{H_3}$. Under  this orthogonal
transformation, the  weak states are  related to the  mass eigenstates
through
\begin{equation}  
 \label{OH}
\left( 
\begin{array}{c}
\phi_1 \\ 
\phi_2 \\ 
\phi_S
\end{array}
\right)\ =\ O^H\, \left( 
\begin{array}{c}
H_1 \\ 
H_2 \\ 
H_3
\end{array}
\right)\, .
\end{equation}
The entries  of $O^H$  can be calculated  analytically by  solving the
third-order  characteristic  equation  of  $M^2_S$. The  procedure  of
deriving  analytic  expressions for  the  elements  of  $O^H$ is  very
similar to the one presented in Appendix B of \cite{CEPW}, and we will
not repeat it here.

\subsection{The Higgsino sector}

In addition to  the Higgs sector, the Higgsino  (or neutralino) sector
of the MSSM  gets minimally extended in the MNSSM  due to the presence
of the  neutral SUSY  partner of the  complex scalar singlet  $S$, the
singlino  $\tilde{s}$.   Instead, the  tree-level  chargino sector  is
identical to that of the MSSM. In the weak basis
\begin{equation}  \label{Psi}
\Psi^T_0\ =\ \Big( \widetilde{B},\ \widetilde{W}_3,\ \tilde{h}_1,\ 
\tilde{h}_2\,,\ \tilde{s}\, \Big)\,,
\end{equation}
the Lagrangian describing the neutralino mass matrix in the MNSSM is given
by 
\begin{equation}  \label{Lneutr}
{\cal L}^0_{{\rm mass}}\ =\ -\,\frac{1}{2}\, \Psi^T_0 {\cal M}_0 \Psi_0\
+\ {\rm h.c.}\,,
\end{equation}
where 
\begin{equation}  
  \label{M0}
{\cal M}_0\ =\ \left(\! 
\begin{array}{ccccc}
m_{\widetilde{B}} & 0 & - M_Z s_w c_\beta & M_Z s_w s_\beta & 0 \\ 
0 & m_{\widetilde{W}} & M_Z c_w c_\beta & -M_Z c_w s_\beta & 0 \\ 
-M_Z s_w c_\beta & M_Z c_w c_\beta & 0 & -\mu & -\frac{\displaystyle v}{
\displaystyle v_S}\, s_\beta \mu \\ 
M_Z s_w s_\beta & -M_Z c_w s_\beta & -\mu & 0 & -\frac{\displaystyle v}{
\displaystyle v_S}\, c_\beta \mu \\ 
0 & 0 & -\frac{\displaystyle v}{\displaystyle v_S}\, s_\beta \mu & 
- \frac{\displaystyle v}{\displaystyle v_S}\, c_\beta \mu & 0
\end{array}
\!\right)\, ,
\end{equation}
with  $c_w  =  \sqrt{1  -  s^2_w} =  M_W/M_Z$.  In  Eq.\  (\ref{Psi}),
$\widetilde{B}$ and  $\widetilde{W}_3$ are the  U(1)$_Y$ and SU(2)$_L$
neutral gauginos,  respectively, and $\tilde{h}_1$,  $\tilde{h}_2$ and
$\tilde{s}$  are  the  corresponding  Higgsino states  of  the  chiral
multiplets $\widehat{H}_1$, $ \widehat{H}_2$ and $\widehat{S}$.

The  neutralino mass  matrix of  the  MNSSM given  in Eq.\  (\ref{M0})
predicts  a relatively light  state, with  mass smaller  than 70  GeV. 
Since  the  neutralino  mass  matrix  is  identical  to  that  of  the
PQ-symmetric extension  of the  MSSM, we call  this light  state axino
$\tilde{a}$. The axino is predominantly  a singlet field for values of
the  $\mu$-parameter in the  phenomenologically relevant  range, i.e.\ 
for $|\mu | \stackrel{>}{{}_\sim} 120$  GeV.  In order to have a first
estimate of the axino mass, we assume that the gaugino mass parameters
$m_{\widetilde{B}}$ and  $m_{\widetilde{W}}$ are very  large, e.g.\ of
order 500 GeV and higher,  such that the gauginos $ \widetilde{B}$ and
$\widetilde{W}_3$ decouple practically from the neutralino sector. The
reduced $3\times  3$ Higgsino-mass matrix,  which is expressed  in the
subspace spanned by $\tilde{h}_1$, $\tilde{h}_2$ and $ \tilde{s}$, can
then be expanded in terms of $v/v_S$, thus yielding the axino mass
\begin{equation}  \label{axino}
m_{\tilde{a}}\ \approx\ \frac{v^2}{v^2_S}\, |\mu\, \sin 2\beta |\ =\ \frac{
2\lambda^2}{g^2_w}\, \frac{M^2_W}{|\mu |}\, |\!\sin 2\beta |\, .
\end{equation}
This  last formula  proves  to be  a  good approximation  for $|\mu  |
\stackrel{>}{ {}_\sim} 200$ GeV.

There  are strict  collider  limits on  the axino-related  parameters,
which come  from LEP2 and especially  from the invisible  width of the
$Z$ boson \cite{PDG}, in which  case a new invisible decay channel for
the  $Z$   boson  into  axino   pairs  opens  up   kinematically  when
$m_{\tilde{a}}   \stackrel{<}{{}_\sim}  M_Z/2$.   Assuming   that  the
gauginos are decoupled from the  neutralino mass matrix $ {\cal M}_0$,
we find  numerically that the  axino mass is  smaller than 45  GeV for
values of $|\mu |  \stackrel{>}{{}_\sim} 150$~GeV and $\lambda \approx
g_w  \approx 0.65$.  Of  course, such  a numerical  estimate crucially
depends on the  values of $m_{\widetilde{B}}$ and $m_{\widetilde{W}}$.
For  example, for  relatively  low values  of $m_{\widetilde{B}}$  and
$m_{\widetilde{W}}$ in the range 200--300 GeV, gaugino-Higgsino mixing
effects  can  no longer  be  neglected, and  the  upper  limit on  the
$\mu$-parameter is estimated to increase by 40--50 GeV.

On  the other  hand, the  upper bound  on the  branching ratio  of the
$Z$-boson invisible width due to  a new-physics decay mode imposes the
constraint~\cite {PDG}
\begin{equation}  
  \label{Zinv}
B( Z \to \tilde{a}\tilde{a} )\ =\ \frac{\alpha_w}{24\, c^2_w}\, \frac{M_Z}{
\Gamma_Z}\, |g_{\tilde{a}\tilde{a}Z}|^2\ <\ 1.\times 10^{-3}\,,
\end{equation}
at the 90\% CL, where  $\alpha_w = g^2_w/(4\pi)$ is the SU(2)$_L$ weak
fine structure constant  and $\Gamma_Z = 2.49$ GeV  is the total width
of  the $Z$  boson.  Moreover, in  the  seesaw-type approximation  the
$\tilde{a}\tilde{a}Z$ -coupling is readily found to be
\begin{equation}  \label{aaZ}
g_{\tilde{a}\tilde{a}Z}\ \approx\ \frac{v^2}{v^2_S}\, (s^2_\beta -
c^2_\beta)\ =\ \frac{2\lambda^2}{g^2_w}\, \frac{M^2_W}{\mu^2}\, (s^2_\beta -
c^2_\beta)\,.
\end{equation}
The constraint  in Eq.\ (\ref{Zinv}), together  with Eq.\ (\ref{aaZ}),
leads to
\begin{equation}  
  \label{mulimit}
\frac{2\lambda^2}{g^2_w}\, \frac{M^2_W}{\mu^2}\,  |\!\cos 2\beta |\ <\
0.122\ .
\end{equation}
This last inequality can be translated into the following bound on the
$\mu$-parameter:
\begin{equation}  \label{muexp}
|\mu |\ \stackrel{>}{{}_\sim}\ 250\ {\rm GeV}\,,
\end{equation}
for  $\lambda  \approx  g_w$  and  $\tan\beta \approx  2$.  The  above
exercise  shows that in  the MNSSM  the LEP  limits on  the $Z$-boson
invisible width  give rise to a  new exclusion range  of $\mu$ values:
$200 \stackrel{<}{{}_\sim} |\mu  | \stackrel{<}{{}_\sim} 250$ GeV, for
$\lambda \approx  0.65$. However,  this additional exclusion  range of
$\mu$  exhibits a quadratic  dependence on  $ \lambda$  and completely
disappears for values of $\lambda \stackrel{<}{ {}_\sim} 0.45$.

\subsection{Effective Higgs-boson couplings}

Apart  from the  Higgs-boson masses,  the effective  couplings  of the
CP-even and  CP-odd Higgs  scalars to the  $W^\pm$ and $Z$  bosons are
very essential for our phenomenological discussion in Section 5. These
effective couplings are given by the interaction Lagrangians
\begin{eqnarray}  
  \label{HAZ}
{\cal L}_{{\rm HVV}} &=& g_w\, M_W\, \sum_{i = 1}^3\, g_{H_iVV}\, \Big(\,
H_i W^+_\mu W^{-, \mu}\ +\ \frac{1}{c^2_w}\, H_i Z_\mu Z^\mu\, \Big)\, , \\
{\cal L}_{HAZ} &=& \frac{g_w}{2\,c_w}\, \sum\limits_{i=1}^3\,
\sum\limits_{j=1}^2\, g_{H_iA_jZ}\, ( H_i\, \!\! \stackrel{\leftrightarrow}{
\vspace{2pt}\partial}_{\!\mu} A_j )\,Z^\mu\, ,
\end{eqnarray}
where   $\stackrel{\leftrightarrow}{\vspace{2pt}  \partial}_{\!  \mu}\
\equiv\   \stackrel{\rightarrow}{\vspace{2pt}  \partial}_{\!   \mu}  -
\stackrel{\leftarrow }{\vspace{2pt} \partial}_{\! \mu}$ and
\begin{eqnarray}  
 \label{gHHZ}
g_{H_iVV} &=& c_\beta\, O^H_{1i}\ +\ s_\beta\, O^H_{2i}\,, \\
 \label{gHAZ}
g_{H_i A_j Z} &=& O^A_{1j}\, \Big(\, c_\beta\, O^H_{2i}\ -\ s_\beta\,
O^H_{1i}\, \Big)\, .
\end{eqnarray}
Here,  we wish to  remind the reader that  the orthogonal matrix $O^A$
($O^H$)  is related to the mixing  of the CP-odd (CP-even) scalars and
is defined in Eq.\ (\ref{OA}) ((\ref{OH})).

It is now worth remarking  that the effective couplings $H_i VV$ (with
$V=Z,W$ ) and $H_i A_j Z$ satisfy the unitarity relations \cite{JFG}
\begin{equation}  
  \label{unit}
\sum_{i = 1}^3\, g^2_{H_iVV}\ =\ 1\, ,\qquad \sum_{i = 1}^3\,\sum_{j =
1}^2\, g^2_{H_iA_j Z}\ =\ 1\, .
\end{equation}
In particular,  in  the limit in  which $A_2$  and  $H_3$  decouple as
singlets, which  is  obtained  for large  $|t_S|$   with the remaining
parameters kept fixed,  one recovers  the  known MSSM  complementarity
relations among the effective Higgs-boson couplings \cite{GT}:
\begin{equation}
  \label{compl}
g^2_{H_1VV}\ =\ g^2_{H_2A_1Z}\quad {\rm and}\quad 
g^2_{H_2VV}\ =\ g^2_{H_1A_1Z}\, .
\end{equation}
As an obvious consequence of the above  decoupling limit, all couplings
of the heavy Higgs scalars $A_2$  and $H_3$ to the  $W$ and $Z$ bosons
go to zero.

Another very important relation which involves the CP-even Higgs-boson
masses and the respective couplings to the $W$ and $Z$ bosons is
\begin{eqnarray}  
  \label{sumcoupl}
\sum_{i = 1}^3\, g^2_{H_iVV}\, M^2_{H_i} &=& c^2_\beta\,(M^2_S)_{11}\
+\ 2s_\beta c_\beta\, (M^2_S)_{12}\ +\ s^2_\beta (M^2_S)_{22}\nonumber\\
&=& M^2_Z\ \bigg(\, \cos^2 2\beta\:
+\: \frac{2\lambda^2}{g^2_w + g^{\prime 2}}\, \sin^2 2\beta\, \bigg) 
\nonumber \\
&&\hspace{-1.5cm}+\, \frac{3h^4_t v^2 s^4_\beta}{16\pi^2}\, \bigg[\, \ln
\bigg(\frac{m^2_{\tilde{t}_1} m^2_{\tilde{t}_2}}{m^4_t}\bigg)\: +\: \frac{
2X^2_t}{m^2_{\tilde{t}_1} - m^2_{\tilde{t}_2} }\, \ln\bigg(\frac{m^2_{\tilde{
t}_1}}{m^2_{\tilde{t}_2}}\bigg)\: -\: \frac{X^4_t\, g(m^2_{\tilde{t}_1},
m^2_{\tilde{t}_2})} {(m^2_{\tilde{t}_1} - m^2_{\tilde{t}_2})^2 }\, \bigg]\,
.\qquad
\end{eqnarray}
This mass-coupling  sum rule is very  analogous to the  one derived in
\cite{MOQ} for the MSSM, where the RHS of Eq.\ (\ref{sumcoupl}) is the
squared lightest Higgs-boson  mass in the decoupling limit  of a heavy
charged Higgs  boson (see also  Eq.\ (\ref{Hpldec})).  In  this limit,
only the  $H_1$ boson has non-vanishing  couplings to the  $W$ and $Z$
bosons  \cite{GT}.  As  can be  seen from  Eq.\  (\ref{sumcoupl}), the
mass-coupling sum rule is independent of the charged Higgs boson mass,
while  it only weakly  depends on  $\mu$ at  the one-loop  order.  The
relations (\ref{unit}) and (\ref{sumcoupl}), which are obviously valid
for  the case of  the NMSSM  as well,  are very  useful to  reduce the
number of independent effective Higgs-boson couplings and so achieve a
better control on the numerical predictions for the Higgs-boson masses
and couplings.

\setcounter{equation}{0}

\section{MNSSM versus NMSSM}

Here,  we shall compare  the generic  predictions for  the Higgs-boson
mass   spectrum    in   the   NMSSM,   which    includes   the   cubic
singlet-superfield  coupling, with  those obtained  in the  MNSSM. For
this purpose, we  shall only focus on the  tree-level structure of the
Higgs sector of the  NMSSM, as the dominant stop/top-radiative effects
are  identical for  both  models  and have  already  been computed  in
Section 3.2.

The often-discussed NMSSM is based on the Higgs superpotential 
\begin{equation}
  \label{Wprime}  
W_{{\rm Higgs}}\ =\ \lambda \,\widehat{S}\widehat{H}_{1}^{T}i\tau _{2}
\widehat{H}_{2}\:+\:\frac{\kappa }{3}\,\widehat{S}^{3}\,.  
\end{equation}
As  usual, the  complete  tree-level Higgs  potential  is obtained  by
adding  the  relevant $F$-  and  $D$-term  contributions  to the  soft
SUSY-breaking terms induced by the superpotential:
\begin{eqnarray}
  \label{LVprime} 
-\,{\cal L}_{V}^{0} &=&m_{1}^{2}\,\Phi _{1}^{\dagger }\Phi
_{1}\:+\:m_{2}^{2}\,\Phi _{2}^{\dagger }\Phi
_{2}\:+\:m_{S}^{2}\,S^{*}S\:+\:\Big( \lambda A_{\lambda }\,S\Phi
_{1}^{\dagger }\Phi _{2}\:+\:{\rm h.c.}\Big)\:+\: \Big(\,\frac{\kappa
}{3}A_{\kappa }\,S^{3}\:+\:{\rm h.c.}\Big) \nonumber\\
&&-\,\lambda _{1}(\Phi _{1}^{\dagger }\Phi _{1})^{2}\:-\:\lambda _{2}\,(\Phi
_{2}^{\dagger }\Phi _{2})^{2}\:-\:\lambda _{3}\,(\Phi _{1}^{\dagger }\Phi
_{1})(\Phi _{2}^{\dagger }\Phi _{2})\:-\:(\lambda _{4}-\lambda ^{2})\,(\Phi
_{1}^{\dagger }\Phi _{2})(\Phi _{2}^{\dagger }\Phi _{1})  \nonumber \\
&&+\,\lambda ^{2}\,S^{*}S\Big( \Phi _{1}^{\dagger }\Phi _{1}\:+\:\Phi
_{2}^{\dagger }\Phi _{2}\Big)\:+\:\kappa ^{2}\,(S^{*}S)^{2}\:+\:\Big[\,\lambda
\kappa \,S^{*2}(\Phi _{1}^{\dagger }\Phi _{2})\:+\:{\rm h.c.}\,\Big]\,.
\end{eqnarray}
Furthermore, the minimization conditions are determined by requiring that
the following tadpole parameters vanish: 
\begin{eqnarray}
  \label{TSprime1}
T_{\phi _{1}} &\equiv &\bigg<\,\frac{\partial {\cal L}_{V}}{\partial \phi
_{1}}\,\bigg>\ =\ -\,v_{1}\,\bigg[ \,m_{1}^{2}\ +\ 
\bigg(\, \frac{1}{\sqrt{2}}\,\lambda A_\lambda v_S\: +\: 
\frac{1}{2}\,\lambda \kappa\,v_S^2\, \bigg)\,t_\beta\ -\ 
\lambda_1 v_1^2\nonumber\\
&&-\, \frac{1}{2}\,(\lambda_3 + \lambda_4 - \lambda^2)\, v_2^2\ 
+\ \frac{1}{2}\,\lambda^2 v_S^2\,\bigg]\,, \\
  \label{TSprime2}
T_{\phi _{2}} &\equiv &\bigg<\,\frac{\partial {\cal L}_{V}}{\partial \phi
_{2}}\,\bigg>\ =\ -\,v_{2}\,\bigg[\,\,m_{2}^{2}\ +\ 
\bigg(\, \frac{1}{\sqrt{2}}\,\lambda A_\lambda v_S\: +\: 
\frac{1}{2}\,\lambda \kappa\,v_S^2\, \bigg)\,t^{-1}_\beta\ -\ 
\lambda_2 v_2^2\nonumber\\
&& -\, \frac{1}{2}\,(\lambda_3 + \lambda_4 - \lambda^2)\,v_1^2\ +\ 
\frac{1}{2}\,\lambda^2 v_S^2\,\bigg]\,, \\
  \label{TSprimeS}
T_{\phi _{S}} &\equiv &\bigg<\,\frac{\partial {\cal L}_{V}}{\partial \phi
_{S}}\,\bigg>\ =\ -\,v_{S}\,\bigg(\,m_{S}^{2}\ +\ \lambda A_{\lambda }\,
\frac{v_{1}v_{2}}{\sqrt{2}\,v_{S}}\ +\ \frac{1}{2}\,\lambda ^{2}\,v^{2}\ \
+\ \kappa A_{\kappa }\,\frac{v_{S}}{\sqrt{2}}\ +\ \kappa ^{2}\,v_{S}^{2} 
\nonumber \\
&&+\,\lambda \kappa \,v_{1}v_{2}\,\bigg)\,.
\end{eqnarray}
We should remark again that spontaneous CP violation  is absent in the
NMSSM at the tree level \cite{JCR,BB}. Also, CP  appears to be still a
good symmetry of  the NMSSM, even  if (CP-conserving) large  radiative
stop-mixing effects were to be taken into account \cite{BB}.

Considering  the  vanishing of the tadpole   parameters given in Eqs.\ 
(\ref{TSprime1})--(\ref{TSprimeS}), it is not difficult to compute the
charged Higgs-boson  mass, and the CP-odd and   CP-even mass matrices. 
More explicitly, the squared charged Higgs-boson mass is given by
\begin{equation}
  \label{MH++} 
M_{H^{+}}^{2(0)}\ =\ \frac{1}{s_{\beta }c_{\beta
}}\,\bigg(\,\mu A_{\lambda }\:-\: \frac{\kappa }{\lambda }\,\mu
^{2}\,\bigg)\ +\ M_{W}^{2}\ -\ \frac{1}{2} \,\lambda ^{2}\,v^{2}\,,
\end{equation}
where the would-be $\mu$-parameter  is defined in Eq.\ (\ref{mu}). The
entries of the tree-level  CP-odd mass matrix $M_{P}^{2(0)}$ are found
to be
\begin{eqnarray}
(M_{P}^{2(0)})_{11} &=&M_{a}^{2(0)}\,,  \nonumber  \label{CPoddprime} \\
(M_{P}^{2(0)})_{12} &=&(M_{P}^{2(0)})_{21}\ =\ \frac{v}{v_{S}}\,\bigg(
\,s_{\beta }c_{\beta }\,M_{a}^{2(0)}\:+\:3\,\frac{\kappa }{\lambda }\,\mu
^{2}\,\bigg) \,,  \nonumber \\
(M_{P}^{2(0)})_{22} &=&\frac{v^{2}}{v_{S}^{2}}\,s_{\beta }c_{\beta }\,\bigg(
\,s_{\beta }c_{\beta }\,M_{a}^{2(0)}\:-\:3\,\frac{\kappa }{\lambda }\,\mu
^{2}\,\bigg)\:+\:3\,\frac{\kappa }{\lambda }\mu A_{\kappa }\,,
\end{eqnarray}
where $M_{a}^{2(0)}$ is given by Eq.~(\ref{Ma}).  Finally, the entries
of CP-even mass matrix $M_{S}^{2(0)}$ read
\begin{eqnarray}
(M_{S}^{2(0)})_{11}\! &=&\!c_{\beta }^{2}\,M_{Z}^{2}\ +\ s_{\beta
}^{2}\,M_{a}^{2(0)}\,,  \nonumber  \label{CPevenprime} \\
(M_{S}^{2(0)})_{12}\! &=&\!(M_{S}^{2(0)})_{21}\ =\ 
-\,s_{\beta }c_{\beta }\Big( M_{a}^{2(0)}\ +\ M_{Z}^{2}\ 
                                 -\ \lambda ^{2}v^{2}\,\Big)\,,  \nonumber \\
(M_{S}^{2(0)})_{13}\! &=&\!(M_{S}^{2(0)})_{31}\ =\ -\,\frac{v}{v_{S}}\,\Big( 
s_{\beta }^{2}c_{\beta }\,M_{a}^{2(0)}\ -\ 2c_{\beta }\mu ^{2}\ -\ \frac{
\kappa }{\lambda }\,s_{\beta }\mu ^{2}\,\Big)\,,  \nonumber \\
(M_{S}^{2(0)})_{22}\! &=&\!s_{\beta }^{2}\,M_{Z}^{2}\ +\ c_{\beta
}^{2}\,M_{a}^{2(0)}\,,  \nonumber \\
(M_{S}^{2(0)})_{23}\! &=&\!(M_{S}^{2(0)})_{32}\ =\ -\,\frac{v}{v_{S}}\,\Big( 
s_{\beta }c_{\beta }^{2}\,M_{a}^{2(0)}\ -\ 2s_{\beta }\mu ^{2}\ -\ \frac{
\kappa }{\lambda }\,c_{\beta }\mu ^{2}\,\Big)\,,  \nonumber \\
(M_{S}^{2(0)})_{33}\! &=&\!\frac{v^{2}}{v_{S}^{2}}\,s_{\beta }c_{\beta }\,
\Big(\,s_{\beta }c_{\beta }\,M_{a}^{2(0)}\ +\ \frac{\kappa }{\lambda }\,\mu
^{2}\,\Big)\ -\ \frac{\kappa }{\lambda }\,\mu A_{\kappa }\ +\ 4\,\frac{
\kappa ^{2}}{\lambda ^{2}}\,\mu ^{2}\ .
\end{eqnarray}
{}From  the above  analytic expressions for  $M^{2(0)}_{H^+}$, and the
CP-odd and   CP-even  Higgs-boson  mass  matrices,  $M^{2(0)}_P$   and
$M^{2(0)}_S$, it is  now evident that the  MSSM limit of  the NMSSM is
obtained for $\kappa ,\ \lambda \to 0$, while holding $\kappa /\lambda
,\ \mu ,\ A_{\lambda }$ and $A_{\kappa }$ fixed.

Parenthetically, we  should remark that   the Higgsino sector  of  the
NMSSM is   also different from  the  corresponding one  in  the MNSSM. 
Because      of     the      presence      of      the        operator
$\frac{\kappa}{3}\,\widehat{S}^3$    in        the      superpotential
(\ref{Wprime}), the  $\{ 55 \}$-matrix  element of the neutralino mass
matrix in Eq.\ (\ref{M0}) receives the additional contribution:
\begin{equation}
  \label{M55}
({\cal M}_0)_{55}\ =\ -\, 2\,\frac{\kappa}{\lambda}\ \mu\, .
\end{equation}
Note  that if $({\cal M}_0)_{55} < 0$ with $\mu < 0$,\footnote[4]{Our
  choice of a negative $\mu$-parameter is  mainly dictated by the fact
  that   $b\to s\gamma$  imposes  a  stronger  lower limit on positive
  values   of  $\mu$ \cite{Masiero}   for  relatively   small  charged
  Higgs-boson masses, close to  the present experimental  bound, i.e.\ 
  for $M_{H^+} \sim 80$~GeV~\cite{PDG}.   Instead, for negative values
  of $\mu$,  the bound on $\mu$ can  be dramatically relaxed up to the
  present   LEP2     limit:  $|\mu|    \stackrel{>}{{}_\sim}   90$~GeV
  \cite{PDG}.}   this additional  contribution  to  the  predominantly
singlet state in the NMSSM is constructive,  rendering its mass larger
than the axino mass in the MNSSM.   However, for small positive values
of $\kappa$, e.g.\ $\kappa  \stackrel{<}{{}_\sim}    0.1$, and  $|\mu|
\stackrel{<}{{}_\sim}  200$~GeV,  with   $\mu <0$  and $\lambda\approx
0.65$, the $\{  55 \}$-matrix element  $({\cal M}_0)_{55}$ is positive
and   its  contribution to   the would-be  axino  mass is destructive,
leading   to light singlet  masses     smaller than $m_{\tilde{a}}$.   

It is now important   to   notice that  unlike  the MNSSM   case,  the
decoupled CP-even and CP-odd scalar singlets  are no longer degenerate
in the MSSM limit of  the NMSSM.  This fact is  a manifestation of the
violation  of  the mass  sum rule (\ref{sumrule})  in  the case of the
NMSSM.  Specifically, in the NMSSM we find that
\begin{equation}  
  \label{sumrul2}
\sum_{i=1}^3 M^{2(0)}_{H_i}\: -\: \sum_{i=1}^2 M^{2(0)}_{A_i}\: 
-\: M^2_Z\ =\ 4\,\frac{\kappa}{\lambda}\, \mu^2\, 
\bigg( \, \frac{v^2}{v^2_S}\, s_\beta c_\beta\: +\: 
\frac{\kappa}{\lambda}\: -\: \frac{A_\kappa}{\mu} \,\bigg)\, .
\end{equation}
It is obvious  that the mass sum rule  (\ref{sumrule}) can be sizeably
violated in  the NMSSM for  relatively large values of  $|\kappa|$ and
$|\mu|$ or $|A_\kappa|$. In such  cases, the violation of the mass sum
rule  becomes  much larger  than  the  one  caused by  radiative  stop
effects.

The  analytic  expressions of  the  Higgs-boson  masses  in the  NMSSM
coincide with those of the MNSSM only in the PQ-symmetric limit, where
$\kappa /\lambda,\ t_{S},\ m_{12}^{2}\to 0$. Although  this limit is
unphysical as it leads to a theory with a visible axion, its vicinity,
however, could  define an acceptable  region of parameter  space where
the predictions  of the two  models exhibit reasonable  agreement.  

An  interesting  property  of   the  tree-level  CP-even  mass  matrix
$M_{S}^{2(0)}$ in the  PQ-symmetric limit is that the  interval of the
allowed  $\mu^2$  values  is  rather  small.   This  interval  may  be
determined by requiring that the determinant of $M_{S}^{2(0)}$,
\begin{eqnarray}
  \label{detMS}             
\det (M_{S}^{2(0)}) &=& -\,
\frac{v^{2}}{v_{S}^{2}}\,\bigg\{ \, 4\,
\bigg[\, M^{2}\: +\: \bigg(\, \frac{1}{2}\,\lambda^2 v^2\: -\: M_Z^2\, 
\bigg) \cos^2 2\beta\, \bigg]\, \mu^4\ -\ 
2 \sin^2 2\beta\, M^2\,M_{a}^{2(0)}\mu^2\   \nonumber\\  
&&+\, \frac{1}{4}\, \sin^4 2\beta\, M^2\,M_{a}^{2(0)}
\bigg(\, M_a^{2(0)}\: -\: \frac{1}{2}\,\lambda^2 v^2\,\bigg)\,\bigg\}\,,
\end{eqnarray}
with    $M^2  =  M_a^{2(0)}  + M_Z^2     - \frac{1}{2}\lambda^2  v^2 =
M_{H^+}^{2(0)}   +  M_Z^2 -  M_W^2$,  be  positive.   Neglecting terms
proportional   to  $(\frac{1}{2}\lambda ^{2}v^{2}-\  M_{Z}^{2}) \cos^2
2\beta$ next to $M^{2}$ in Eq.\  (\ref{detMS}), we may approximate the
determinant $\det (M_S^{2(0)})$ as
\begin{equation}         
\det (M_{S}^{2(0)})\ \approx\ -\,
\frac{v^2}{v_S^2}\, M^2\,\bigg[\, 4\mu^4\ -\ 
            2\sin^2 2\beta\,M_{a}^{2(0)}\mu^2\   +\ 
                   \frac{1}{4}\, \sin^4 2\beta\, M_a^{2(0)}
\bigg(\, M_a^{2(0)}\: -\: \frac{1}{2}\,\lambda^2 v^2\,\bigg)\,\bigg]\, .
\end{equation}  
Requiring now that $\det (M_{S}^{2(0)})$ be positive gives the allowed
$\mu^2$ interval:
\begin{equation}
  \label{muint}   
\frac{1}{4}\, \sin^2 2\beta\, M_a^{2(0)}\, 
\left(1-\delta \right)\ \stackrel{<}{{}_\sim}\  \mu^2\ \stackrel{<}{{}_\sim}\
\frac{1}{4}\, \sin^2 2\beta\, M_{a}^{2(0)}\, \left( 1+\delta \right)
\end{equation}  
with   
\begin{equation}  
  \label{delta}
\delta\ =\ \sqrt{ \frac{\lambda^2 v^2}{2M_a^{2(0)}} }\ .
\end{equation} 
Here,  it  is  understood  that  $\delta  \leq  1$  or,  equivalently,
$M_{H^+}^{2(0)}\geq   M_W^2$.    Especially   for  $M_{H^+}^{2(0)}   =
M_{W}^{2}$,  for   which  the  terms   proportional  to  $(\frac{1}{2}
\lambda^2 v^2  - M_Z^2) \cos^2  2\beta$ are no longer  negligible with
respect to $M^2 = M^2_Z$, the allowed range of $\mu^2$ becomes
\begin{equation}
  \label{mubound}     
0\ <\ \mu^2\ <\ \frac{1}{2}\, \sin^2 2\beta\,M_a^{2(0)}\,
\bigg(\, 1\ +\ \frac{\frac{1}{2}\,\lambda^{2}v^{2}\:
-\: M_{Z}^{2}}{M_{Z}^{2}}\, \cos^2 2\beta\ \bigg)^{-1}.
\end{equation} 
Further insight into the predictions  of the PQ-symmetric limit may be
gained by  analyzing the kinematic  situation where $M^{2(0)}_{H^+}\gg
M^2_W$ (i.e.~$\delta \ll 1$) and  $\mu^2 = \mu^2_{\rm mid} = s^2_\beta
c^2_\beta M^{2(0)}_a$  which is approximately the middle  point of the
allowed  $\mu^2$-interval.    In  this  case,  we   obtain  (see  also
Appendix~B)
\begin{eqnarray}
  \label{PQmass}
M^{2(0)}_{H_1} \!\!&\approx&\!\! 
\frac{1}{2}\,\lambda^2 v^2 \sin^2 2\beta\,,\qquad
M^{2(0)}_{H_2}\ \approx\ M^2_Z \cos^2 2\beta\: +\: 
\frac{1}{2}\, \lambda^2 v^2 \sin^2 2\beta\,,\nonumber\\
M^{2(0)}_{H_3} \!\!&\approx&\!\! M^{2(0)}_a\: +\: \frac{1}{2}\,
\lambda^2 v^2\cos^2 2\beta\: -\: \bigg(\, \frac{1}{2}\,\lambda^2 v^2
- M^2_Z\,\bigg)\, \sin^2 2\beta\,,\nonumber\\
M^{2(0)}_{A_1} \!\!&\approx &\!\! 0\,,\quad
M^{2(0)}_{A_2}\ \approx\ M^{2(0)}_a \: +\: \frac{1}{2}\,\lambda^2 v^2
\end{eqnarray}
and 
\begin{equation}
  \label{ZZHPQ}
g^2_{H_2ZZ}\ \approx\ g^2_{H_3A_2Z}\ \approx\ 1\, .
\end{equation}
Thus,  the light  $H_1$  and  $A_1$ scalars  decouple  from the  gauge
bosons,  whilst the  $H_2$ boson  couples  maximally to  them with  SM
strength.    Moreover,  according  to   the  mass-coupling   sum  rule
(\ref{sumcoupl}), the $H_2$-boson mass  saturates the mass upper bound
obeyed  by the  SM-like Higgs  boson.  Given  that the  length  of the
allowed $\mu^2$-interval  is very small relative  to $\mu^2_{\rm mid}$
for  $\delta\ll 1$,
one does not expect serious changes
regarding the heaviest
Higgs-boson masses $M_{H_3}$ and $M_{A_2}$ and the qualitative
features of the Higgs to gauge-boson couplings as $\mu$ takes all other
allowed values.

A minimal deviation from the  PQ-symmetric limit,  in which the  NMSSM
could easily be compared with the MNSSM, is  the limit $\kappa \to 0$,
with  $\lambda ,\ \mu ,\ A_\lambda$  and $\kappa A_\kappa$ held fixed.
In fact, in  this limit, the coupling $\lambda$  could be the largest,
thereby allowing for   the  largest possible value  for   the lightest
Higgs-boson mass  $M_{H_1}$.      By the  same  token,  the   unwanted
U(1)$_{\rm   PQ}$   symmetry  gets broken   by    the trilinear   soft
SUSY-breaking self-coupling $\kappa  A_\kappa$ of the  singlet~$S$.  A
corresponding   limit  of the  MNSSM,  which  has  the  same number of
independent parameters as in the NMSSM, is the one with $m_{12}^{2}\to
0$, but $\lambda ,\ \mu ,\ A_\lambda$ and $t_S$ fixed.  We should also
bear in mind that vanishing of $t_S$ entails vanishing of $m_{12}^{2}$
as well   which  eliminates  the  possibility    of $t_S\to  0$,  with
$m_{12}^{2}$  fixed.  In this way,  we  compare essentially two models
which only differ in soft  operators of dimensionality  $d\leq 3$.  An
additional reason that  renders such a  comparison very interesting is
the fact that   the dimensionful parameters,  such  as  $A_\kappa$ and
$t_S$,  remain  unconstrained  by perturbativity  arguments, and hence
could severely affect the structure of the mass matrices.

The aforementioned physical  limit allows for more  direct comparisons
of the NMSSM  with    the MNSSM.   Equating  the  tadpole   parameters
$T_{\phi_1}$,  $T_{\phi_2}$ and   $T_{\phi_S}$  pertinent to the   two
models yields, in this limit, the simple relation
\begin{equation}
  \label{tadrel}
\frac{\lambda t_S}{\mu} \ =\ \frac{\kappa}{\lambda}\, \mu A_\kappa \, .
\end{equation}
Moreover,  in  the  same  limit, except  for  $(M_P^{2(0)})_{22}$  and
$(M_S^{2(0)})_{33}$, all other elements  of the mass matrices coincide
as  well.   In  the   MNSSM  $\lambda  t_S/\mu$  enters  the  elements
$(M_P^{2(0)})_{22}$ and  $(M_S^{2(0)})_{33}$ in exactly  the same way,
whereas   in  the  NMSSM   the  corresponding   parameter  $(\kappa\mu
A_\kappa)/\lambda$ appears in these two matrix elements with different
coefficients,  and most importantly,  with different  signs reflecting
the violation  of the tree-level  mass sum rule  (\ref{sumrule}).  The
fact that the determinants of  the CP-odd Higgs-boson mass matrices in
the  MNSSM and  NMSSM  are proportional  to  $\lambda t_S  / \mu$  and
$(\kappa\mu A_\kappa  )/\lambda$ necessitates that  the two parameters
must  be both  positive.   As a  result  of this,  the matrix  element
$(M_{S}^{2(0)})_{33}$ will  be enhanced in  the MNSSM, but  reduced in
the  NMSSM.   In  addition,  it  is  not difficult  to  see  that  the
determinant   of   the   tree-level   CP-even   mass   matrix,   $\det
(M_S^{2(0)})$,   is    a   monotonically   increasing    function   of
$(M_S^{2(0)})_{33}$  and is  already  negative if  $(M_S^{2(0)})_{33}$
vanishes.  This last  property relies on the fact  that the upper-left
$2\times 2$ submatrix is positive definite.  Therefore, the larger the
mass $M_{A_1}$ of  the lightest CP-odd scalar is  the larger (smaller)
$\det (M_S^{2(0)})$  is in  the MNSSM (NMSSM).   On the other  hand, a
very   small   value   for   $\lambda   t_S   /\mu$   or   $(\kappa\mu
A_\kappa)/\lambda$, which  amounts to having a very  light $A_1$, does
not seriously  affect $M_S^{2(0)}$, and hence  no essential difference
in the predictions for the  Higgs spectrum between the MNSSM and NMSSM
can  be  observed.  In  this  region, both  models  are  close to  the
PQ-symmetric  limit. However,  the difference  between the  two models
becomes  appreciable,  once   the  parameters  $\lambda  t_S/\mu$  and
$(\kappa\mu A_\kappa ) /\lambda$ become large. The first parameter has
no upper  bound, whereas the  second one is  limited by the  fact that
$(M_{S}^{2(0)})_{33}$  should be  positive.  Thus,  only in  the MNSSM
case a significant departure  from the PQ-symmetric limit is possible,
which  may   change  the   situation  drastically.   For   example,  a
$\mu$-independent  contribution to  $(M_{S}^{2(0)})_{33}$,  say $T^2$,
changes the coefficient of $\mu^2$ in the expression (\ref{detMS}) for
$\det (M_{S}^{2(0)})$, and as  a consequence, the allowed interval for
$\mu^2$  can  now  expand  (or  further  shrink)  for  $T^2$  positive
(negative).  For  the particular  case that $M_{H^+}^{2(0)}  = M_W^2$,
the  interval  of $\mu^2$  increases  (decreases)  by  a factor  $1  +
T^2/(\lambda^2  v^2 \sin^2  2\beta)$.  As  we just  observed,  such an
unconstrained   (constrained)  positive  (negative)   contribution  is
available  in the MNSSM  (NMSSM), i.e.\  $T^2 \equiv  \lambda t_S/\mu$
($T^2   \equiv   -(\kappa\mu   A_\kappa   )/\lambda$),   where   $t_S$
($A_\kappa$) should be regarded as a $\mu$-dependent parameter.

At this point, it should be  stressed that our discussion of the NMSSM
in the limit  $\kappa \to 0 $, with $\lambda ,\  \mu ,\ A_\lambda$ and
$\kappa A_\kappa$ being kept fixed,  by no means exhausts all possible
predictions that  the model offers for viable  scenarios.  Being close
to  the above  limit requires  that  $|\kappa /  \lambda |  \ll 1  \ll
|A_\kappa /\mu |$.  However, it is possible to considerably depart from
this limit, even if $|\kappa /  \lambda |$ is very small but non-zero,
while avoiding  the known  problem associated with  the presence  of a
visible axion.   In order to better  investigate alternative scenarios
that  avoid the  presence of  a visible  axion, we  compute  the exact
determinant of the CP-odd Higgs-scalar mass matrix
\begin{equation}
  \label{detMP}
\det (M_{P}^{2(0)})\ =\ 3\,\frac{\kappa }{\lambda }\,\bigg(\, 
\mu A_\kappa \ -\ \frac{3}{4}\, \sin 2\beta\, \lambda^2 v^2\ 
-\ 3\frac{\kappa }{\lambda }\,\delta^2\mu^2\,\bigg)\,M_{a}^{2(0)}\,.
\end{equation}
We shall  now examine other possible deviations  from the PQ-symmetric
limit, for  which the  $\mu$ values, however,  are not  very different
from   those   determined   by   the   allowed   $\mu$   interval   in
Eq.~(\ref{muint}).  Under this assumption and the fact that $|\kappa /
\lambda|$ is considered to be  adequately small, the third term on the
RHS  of Eq.\  (\ref{detMP})  remains always  subdominant; it  actually
diminishes  the size  of $\det  (M_{P}^{2(0)})$ irrespectively  of the
sign of $\kappa / \lambda$.  For  $|A_\kappa / \mu | \gg 1$, the first
term $\mu A_\kappa$ on the RHS of Eq.\ (\ref{detMP}) becomes dominant.
In this case, this term should have the same sign as $\kappa/\lambda$,
in   compliance  with   our  earlier   requirement   that  $(\kappa\mu
A_\kappa)/\lambda$  be positive.   However, as  $|A_\kappa /\mu  |$ is
getting  smaller, the  second term  $\frac{3}{4}\sin  2\beta \lambda^2
v^2$ within the parentheses on the RHS of Eq.\ (\ref{detMP}) will then
start playing an important  r\^ole.  For $\kappa /\lambda >0$ ($\kappa
/ \lambda  <0$), this  second term provides  a lower (upper)  bound on
$\mu A_\kappa$, which  should not be saturated.  In  fact, the mass of
the lightest CP-odd scalar depends crucially on the difference between
these two  first terms on the  RHS of Eq.\ (\ref{detMP}).   It is then
obvious that if $\kappa /\lambda$ is negative, $\mu A_\kappa$ could be
negative,  zero or  even a  positive  quantity which  is bounded  from
above.

Having  gained some  insight from  the  above discussion,  let us  now
consider  the   most  general  case  without   resorting  to  specific
assumptions or  kinematic approximations.  Then,  the requirement that
$\det  (M_{P}^{2(0)})$ in  Eq.~(\ref{detMP}) be  positive  implies the
constraint
\begin{equation}
  \label{mugen}
\mu_- \ <\ \mu\ <\ \mu_+\
\end{equation}
with
\begin{equation}
  \label{mulimits}
\mu_\pm\ =\ \bigg( 6\frac{\kappa }{\lambda }\,\delta^2\bigg)^{-1}\,
\bigg[\, A_\kappa\ \pm\ {\rm sign}\bigg(\frac{\kappa }{\lambda }\bigg)\,
\sqrt{A_\kappa^2\: -\: 9\sin 2\beta\, 
\delta^2 \kappa\lambda v^2 }\ \bigg]\, ,\nonumber
\end{equation}
where $A_\kappa^2 > 9\sin 2\beta\,   \delta^2 \kappa\lambda v^2$.   We
see again that  $A_\kappa = 0$ is  only allowed for $\kappa  / \lambda
<0$. In this case, $\mu^2$ is constrained to be in the range:
\begin{equation}
  \label{muappr}
0\ <\ \mu^2\ <\ -\,\frac{\lambda }{2\kappa }\, \sin 2\beta\, M_{a}^{2(0)}\,.
\end{equation}
Here, it is also important to  reiterate the fact that the requirement
for    a positive $\det  (M_{P}^{2(0)})$     constrains by itself  the
$\{33\}$-element of $M_{S}^{2(0)}$:
\begin{equation}
  \label{MS33}
(M_{S}^{2(0)})_{33}\ <\ \frac{v^2}{v_S^2}\, s_\beta c_\beta\,
\bigg(\, s_\beta c_\beta\, M_a^{2(0)}\ -\ 
2\,\frac{\kappa }{\lambda }\,\mu^2\,\bigg)\ +\ \left( 4\:
-\: 3\delta^2\right)\, \frac{\kappa^2}{\lambda^2}\,\mu^2\,.
\end{equation}
The constraint in Eq.\ (\ref{MS33}) seems to favour negative values of
$\kappa /\lambda$, as the   upper limit on $(M_{S}^{2(0)})_{33}$  gets
larger in  this case.  Furthermore,  saturation of the  upper bound in
Eq.\ (\ref{MS33}) leads to $M_{A_1}^{2(0)}=0$.

As the key parameter $|\kappa / \lambda |$ increases, the situation is
getting more   involved since new terms start    playing a r\^ole.  In
particular,   a term  which  deserves  special   attention is  the one
proportional     to  $\kappa^2\mu^2/\lambda^2$     that   occurs    in
Eq.~(\ref{MS33}).  This term  becomes very important for larger values
of $M_{H^+}^{2(0)}$  which lead to smaller values   of $\delta$ and to
larger values of   $\mu^2$ in accordance  with  Eqs.~(\ref{muint}) and
(\ref{delta}).  In such a case, we may hope  for an enlargement of the
allowed interval of $\mu^2$ values, for which $\det (M_{S}^{2(0)})$ is
positive.  Therefore, it would be  interesting to investigate to which
extent such a  situation can  indeed  be realized, especially for  low
values of  $M_{H^+}^{2(0)}$ for which  $\delta$ is  not very small and
$(M_S^{2(0)})_{33}$ appears to be  more severely constrained.  To this
end, we shall consider the special case where $M_{H^+}^{2(0)}= M_W^2$,
i.e.\ $\delta = 1$. Then, after taking into  account the constraint in
Eq.\ (\ref{MS33}) and making use of the  fact that $\det (M_S^{2(0)})$
increases   monotonically    with  $(M_S^{2(0)})_{33}$,  the following
inequality may be derived:
\begin{equation}
  \label{MS20}
\det (M_S^{2(0)})\ <\ -\, \lambda^2 v^2\, M_Z^2\,
\bigg[\, 2\bigg(\, 1\: +\: \frac{\kappa }{\lambda }\,\sin 2\beta\: +\:
\frac{\frac{1}{2}\,\lambda^2 v^2 \: -\: M_Z^2}{M_Z^2}\,
\cos^2 2\beta\,\bigg) 
\mu^2\ -\ \sin^2 2\beta\, M_a^{2(0)}\,\bigg]\,.
\end{equation}
Assuming that  the corresponding upper bound   in Eq.\ (\ref{MS33}) is
saturated (i.e.\ $M_{A_{1}}^{2(0)}=0$), then Eq.\ (\ref{MS20}) and the
fact that $\det (M_{S}^{2(0)})>0$ lead to
\begin{equation}
  \label{mu2bound}
0\ <\ \mu^2\ <\ \frac{1}{2}\, \sin^2 2\beta\, M_a^{2(0)}\, 
\bigg(\, 1\: +\: \frac{\kappa }{\lambda }\,\sin 2\beta\: +\:
\frac{\frac{1}{2}\,\lambda^2 v^2 \: -\: 
M_Z^2}{M_Z^2}\, \cos^2 2\beta\,\bigg)^{-1}\,.
\end{equation}
It is easy to  see that for $|\kappa /  \lambda |  \ll 1$, the  double
inequality in Eq.\  (\ref{mu2bound})  reduces to our  previous  result
found in  Eq.\  (\ref{mubound}).  We   observe  now that for   $\kappa
/\lambda  >  0$,  the  allowed   interval of   $\mu^2$ given  by  Eq.\
(\ref{mu2bound}) shrinks as $|\kappa  /\lambda |$ increases.  Instead,
for $\kappa /\lambda  < 0$ with  $|\kappa /\lambda |$ increasing,  the
allowed interval gets larger and, especially for  values of $|\kappa /
\lambda|$  close to unity, it  may  even become  infinitely large.  Of
course, at this critical kinematic  region, radiative corrections  are
expected to play the  dominant r\^ole.  Notwithstanding this fact, our
tree-level   results   should still  be   indicative   of the  various
tendencies which govern the kinematic parameters of  the theory. As we
will see below, however, values of $| \kappa /\lambda| \sim 1$ are not
compatible with  the largest possible  value  for $\lambda$  and hence
with the largest value of the lightest Higgs-boson mass $M_{H_1}$.

The Yukawa-type couplings $\kappa$ and $\lambda$ cannot be arbitrarily
large,  if  we wish to    preserve the  good   property of  SUSY  that
perturbation  theory be applicable up  to the  gauge unification scale
$M_{{\rm U}} \sim 10^{16}$ GeV \cite{CPW}.  Therefore, upper limits on
$|\lambda|$   and $|\kappa|$   can  be  obtained    by studying  their
renormalization-group (RG) evolution along with the corresponding ones
of the  strong   coupling constant $g_s$   and the  $t$  -quark Yukawa
coupling $h_t$ \cite{BS,EKW,GKY}:
\begin{eqnarray}  
  \label{RG}
16\pi^2\, \frac{d g_s}{dt} &=& -\, \frac{3}{2}\, g^3_s\, ,  \nonumber \\
16\pi^2\, \frac{d h_t}{dt} &=& h_t\, \bigg(\,3\,h^2_t\: +\: \frac{1}{2}\,
\lambda^2\: -\: \frac{8}{3}\, g^2_s\, \bigg)\,,  \nonumber \\
16\pi^2\, \frac{d\lambda}{dt} &=& \lambda\, \bigg(\,\kappa^2 \: +\:
2\lambda^2\: +\: \frac{3}{2}\, h^2_t\, \bigg)\, ,  \nonumber \\
16\pi^2\, \frac{d\kappa}{dt} &=& 3\kappa\, (\kappa^2 \: +\: \lambda^2)\, ,
\end{eqnarray}
where $t = \ln ( Q^2/M^2_t )$. In writing the RG equations (\ref{RG}),
we have ignored possible mass  threshold effects of the SUSY particles
while  running from  the $t$-quark-pole  mass  $M_t =  175$~GeV up  to
$M_{{\rm U}} \sim  10^{16}$ GeV. In the RG analysis,  we use the value
for  the  strong  fine-structure  constant  $\alpha_s  (M_t)  =  g^2_s
(M_t)/(4\pi)  \approx  0.109$.   Furthermore,  the  running  $t$-quark
Yukawa coupling $h_t$ is determined by
\begin{equation}  \label{htMt}
h_t (M_t)\ =\ \frac{m_t (M_t)}{v (M_t)\, s_\beta (M_t)}\ ,
\end{equation}
where $v(M_t) = 174.1$~GeV and 
\begin{equation}  \label{mtMt}
m_t (M_t) \ =\ \frac{M_t}{1\: +\: \frac{4}{3\pi}\, \alpha_s (M_t)}
\end{equation}
is the known relation  between the on-shell $\overline{{\rm MS}}$ mass
and $M_t$. For $3 \stackrel{<}{{}_\sim} \tan\beta (M_t) < 10$, we find
the approximate upper bounds
\begin{equation}  
  \label{lk1}
|\lambda (M_t)|\ \stackrel{<}{{}_\sim}\ 0.70,\ 0.63,\ 0.57,\ 0.44,\ 0.22,\quad 
\mbox{for}\quad |\kappa (M_t)| = 0,\ 0.3,\ 0.4,\ 0.5\ \mbox{and}\ 0.6,
\end{equation}
respectively. Correspondingly, for $\tan\beta \approx 2$, we obtain 
\begin{equation}  \label{lk2}
|\lambda (M_t)|\ \stackrel{<}{{}_\sim}\ 0.65,\ 0.59,\ 0.54,\ 0.42,\ 0.21\, .
\end{equation}
The results in Eqs.\ (\ref{lk1}) and (\ref{lk2}) are in good agreement
with Ref.\  \cite{EKW,GKY}.  {}From the above analysis,  it is obvious
that  the  largest  value  for  $|\lambda (M_t)|$  is  more  naturally
attained in  the MNSSM  (corresponding to $\kappa  (M_t) =  0$) rather
than in the NMSSM, as one would generically expect $\lambda (M_t) \sim
\kappa  (M_t) \neq 0$.  This is  another important  difference between
these  two  models.  In   particular,  this  implies  that  the  MNSSM
generically predicts  higher masses for the lightest  Higgs boson than
the NMSSM.

In the next section, we shall study the Higgs sector of the MNSSM more
quantitatively and also compare our numerical predictions with those
obtained in the NMSSM.

\setcounter{equation}{0}
\section{Phenomenological discussion}

In this  section,  we shall  discuss  the phenomenology  of  the Higgs
bosons in  the  MNSSM, and  make comparisons of   our predictions with
those obtained in the NMSSM.

At  LEP2,  the  CP-even  and   CP-odd Higgs scalars,  $H_{1,2,3}$  and
$A_{1,2}$,  are mainly  produced  through the  Higgs-strahlung process
$e^+e^-\to Z^* \to Z H_i$ or in pairs via $e^+e^-\to Z^* \to H_i A_j$.
Analogous  Higgs-boson   production mechanisms  can   take  place   at
Fermilab, where  instead of electrons the  initial  states are the $u$
and $d$ quarks at the  quark-parton level \cite{revII}. Therefore, the
necessary ingredients for our numerical discussion following below are
the  analytic expressions   for the radiatively-corrected  Higgs-boson
masses  and the effective  Higgs-boson couplings to  the gauge bosons. 
These analytic expressions pertaining to the MNSSM and NMSSM have been
presented in Sections 3 and 4, respectively.

There are  several possible  combinations in choosing  the independent
kinematic  parameters for  the  two supersymmetric  extensions of  the
MSSM, the MNSSM and the  NMSSM.  For definiteness, for the MNSSM case,
we consider
\begin{equation}
  \label{param1}
t_\beta\,,\quad M^2_{H^+}\,,\quad \mu\,,\quad \lambda\,,\quad 
\frac{\lambda t_S}{\mu}\quad {\rm and}\quad m^2_{12}\,,
\end{equation}
as free phenomenological parameters of   the Higgs sector. As for  the
NMSSM, we take as input parameters
\begin{equation}
  \label{param2}
t_\beta\,,\quad M^2_{H^+}\,,\quad \mu\,,\quad \lambda\,,\quad
\kappa\quad {\rm and}\quad \frac{\kappa\mu A_\kappa}{\lambda}\ .
\end{equation}
For both SUSY  models, the stop-related parameters are  chosen to have
the typical values:
\begin{equation}
\widetilde{M}_Q\ =\ \widetilde{M}_t\ =\ 0.5\ {\rm TeV}\,,\qquad 
A_t\ =\ 1\ {\rm TeV}\,.
\end{equation}
Here, we should remark that $m^2_{12}$ in Eq.\ (\ref{param1}) could in
principle be  absent, without  spoiling  the renormalizability of  the
theory.  In this case, the U(1)$_{\rm PQ}$ symmetry  of the MNSSM gets
broken explicitly by    the effectively  generated tadpole   parameter
$t_S$, which is  a term  of  the lowest possible dimension,  namely of
dimension 1.  Such a reduction of the renormalizable parameters is not
possible in the NMSSM because of the presence of $\widehat{S}^3$ which
violates U(1)$_{\rm PQ}$  hardly.  Therefore, it  is  fair to conclude
that  in  the admissible   limit  $m^2_{12}\to  0$,  the   MNSSM under
investigation  represents  the most economic,  renormalizable scenario
among the proposed non-minimal supersymmetric standard models.

In Fig.\  \ref{fig:nmssm1} we display  the dependence of  the lightest
Higgs boson  $H_1$ in the  MNSSM with $m^2_{12}  = 0$ on  the would-be
$\mu$-parameter, for different values of the charged Higgs-boson mass,
i.e.\  for  $M_{H^+}   =  0.1,\  0.3,\  0.7$  and   1~TeV.   In  Fig.\ 
\ref{fig:nmssm1}(a),  we choose  the tadpole-parameter  value $\lambda
t_S/\mu = 1$~TeV$^2$.  As we  are interested in maximal values for the
lightest  Higgs-boson mass  $M_{H_1}$ which  occur for  low  values of
$\tan\beta$,  i.e.\  for $\tan\beta  =  2$,  we  consider the  largest
allowed  coupling  $\lambda  =   0.65$,  for  which  the  MNSSM  stays
perturbative  up  to  the  gauge  unification scale  $M_{\rm  U}  \sim
10^{16}$ GeV (see also discussion in  Section 4).  As can be seen from
Fig.\ \ref{fig:nmssm1}(a), the $H_1$-boson mass varies between 120 and
145 GeV depending on $M_{H^+}$ for a wide range of $\mu$ values, which
is significantly  larger than the current experimental  lower bound of
113.3~GeV  on the  SM-type Higgs  boson.  Furthermore,  we  observe an
asymmetry   of   order  5   GeV   in   $M_{H_1}$   for  large   $|\mu|
\stackrel{>}{{}_\sim} 300$ GeV between positive and negative values of
$\mu$.   This  is  because  stop-radiative effects  on  $M_{H_1}$  get
enhanced for larger values of  the stop-mixing parameter $|X_t| = |A_t
-\mu/\tan\beta|$ which obviously result  from large negative values of
$\mu$, provided  $|X_t/{\rm max}\, (\widetilde{M}_Q, \widetilde{M}_t)|
\stackrel{<}{{}_\sim} \sqrt{6}$ (cf.\ Eq.\ (\ref{sumcoupl})).

In  Fig.\ \ref{fig:nmssm1}(b)  we  consider a  smaller  value for  the
tadpole parameter, i.e.\ $\lambda t_S/\mu = 0.04$~TeV$^2$. As in Fig.\ 
\ref{fig:nmssm1}(a), we present numerical  estimates of $M_{H_1}$ as a
function  of  $\mu$, for  the  same  discrete  values of  the  charged
Higgs-boson mass: $M_{H^+} = 0.1,\ 0.3,\ 0.7$ and 1~TeV.  We find that
the  allowed range  of $\mu$  becomes  much smaller,  but the  maximum
values of  $M_{H_1}$ are still very  close to those  obtained in Fig.\ 
\ref{fig:nmssm1}(a).  Most  interestingly, we observe  that the maxima
of $M_{H_1}$ are located at almost the same $\mu$ values found for the
tadpole parameter of 1~TeV$^2$; the maxima are practically independent
of the tadpole parameter, for  all relevant values of $\lambda t_S/\mu
=  0.01$--1~TeV$^2$. This  feature  that the  allowed  range of  $\mu$
values shrinks as $\lambda t_S/\mu$  gets smaller is in good agreement
with our  discussion in Section  4 concerning the  CP-even Higgs-boson
mass matrix in the PQ-symmetric limit.  Specifically, for small values
of  $\lambda t_S/\mu$,  the allowed  $|\mu|$-ranges can  be accurately
determined by Eqs.\ (\ref{muint}) and (\ref{mubound}).  In particular,
the mean values of the allowed $|\mu|$-ranges, which are approximately
given  by $s_\beta  c_\beta  M_{H^+}$ and  are  almost independent  of
$\lambda t_S/\mu$, appear to describe  well the location of the maxima
of $M_{H_1}$.

It is now very interesting to analyze a scenario within the context of
the MNSSM, in which the charged Higgs boson has a relatively low mass,
in the range $M_{H^+}=80$--160 GeV,  and may be accessed in next-round
experiments  at LEP2 and/or  at the  upgraded Tevatron  collider.  For
this purpose, in Fig.\ \ref{fig:nmssm2} we display numerical estimates
of  the  two   lightest  Higgs  bosons  $H_1$  and   $H_2$  and  their
corresponding squared couplings  to the $Z$ boson as  functions of the
parameter  $\mu$,  for $M_{H^+}=80,\  120$  and  160  GeV.  The  other
kinematic  parameters  are   chosen  to  be  the  same   as  in  Fig.\ 
\ref{fig:nmssm1}(a): $\tan\beta  = 2$,  $\lambda = 0.65$  and $\lambda
t_S/\mu = 1$~TeV$^2$.  Let us first consider the lowest experimentally
allowed  value for  the charged  Higgs-boson mass  $M_{H^+}  = 80$~GeV
\cite{ADLO,PDG}.   Then, in Fig.\  \ref{fig:nmssm2}(a) we  notice that
the $H_1$-boson  mass cannot  become larger than  105 GeV,  whilst the
next-to-lightest $H_2$  boson can  be as heavy  as 146~GeV. As  can be
seen from  Fig.\ \ref{fig:nmssm2}(b), such a scenario  is not excluded
experimentally,  since  the  $H_1ZZ$-coupling gets  suppressed,  i.e.\ 
$g^2_{H_1ZZ}      \stackrel{<}{{}_\sim}      0.2$,     for      $|\mu|
\stackrel{<}{{}_\sim}  350$ GeV.   In this  scenario, the  $H_2$ boson
becomes SM  type ($H_2\equiv  H_{\rm SM}$), and  is much  heavier than
$H^+$.  This novel  prediction of the MNSSM for  viable scenarios with
$M_{H^+}  \stackrel{<}{{}_\sim}  M_{H_{\rm  SM}}$  and  low-values  of
$\tan\beta  <  5$  cannot  be   realized  within  the  MSSM,  even  if
CP-violating loop effects are included in the Higgs sector of the MSSM
\cite{APRD,CEPW}.\footnote[5]{Using  the  code  {\tt  cph}  \cite{cph}
  based  on \cite{CEPW},  one finds  that only  for extreme  values of
  $|\mu| \stackrel{>}{{}_\sim} 5$ TeV and for $\tan\beta > 20$, such a
  scenario might be  made viable \cite{Carlos}.}  In fact,  as we will
see later  on, neither the  NMSSM can naturally  accommodate scenarios
with   $M_{H^+}  \stackrel{<}{{}_\sim}   M_{H_{\rm   SM}}$,  for   the
experimentally allowed values  of $|\mu| \stackrel{>}{{}_\sim} 90$~GeV
\cite{PDG}.

As  the charged Higgs  boson becomes  heavier  in the MNSSM, the $H_1$
boson also gets heavier and resembles the SM Higgs boson $H_{\rm SM}$.
Thus, from Fig.\ \ref{fig:nmssm2} we see that for $M_{H^+} = 120$~GeV,
$M_{H_1} \stackrel{<}{{}_\sim}  132$~GeV, with $g^2_{H_1ZZ} \sim 0.5$,
while for  $M_{H^+} = 160$~GeV,   it is $M_{H_1} \stackrel{<}{{}_\sim}
142$~GeV, with   $g^2_{H_1ZZ}  \approx 1$.    Furthermore,  as we have
already   discussed in Section  3.1,   for  the considered  values  of
$M_{H^+}$ much smaller than $\lambda t_S/\mu$,  the Higgs states $A_2$
and $H_3$ decouple and are  almost degenerate with $M^2_{A_2}  \approx
M^2_{H_3}  \approx \lambda  t_S/\mu$.   In  particular,  our numerical
estimates confirm the  relations: $M^2_{A_1} \approx M^2_{H^+} - M^2_W
+ \frac{1}{2} \lambda^2 v^2$ (cf.~Eq.~(\ref{tadec})), and $g^2_{H_1ZZ}
\approx  g^2_{H_2A_1Z}$   and   $g^2_{H_2ZZ} \approx    g^2_{H_1A_1Z}$
(cf.~Eq.~(\ref{compl})),  which are only valid  in  the above specific
decoupling regime of the $A_2$ and $H_3$ bosons  in the MNSSM.  {}From
Fig.\    \ref{fig:nmssm2}(a), we  see   finally  that   for $M_{H^+} =
160$~GeV, the $H_2$-boson  mass is nearly $\mu$-independent and equals
the $A_1$-boson mass $M_{A_1} \approx 179$~GeV.  This result is just a
consequence of  the  expected decoupling property   of a heavy charged
Higgs boson, with $M_{H^+} = 160~{\rm GeV} \gg M_Z$.

In  Fig.~\ref{fig:nmssm11} we display  predicted  values for $M_{H_1}$
and $M_{H_2}$, as  well as for $g^2_{H_1ZZ}$  and $g^2_{H_2ZZ}$ in the
MNSSM, using the same input   parameters as in  Fig.~\ref{fig:nmssm2},
but with $\tan\beta = 20$, i.e.\ $\lambda = 0.65$ and $\lambda t_S/\mu
= 1$~TeV$^2$.  We encounter  a functional dependence of  the evaluated
kinematic  parameters  on $\mu$,  for  $M_{H^+}=80,\ 120$ and 160 GeV,
which is qualitatively      similar    to the  one      presented   in
Fig.~\ref{fig:nmssm2}.  Again, we see   that the charged  Higgs  boson
$H^+$  can be lighter than  the $H_1$ boson,  even for large values of
$\tan\beta$.  Yet, we observe that for  larger $H^+$-boson masses, the
squared $H_1$-boson coupling  to  the $Z$  boson,  $g^2_{H_1ZZ}$, goes
more  rapidly to  unity than in  the $\tan\beta  = 2$ case.   Here, we
should emphasize again   that as $M_{H^+}$  becomes much   larger than
$M_Z$, the  predicted values for  $M_{H_1}$ approach the  one given by
the square root of the RHS of  Eq.~(\ref{sumcoupl}), where, of course,
the term   proportional to  $\lambda^2 \sin^22\beta$  is   negligible. 
Therefore,  only in this kinematic regime  where  both $\tan\beta$ and
$M_{H^+}$ are  large, the predictions of the  MNSSM will coincide with
those of the MSSM.

In  our numerical analysis in  connection with Fig.\ \ref{fig:nmssm2},
we have  already observed that  for large values of  $\lambda t_S/\mu$
but  small  values  of $M_{H^+}$, the   $H_1$ boson   does  not couple
strongly to the $Z$ boson, but  it is rather  the $H_2$ boson which is
SM-type.   Actually, this kind  of  behaviour is encountered even  for
larger values of $M_{H^+}$, provided $\lambda t_S/\mu$ is sufficiently
small.  As was already discussed in Section 4, the latter reflects the
fact that the  model approaches the PQ-symmetric  limit  in this case.
In  Fig.~\ref{fig:nmssm7}, we present numerically the $\mu$-dependence
of the  two    lightest  CP-even Higgs-boson   masses,  $M_{H_1}$  and
$M_{H_2}$,     and their respective    couplings  to   the $Z$  boson,
$g^2_{H_1ZZ}$ and $g^2_{H_2ZZ}$, in the MNSSM with $m^2_{12} = 0$, for
$M_{H^+} = 200,\ 400,\ 600$ and 800~GeV. In addition, we have selected
the value of the tadpole parameter $\lambda t_S/\mu =0.01328$~TeV$^2$.
For this specific value  of the tadpole  parameter and for  $M_{H^+} =
400$~GeV, we see that there is a value of $\mu$ where the Higgs states
$H_1$ and  $H_2$ interchange their couplings  to  the $Z$ boson, while
being nearly degenerate having   a mass close to   the upper bound  of
$M_{H_1}$.  We shall denote by $\mu_*$ this specific value of $\mu$ at
which a level crossing  in  the couplings  of $H_1$ and  $H_2$ occurs.
Thus, for values of $|\mu|$ smaller than $|\mu_*|$, it is $g^2_{H_1ZZ}
> g^2_{H_2ZZ}$, while  this inequality of  the  squared couplings gets
inverted for $|\mu| > |\mu_*|$.  If we now consider smaller values for
$M_{H^+} = 200$ GeV  for  the chosen value   of $\lambda t_S/\mu$,  we
observe    from   Fig.\  \ref{fig:nmssm7}    that   as $|\mu|$  grows,
$g^2_{H_1ZZ}$ starts higher than $g^2_{H_2ZZ}$, and the crossing point
of these two squared couplings is before $M_{H_1}$ reaches its highest
value.  If  we now take larger  values for $M_{H^+}$, e.g.\ $M_{H^+} =
600$, 800~GeV, we see  that $g^2_{H_1ZZ}$ starts again higher, becomes
almost  unity  with $M_{H_1}$  close to  its   largest allowed  value,
according  to the    mass-coupling sum   rule  (\ref{sumcoupl}),   and
intersects $g^2_{H_2ZZ}$  at a smaller  value of  $M_{H_1}$.  The very
special value of $\lambda t_S/\mu$,  for a  given value of  $M_{H^+}$,
for which $M_{H_1}$ and $M_{H_2}$ become equal at the highest possible
value  for $M_{H_1}$ and  $g^2_{H_1ZZ} \approx g^2_{H_2ZZ}\approx 0.5$
should be   regarded as a critical   point.  Generically speaking, for
values of $\lambda t_S/\mu$  lower than the  one corresponding to  the
critical point,  the $H_2$  boson   couples predominantly  to the  $Z$
boson.   Instead, if  $\lambda  t_S/\mu$ is higher  than its  critical
value, it is  then the $H_1$ boson that  couples with SM strength.  In
addition, in Fig.~\ref{fig:nmssm7} we see that almost independently of
$M_{H^+}$,  the  squared   couplings $g^2_{H_1ZZ}$  and  $g^2_{H_2ZZ}$
remain comparable for a wide range of $\mu$ values.   The latter is an
indication of the  fact that the  critical value of $\lambda  t_S/\mu$
depends only weakly on the  charged Higgs-boson mass $M_{H^+}$ and has
a value close to 0.01~TeV$^2$, for $0.3~{\rm TeV}\stackrel{<}{{}_\sim}
M_{H^+} \stackrel{<}{{}_\sim} 1$~TeV,  where the remaining independent
kinematic parameters are held fixed.

We shall now analyze the predictions of the MNSSM for relatively small
values   of    the   tadpole   parameter    $\lambda   t_S/\mu$.    In
Fig.~\ref{fig:nmssm8}, we display numerical estimates of $M_{H_1}$ and
$M_{H_2}$, as well as of $g^2_{H_1ZZ}$ and $g^2_{H_2ZZ}$, as functions
of the $\mu$-parameter, for $\lambda t_S/\mu = 0.0026$~TeV$^2$. As for
charged Higgs-boson  masses, we choose  $M_{H^+} = 0.3$, 0.5,  0.7 and
1~TeV.   It  is  easy  to  see  that, to  a  good  approximation,  the
functional dependence of the masses  of the two lightest CP-even Higgs
bosons $H_1$ and $H_2$ are  insensitive to the value of $M_{H^+}$.  In
this scenario of  the MNSSM, the $H_2$ boson  has always the strongest
coupling   to   the   $Z$    boson.    Although   not   displayed   in
Fig.~\ref{fig:nmssm8}, the mass of the lightest CP-odd scalar $A_1$ is
found  to be  $M_{A_1} \approx  50$~GeV and  is almost  independent of
$M_{H^+}$.  In addition, the  CP-odd Higgs scalar $A_1$ has suppressed
couplings  to   the  $Z$   and  $H_1$  bosons,   i.e.\  $g^2_{H_1A_1Z}
\stackrel{<}{{}_\sim} 10^{-2}$, and  therefore can escape detection at
LEP2.   In Fig.~\ref{fig:nmssm8}  we notice  finally that  the allowed
intervals of $\mu$ values become  even shorter than those found in the
previous scenarios  of the MNSSM.  These results  are all consequences
of our  choice of a relatively  small value for  the tadpole parameter
and are in good qualitative agreement with our discussion in Section~4
pertaining to the PQ-symmetric limit.

It is very interesting to examine the consequences  of the presence of
a  non-vanishing   effective  $F_S$-tadpole   term $m^2_{12}$   on the
Higgs-boson   mass spectrum    of     the  MNSSM.    Therefore,     in
Fig.~\ref{fig:nmssm9}      we plot the     dependence   of the CP-even
Higgs-boson masses $M_{H_1}$  and $M_{H_2}$ and the squared  couplings
$g^2_{H_1ZZ}$ and  $g^2_{H_2ZZ}$, as functions of the $\mu$-parameter,
for $t_S = -1$~TeV$^3$ and $m^2_{12} = 0.325$~TeV$^2$.  Because of the
close relationship between $t_S$ and $m^2_{12}$,  we are now compelled
to treat $t_S$ as a $\mu$-independent constant.   In fact, for $M_{\rm
SUSY} = 1$~TeV, we can easily compute  from Eq.\ (\ref{Lpar}) that the
adopted values  for $t_S$ and   $m^2_{12}$  correspond to the  typical
values of  $\xi_S$ and $\xi_F$: $\xi_S  = -1$ and   $\xi_F = 1/2$.  To
enable a direct comparison  with Fig.~\ref{fig:nmssm2}, we  choose the
same values  as  in Fig.~\ref{fig:nmssm2} for  the remaining kinematic
parameters of  the theory.  {}From  Fig.~\ref{fig:nmssm9}, we see that
the presence of a  non-vanishing, positive tadpole term $m^2_{12}$ can
shift the  maxima of $M_{H_1}$ and  $M_{H_2}$ towards larger values of
$|\mu|$, whereas all other features found in Fig.~\ref{fig:nmssm2} are
retained.   In Fig.~\ref{fig:nmssm9}(a),   we have  also displayed the
dependence of the mass $M_{A_1}$ of the  lightest CP-odd scalar $A_1$,
as a function of $\mu$. We see that $M_{A_1}$  decreases with $|\mu |$
decreasing. This kinematic behaviour originates from the fact that the
contribution of the off-diagonal    terms to the CP-odd  mass   matrix
becomes rather significant  for smaller values  of $|\mu |$.  Instead,
for larger values of  $|\mu|$, the  corresponding contribution of  the
off-diagonal   terms  is smaller, and    leads  to the mass  relation
$M_{A_1}\approx M_a$.

Unlike the  MSSM, the charged Higgs-boson $H^+$  cannot be arbitrarily
heavy  in  the  MNSSM  for  fixed  given  values  of  $\tan\beta$  and
$\lambda$, and for  natural choices of $\lambda t_S/\mu$  and the soft
squark  masses,  i.e.\   for  $\lambda  t_S/\mu,\  \widetilde{M}^2_Q,\ 
\widetilde{M}^2_t        \stackrel{<}{{}_\sim}       1$~TeV$^2$.       
Figure~\ref{fig:nmssm3} displays the dependence  of the maximum of the
lightest Higgs-boson  mass, ${\rm  max}\,(M_{H_1})$, as a  function of
$M_{H^+}$, for $\tan\beta = 2$, $\lambda = 0.65$ and for two different
values  of  the  tadpole  parameter:  $\lambda  t_S/\mu  =  0.04$  and
1~TeV$^2$.  The coupling of the  $H_1$ scalar to the $Z$ boson becomes
SM-type,   for   $M_{H^+}   \stackrel{>}{{}_\sim}  150$~GeV.    {}From
Fig.~\ref{fig:nmssm3}, it  is then easy  to see that the  current LEP2
lower bound  on ${\rm  max}\,(M_{H_1})$ implies the  approximate upper
limit  on $M_{H^+}$:  $M_{H^+} \stackrel{<}{{}_\sim}  2.7$~TeV, almost
independently of $\lambda t_S/\mu$.   This result may be understood as
follows.   As the  mass  of  the charged  Higgs  boson increases,  the
maximum of $M_{H_1}$  occurs for larger values of  $|\mu|$, which is a
consequence  of the  tree-level structure  of the  CP-even Higgs-boson
mass matrix in Eq.~(\ref{CPeven0}). On  the other hand, the larger the
value of $|\mu|$ becomes the larger the stop-mixing parameter $|X_t| =
|A_t   -  \mu/t_\beta   |$   is  getting.    Thus,  when   $|X_t|/{\rm
  max}\,(\widetilde{M}_Q,\widetilde{M}_t    )    \stackrel{>}{{}_\sim}
\sqrt{6}$,  the contributions  of  the stop-radiative  effects to  the
lightest  Higgs-boson mass  $M_{H_1}$  become negative,  and so  drive
${\rm  max}\,(M_{H_1})$  to  unphysical  values.  For  the  very  same
reasons, a  similar dependence of ${\rm  max}\,(M_{H_1})$ on $M_{H^+}$
is found to apply to the NMSSM case as well.

For    comparison, we shall    now investigate   a  few representative
scenarios within  the context of  the  NMSSM.  As  a first example, we
consider the scenario with $\tan\beta =  2$, $\lambda = 0.65$, $\kappa
= 0.01$  and $(\kappa  \mu    A_\kappa )/\lambda =   0.0026$~TeV$^2$.  
Figure~\ref{fig:nmssm5} exhibits the numerical predictions for the two
lightest  Higgs-boson masses,   $M_{H_1}$ and  $M_{H_2}$,   and  their
corresponding  squared couplings to  the  $Z$ boson, $g^2_{H_1ZZ}$ and
$g^2_{H_2ZZ}$, as functions of the $\mu$-parameter.   We also vary the
charged Higgs-boson mass in a discrete  manner, i.e.\ $M_{H^+} = 0.3$,
0.5, 0.7  and  1~TeV.  We  observe  that $M_{H_1}$  and  $M_{H_2}$ are
practically  independent  of $M_{H^+}$,   with $M_{H_1}$  consistently
below  80~GeV.  Such low  values of $M_{H_1}$  are still acceptable at
LEP2, in    the    range   of   $\mu$    values    where  $g^2_{H_1ZZ}
\stackrel{<}{{}_\sim} 0.07$.  In this scenario, the  $H_2$ boson has a
SM-type  coupling to  the $Z$  boson.   Also, the mass of  the lighest
CP-odd Higgs scalar $M_{A_1}$  is almost independent of $M_{H^+}$  and
comes out  to be slightly    higher than $M_{H_1}$.  The NMSSM   under
discussion, with the chosen low value of $\kappa \approx 0.01$, may be
considered   to   adequately  describe  the  limiting   scenario where
$\kappa\to  0$ and  $\kappa A_\kappa$ is  held fixed.   This last fact
enables one to directly compare the present scenario of the NMSSM with
the MNSSM where the tadpole parameter $\lambda t_S/\mu$  is set to the
same   value with  that of  $(\kappa  \mu  A_\kappa )/\lambda$,  i.e.\ 
$\lambda t_S/\mu = 0.0026$~TeV$^2$.    Such a scenario  in  connection
with  the   MNSSM   has    already   been analyzed above      in Fig.\ 
\ref{fig:nmssm8}.  Thus, if we  now compare Fig.~\ref{fig:nmssm5} with
Fig.\  \ref{fig:nmssm8}, we  observe resembling numerical  predictions
for the Higgs-boson masses and couplings in the  two models.  The only
visible difference   between them is that in   the MNSSM, the lightest
CP-even Higgs   boson $H_1$ is  consistently 30   GeV heavier than the
corresponding  one  in the  NMSSM, while   the  mass $M_{A_1}$  of the
lightest CP-odd scalar is about  30~GeV lower.  These findings are  in
excellent agreement with our discussion in Section~4.

We shall now analyze in Fig.~\ref{fig:nmssm4} a second scenario of the
NMSSM, in which the Yukawa-type  coupling $\kappa$ is larger, but with
the expression  $(\kappa \mu  A_\kappa   )/\lambda$ being held   fixed
again, i.e.\   $\kappa = 0.1$ and  $(\kappa  \mu A_\kappa  )/\lambda =
0.0026$~TeV$^2$. In Fig.~\ref{fig:nmssm4},   we also vary  the charged
Higgs-boson mass in the same way as in Fig.~\ref{fig:nmssm5}: $M_{H^+}
= 0.3$,  0.5, 0.7 and  1~TeV.  In this  scenario, the ratio $|A_\kappa
/\mu |$ varies from 1.69 for $|\mu | = 100$~GeV up  to 0.096 for $|\mu
| = 420$~GeV,  namely the ratio $|A_\kappa /\mu  |$ is no  longer much
larger than  1 for all  relevant  values of $|\mu|$.   Furthermore, as
$|\mu  |$  increases,  the strong inequality   $|A_\kappa  /\mu |  \gg
|\kappa /\lambda  |$   gets  gradually   violated  as  well.    As   a
consequence,  as the charged   Higgs-boson mass $M_{H^+}$ takes higher
values, the   picture     starts  changing   in     comparison    with
Fig.~\ref{fig:nmssm5}.  To be precise, as $M_{H^+}$ becomes larger, we
observe a  progressive enhancement of the  maximum  of the $H_1$-boson
mass $M_{H_1}$ and of its respective squared coupling to the $Z$ boson
$g^2_{H_1ZZ}$;  the  values  of $M_{H_1}$  and  $g^2_{H_1ZZ}$ approach
those of  $M_{H_2}$  and $g^2_{H_2ZZ}$, respectively.   In particular,
when $M_{H^+}$ approaches 1~TeV, a level crossing effect in the masses
and couplings of the $H_1$ and $H_2$ bosons takes  place and the $H_1$
boson  becomes SM-type.  In addition, the  mass of the lightest CP-odd
Higgs scalar  $M_{A_1}$ gets very  small, i.e.\ $M_{A_1} \sim 15$~GeV,
resulting from a  partial cancellation of the first  two terms  on the
RHS   of Eq.~(\ref{detMP}).   It    is  obvious that with   increasing
$|\kappa|$ and  $|\mu|$,  the predictions of  the   NMSSM start slowly
resembling  those of  the MNSSM  with $\lambda  t_S/\mu$ being  in the
vicinity of its critical value.

On  the other  hand, as  the  charged Higgs-boson  mass decreases,  we
notice  in  Fig.\ \ref{fig:nmssm4}  that  viable  scenarios occur  for
smaller values  of $|\mu|$.  In  fact, within the specific  NMSSM with
$\kappa =  0.1$ that we  have been considering here,  the experimental
constraint, $|\mu |  \stackrel{>}{{}_\sim} 90$~GeV \cite{PDG}, implies
that  $M_{H^+}$ cannot  be lighter  than 180~GeV.   Of course,  such a
scenario  could be  directly excluded  from  the fact  that for  small
positive  values of  $\kappa \sim  0.1$, the  lightest  singlino state
contributes significantly  to the $Z$-boson invisible  width (see also
discussion  after Eq.~(\ref{M55})).  For  this reason,  we present  in
Fig.~\ref{fig:nmssm10} numerical estimates for a related scenario with
negative $\kappa$,  i.e.\ $\kappa  = -0.1$. We  also choose  a smaller
value  for  $(\kappa  \mu  A_\kappa  )/\lambda$,  i.e.\  $(\kappa  \mu
A_\kappa )/\lambda = - 0.0021$~TeV$^2$, so as to obtain a light CP-odd
Higgs state  $A_1$.  In  Fig.~\ref{fig:nmssm10}, we observe  again the
same  characteristics  as in  Fig.\  \ref{fig:nmssm4},  namely as  the
charged Higgs-boson  mass decreases,  viable scenarios take  place for
smaller values of  $|\mu|$, leading to a similar  lower bound of about
180~GeV on  $M_{H^+}$.  In fact,  after having carefully  explored all
the relevant  parameter space  of the NMSSM  with $\tan\beta =  2$, we
found that this is a general feature of the NMSSM for any perturbative
value  of  $\lambda$  and   $\kappa$  (see  also  discussion  below).  
Consequently, as in the MSSM, the  SM-type Higgs boson in the NMSSM is
also predicted to be lighter than the charged Higgs boson.

To get a better understanding of this last phenomenological feature of
the NMSSM,  it  is very instructive   to analyze a  scenario where the
Yukawa-type couplings $\kappa$ and  $\lambda$ are comparable in size.  
Specifically,   we choose  $\lambda  =   0.5$ and  $\kappa  = -0.45$.  
According to our discussion in Section 4, the parameters of this model
have been chosen in a way such  that the charged  Higgs boson might be
allowed  to  become lighter than   the one  predicted in  the previous
scenarios of the NMSSM.   Furthermore, in order  to obtain the largest
possible values for the masses of the CP-even Higgs scalars, we always
fix  $A_\kappa$ by the  requirement that the  $A_1$ boson be extremely
light of the order  of a few~GeV.\footnote[6]{Despite  its similarity,
  our  scenario differs  from the  one   discussed in \cite{DM}   very
  recently.   In  our case, the  tree-level  values of $A_\lambda$ and
  $A_\kappa$, required for  $M_{A_1}\approx 0$, are  not forced to  be
  suppressed.  The latter turns out to be the  case only within a very
  narrow range of $\mu$ values close to  the upper-end of the interval
  given by Eq.~(\ref{muappr}).}  Having  the above in mind, we present
in  Fig.~\ref{fig:nmssm6}  numerical    estimates   of $M_{H_1}$   and
$M_{H_2}$,  and $g^2_{H_1ZZ}$ and  $g^2_{H_2ZZ}$, as functions of the
$\mu$-parameter, for charged Higgs-boson  masses $M_{H^+} = 120$,  400
and 800~GeV.   We    observe  that for  $|\mu|   \stackrel{>}{{}_\sim}
100$~GeV, the  $H_1$  boson  is   always SM-type.   In  addition,  for
$M_{H^+} =120$~GeV, the mass of the $H_1$ boson has a maximum of $\sim
113$~GeV  at $|\mu|=100$~GeV with  $g^2_{H_1ZZ} = 0.5$, which is close
to the present experimental lower bound  of LEP2~\cite{ADLO}.  In this
scenario, the next-to-lightest CP-even Higgs boson $H_2$ has a smaller
coupling to  the $Z$ boson and  its mass  varies between 120--130~GeV. 
For  larger values of  $M_{H^+}$, the $H_1$   boson is always SM-type,
with $M_{H_1} \approx   120$--130~GeV  for a  wide   range of  $|\mu|$
values,  whilst the $H_2$ boson  is very heavy  and decoupled from the
lightest Higgs sector.

Our numerical analysis  as  presented above  in  Fig.~\ref{fig:nmssm6}
explicitly demonstrates  that    for $M_{H^+}  =  120$~GeV,  the  mass
$M_{H_1}$ of the lightest  CP-even Higgs boson becomes acceptable only
within a very narrow interval of  $|\mu|$, which is, however, close to
its current lowest bound as set by LEP2~\cite{PDG}.  Thus, even within
this optimized scenario of the NMSSM  with $|\kappa/\lambda | \sim 1$,
the $H_1$  boson cannot become heavier than  the charged  Higgs boson. 
Therefore, we   reach the conclusion   that a possible  discovery of a
charged Higgs boson   lighter than 120--130  GeV and  a SM-type  Higgs
boson  heavier than 130--140~GeV can  only be  naturally accounted for
within the MNSSM.

\section{Conclusions}

We    have considered   the    simplest  extension   of  the   minimal
supersymmetric  standard model, in  which the $\mu$-parameter has been
promoted   to a dynamical  variable     by means of a    gauge-singlet
superfield $\widehat{S}$,  with    the linear,   quadratic  and  cubic
singlet-superfield    terms,  $\widehat{S}$,      $\widehat{S}^2$  and
$\widehat{S}^3$,  absent from  the superpotential.   Moreover, we have
assumed that   the  breaking of   SUSY  in the  observable  sector  is
communicated  gravitationally by  a  set of  hidden-sector superfields
which break $N=1$  supergravity spontaneously.  In such a supergravity
scenario, the absence  of harmful destabilizing tadpole divergences at
lower     loop  levels can  be   assured     by  forcing the  complete
superpotential and K\"ahler potential to respect specific discrete $R$
symmetries.  In  particular, we have been  able to show  that with the
imposition of the discrete  $R$  symmetries ${\cal Z}^R_5$  and ${\cal
  Z}^R_7$, the potentially dangerous  tadpole divergences first appear
at the   six-  and seven-loop levels,   respectively,  and  hence  are
naturally  suppressed to the order  of the  electroweak scale, without
destabilizing the gauge hierarchy.

The  MNSSM we  have  been  studying in this   paper  has a number   of
appealing field-theoretic and  phenomenological features, which may be
summarized as follows:
\begin{itemize}
  
\item   The   model provides a   natural    solution to the  so-called
  $\mu$-problem of the MSSM, since  the  value of the  $\mu$-parameter
  can now be  directly set by the VEV  of the gauge-singlet superfield
  $\widehat{S}$ which is of the required order of $M_{\rm SUSY}$, as a
  consequence of the ${\cal Z}^R_5$ and ${\cal Z}^R_7$ symmetries.
  
\item The  presence of the  effectively generated tadpole terms linear
  in $S$ and $F_S$ (or $\widehat{S}$) breaks explicitly the continuous
  U(1)$_{\rm  PQ}$ and its discrete subgroup  ${\cal Z}_3$.  Thus, the
  model  offers   a  natural  solution   to  the   visible  axion  and
  cosmological domain-wall problems.
  
\item  Depending on  the underlying mechanism  of   SUSY breaking, the
  effective tadpole proportional to $F_S$ could in principle be absent
  from  the model.  Such a reduction   of the renormalizable operators
  does not thwart the renormalizability of  the theory.  The resulting
  renormalizable low-energy scenario has  one parameter less than  the
  frequently-discussed NMSSM with  the  cubic  singlet-superfield term
  $\frac{\kappa}{3}\widehat{S}^3$ present; it therefore represents the
  most   economic,   renormalizable  version  among   the  non-minimal
  supersymmetric models proposed in the literature.
  
\item As opposed to the NMSSM, the MNSSM satisfies the tree-level mass
  sum  rule (\ref{sumrule}),   which    is  very analogous   to    the
  corresponding one  of the MSSM \cite{GT}.   This striking analogy to
  the MSSM  allows us to advocate that  the Higgs  sector of the MNSSM
  differs indeed  minimally from  the  one  of   the MSSM,  i.e.\  the
  introduced   model  truly   constitutes the   minimal supersymmetric
  extension of the MSSM.  In the NMSSM, the  violation of the mass sum
  rule can become  much  larger than  the one induced  by the one-loop
  stop/top effects, especially for relatively large values of $|\kappa
  |$, $|\mu|$ and $|A_\kappa|$.
  
\item A generic prediction  of the non-minimal supersymmetric standard
  models is that  for low values of $\tan\beta$,  the lightest CP-even
  Higgs-boson  mass  $M_{H_1}$  increases significantly  with  growing
  $|\lambda|$ [cf.\ Eq.\ (\ref{M0H1})].   Since in the MNSSM $\lambda$
  can take  its maximum allowed  value naturally corresponding  to the
  NMSSM with $\kappa  = 0$, the value of $M_{H_1}$  is predicted to be
  the  highest,  after  the   dominant  stop-loop  effects  have  been
  included, i.e.\ $M_{H_1} \stackrel{<}{{}_\sim} 145$ GeV.  Therefore,
  such  a scenario  can  only  be decisively  tested  by the  upgraded
  Tevatron collider at Fermilab and by the Large Hadron Collider (LHC)
  at CERN.
  
\item  The MNSSM can comfortably predict   viable scenarios, where the
  mass of  the charged Higgs boson $H^+$   is in the  range: 80~GeV~$<
  M_{H^+}    \stackrel{<}{{}_\sim}$~3~TeV,     for  phenomenologically
  relevant   values  of $|\mu   | \stackrel{>}{{}_\sim}  90$  GeV.  In
  particular, numerical estimates in  Section 5 reveal that a possible
  discovery     of  a   charged       Higgs   boson,  with    $M_{H^+}
  \stackrel{<}{{}_\sim}   120$ GeV, and   a  neutral Higgs boson, with
  $M_{H_1} \stackrel{>}{{}_\sim}  130$   GeV, can  only  be  naturally
  accounted  for within the  MNSSM, whereas the  NMSSM would be highly
  disfavoured.  This important phenomenological  feature of the MNSSM,
  which  is  very helpful to   discriminate it from  the  NMSSM,  is a
  reflection of a  new  non-trivial decoupling limit  due to  a  large
  tadpole $|t_S|$,  which  is only attainable in  the  MNSSM (see also
  discussion of the  paragraph that  includes Eq.~(\ref{tadec})).   
  
\item For scenarios with $M_{H^+}  \stackrel{>}{{}_\sim} 200$ GeV, the
  distinction between the MNSSM and  the NMSSM becomes more difficult. 
  In this case, additional experimental information would be necessary
  to distinguish the two SUSY extensions of the MSSM, resulting from a
  precise  determination of the   masses,  the widths, the   branching
  ratios and  the production cross sections  of the CP-even and CP-odd
  Higgs bosons.     Nevertheless, if  the   tadpole parameter $\lambda
  t_S/\mu$ becomes much  larger  than $M^2_{H^+}$  with  the remaining
  kinematic parameters held  fixed, the Higgs  states  $H_3$ and $A_2$
  will  be predominantly  singlets.   As an important phenomenological
  consequence of this,   the  complementarity relations  (\ref{compl})
  between the $H_{1,2}ZZ$- and $H_{2,1}A_1Z$- couplings will then hold
  approximately true in  the MNSSM.  However,  these relations will be
  generically   violated  in the   NMSSM,   as there is   no analogous
  decoupling limit in the latter model,  in which the states $H_3$ and
  $A_2$ could decouple as singlets.

\end{itemize}

The  MNSSM  also  predicts the existence  of   a light neutralino, the
axino.  The  axino  is predominantly    a singlet field,   for  $|\mu|
\stackrel{>}{{}_\sim} 120$ GeV.  LEP limits on the $Z$-boson invisible
width  lead to  the  additional constraint: $200 \stackrel{<}{{}_\sim}
|\mu  | \stackrel{<}{{}_\sim}  250$ GeV, for  $\lambda  \approx 0.65$. 
However, such a constraint disappears completely for smaller values of
$\lambda$, namely  for $\lambda \stackrel{<}{{}_\sim} 0.45$.  In fact,
the axino may become the LSP in the MNSSM.  In this paper we shall not
address the  issues  associated with the  cosmological consequences of
the axino on the reheating temperature of  the Universe \cite{GKR} and
on the dark-matter problem.  A detailed discussion of these issues may
be given elsewhere.

The present study has shown that the MNSSM is a viable scenario, which
departs minimally  from the MSSM, having a   large number of appealing
field-theoretic  and  phenemenological features.   Even though further
refinements  of our  treatment of loop  effects  might be very useful,
such  as the   inclusion  of one-loop $D$-term   contributions  to the
effective  potential    and  the computation    of  two-loop   leading
logarithmic   corrections, our  predictions for   the Higgs-boson mass
spectrum as  well as the results  of our  comparative analysis between
the MNSSM  studied  here and  the  frequently-discussed NMSSM  are not
expected to   modify dramatically.  In particular,  we  find  that the
MNSSM  can  naturally predict viable  scenarios in   which the charged
Higgs boson $H^+$ is  much lighter than  the neutral Higgs  boson with
SM-type coupling to the  $Z$ boson. The  planned colliders, i.e.\  the
upgraded   Tevatron collider     and the  LHC,  have the     potential
capabilities to test  such   interesting scenarios with a   relatively
light $H^+$, as  well  as  probe large  domains of  the   Higgs-sector
structure of this truly  minimal supersymmetric extension of the MSSM,
the MNSSM.

\subsection*{Acknowledgements}

We  thank Carlos  Wagner for  illuminating  discussions and Alexandros
Kehagias for a useful suggestion.

\newpage

\def\theequation{\Alph{section}.\arabic{equation}}
\begin{appendix}
\setcounter{equation}{0}
\section{Non-destabilizing tadpole divergences}

Employing  standard power counting rules  \cite{Bag,SA}, we shall show
the absence of harmful  tadpole divergences up  to a sufficiently high
loop order $n$, i.e.\ $n\le 5$, within the context of the supergravity
scenarios described in Section 2.

It is  useful to briefly review  first  the sufficient conditions that
govern the  absence of harmful tadpole  divergences.  To this end, let
us consider a supergraph with one external leg, i.e.\ a tadpole graph.
The tadpole  graph  may involve   a  number $V_d$  of   superpotential
vertices  of dimension $d+3$, which  are of the form $z^{d+3}/M^d_{\rm
  P}$  where $z$ represents a generic  chiral superfield, and a number
$U_d$ of  K\"ahler-potential vertices of   dimension $d+2$, which have
the  form $z^{d+2}/M^d_{\rm   P}$.  Then,  the   superficial degree of
divergence of the tadpole graph, e.g.\ that of $\widehat{S}$, is given
by
\begin{equation}
  \label{D1}
D\ =\ 1\: +\: \sum\limits_d d\,V_d\: +\: \sum\limits_d d\,U_d\, ,
\end{equation}
which leads to a contribution to the effective potential
\begin{equation}
  \label{Vtad2}
V_{\rm tad}\ \sim \ \frac{1}{(16\pi^2)^n}\
\frac{\Lambda^D\,M^{3-D+ \sum_d d\,V_d\: +\: \sum_d d\,U_d}_{\rm
SUSY}}{M^{\ \sum_d d\,V_d\: +\: \sum_d d\,U_d}_{\rm P} }\, S\ +\ {\rm h.c.}\ \sim\ 
\frac{1}{(16\pi^2)^n}\, M_{\rm P}\, M^2_{\rm SUSY}\, S\ +\ {\rm h.c.}\, ,
\end{equation}
where  $n$ counts the  number of loops and  $M_{\rm SUSY}$ is the soft
SUSY-breaking   scale.   In   obtaining  the   last   step of     Eq.\ 
(\ref{Vtad2}), we have used   $\Lambda \sim M_{\rm   P}$ as a  natural
energy cut-off scale.  This very last step in Eq.\ (\ref{Vtad2}) shows
that a tadpole contribution to the effective potential is proportional
to one power of $M_{\rm P}$ at most.  Such tadpole contributions which
remain proportional to $M_{\rm P}$ will be  referred to as ``harmful''
to  be distinguished from the ``harmless''  ones in  which the cut-off
dependence disappears. In this context,  an additional requirement for
a tadpole graph to be harmful is that $D$  be an even number. Finally,
the degree of divergence can also be determined by the number of loops
$n$ and superpotential vertices $V_d$ through the relation
\begin{equation}
  \label{D2}
D\ =\ 2n\: -\: \sum\limits_d V_d\ .
\end{equation}
In summary,  one finds  that  a set of    vertices produces a  harmful
tadpole divergence if  the following equalities are all simultaneously
satisfied:
\begin{equation}
  \label{Dcond}
D\ =\ 1\: +\: \sum\limits_d d\,V_d\: +\: \sum\limits_d d\,U_d\ =\ 
2n\: -\: \sum\limits_d V_d\ =\ {\rm even}\, ,
\end{equation}
with $D\ge 2$. 

In the next two subsections, we shall apply the power counting rule of
superficial divergences,   stated in Eq.\   (\ref{Dcond}),  to the two
models based on  the discrete $R$-symmetries $Z^R_5$  and
$Z^R_7$.

\subsection{The $Z^R_5$ case}

Here, we shall show that  the potentially harmful tadpole  divergences
are absent up to five loops.  Alternatively, we shall prove that it is
impossible to construct a tadpole   diagram from the sets of  vertices
which satisfy the condition  (\ref{Dcond}) for $n\le 5$, corresponding
to $D\le 10$.  Suppose now that at least  one superpotential vertex is
involved in  a tadpole supergraph.  Based on  Eq.\  (\ref{D2}), we see
that we need  at least two superpotential  vertices to form a  tadpole
graph with $D$  even, i.e.\ $\sum_d V_d  \ge 2$.  Thus, for $n=5$, one
has $D\le 8$,  and by virtue of Eq.\  (\ref{D1}), it is $\sum_d d\,V_d
\le 7$ and $d\le  7$. In the case  that no superpotential vertices are
involved, we have  the  relation $D =  1  +  \sum_d d U_d  \le  10$ or
$\sum_d  d U_d \le  9$ on account of Eq.\  (\ref{D1}).  We consider it
obvious that  it is impossible to  form a tadpole  graph with only one
K\"ahler-potential  vertex    of $d=9$.    This   observation excludes
K\"ahler-potential operators  of  $d=9$.  Furthermore, as we  will see
below,  the  imposition  of  ${\cal Z}^R_5$  on the  complete K\"ahler
potential  does not  permit operators of   $d=1$.   If we now  wish to
satisfy the above constraint  $\sum_d d U_d \le  9$ with two vertices,
we then need  one  operator of  $d=2$  and another  one of $d=7$;  the
latter is the  K\"ahler-potential  term of the highest  dimensionality
that could participate into a  harmful divergent tadpole graph with $n
\le    5$.  Consequently,   we    reach  the   conclusion  that   only
superpotential and  K\"ahler-potential vertices with  $d\le 7$ will be
of relevance here.

We  shall confine  ourselves to  a minimal  model, in  which  only the
superfields  $\widehat{H}_1$,  $\widehat{H}_2$  and $\widehat{S}$  are
present and ignore quark and lepton superfields, as they do not couple
directly to $\widehat{S}$; the inclusion of the fermion superfields is
straightforward and does not alter our results. Moreover, we shall not
include  in the  list of  K\"ahler-potential terms  those  obtained by
multiplying  the  latter  with  any  power  of  $\widehat{H}^\dagger_1
\widehat{H}_1$,         $\widehat{H}^\dagger_2         \widehat{H}_2$,
$\widehat{S}^\ast \widehat{S}$.  The reason is that  the omitted terms
generate graphs of higher loop order than the included ones.

Having the above in  mind, we are now able  to list all superpotential
and K\"ahler-potential terms  of  $d\le  7$, respecting the   discrete
$R$-symmetry ${\cal Z}^R_5$ (cf.\ Eq.\ (\ref{Z5R})):
\begin{eqnarray}
  \label{W}
W:\quad 
W_0 \!&\equiv&\! \widehat{S} (\widehat{H}^T_1 i\tau_2 \widehat{H}_2)\,
\delta (\bar{\theta} )\
     +\ {\rm h.c.},\qquad\  
W_1\ \equiv\ \frac{\widehat{S}^4}{M_{\rm P}}\,
\delta (\bar{\theta} )\ +\ {\rm h.c.},\nonumber\\
W_3 \!&\equiv&\!
\frac{(\widehat{H}^T_1 i\tau_2 \widehat{H}_2)^3}{M^3_{\rm P}}
\, \delta (\bar{\theta} )\ 
+\ {\rm h.c.},\qquad\ \
W_4\ \equiv\ 
\frac{\widehat{S}^3 (\widehat{H}^T_1 i\tau_2 \widehat{H}_2)^2}{M^4_{\rm P}}\,
 \delta (\bar{\theta} )\ +\ {\rm h.c.},\nonumber\\
W_5 \!&\equiv&\!
\frac{\widehat{S}^6 (\widehat{H}^T_1 i\tau_2 \widehat{H}_2)}{M^5_{\rm
    P}}\, \delta (\bar{\theta} )\ +\ {\rm h.c.},\qquad
W_6\ \equiv\ 
\frac{\widehat{S}^9}{M^6_{\rm P}}\, \delta (\bar{\theta} )\ 
+\ {\rm h.c.},\nonumber\\
W_7 \!&\equiv&\! 
\frac{\widehat{S}^2 (\widehat{H}^T_1 i\tau_2
  \widehat{H}_2)^4}{M^7_{\rm P}}\, \delta (\bar{\theta} )\ +\ 
{\rm h.c.}\\[0.3cm]
  \label{K}
K:\quad
K^{(1)}_0 \!&\equiv&\! \widehat{H}^\dagger_1 \widehat{H}_1\,,\qquad
K^{(2)}_0\ \equiv\ \widehat{H}^\dagger_2 \widehat{H}_2\,,\qquad
K^{(3)}_0\ \equiv\ \widehat{S}^\ast \widehat{S}\,,\nonumber\\[0.2cm]
K_2 \!&\equiv&\! \frac{\widehat{S}^2 (\widehat{H}^T_1 i\tau_2
  \widehat{H}_2)}{M^2_{\rm P}}\ +\ {\rm h.c.},\qquad\ \
K^{(1)}_3\ \equiv\ \frac{\widehat{S}^5}{M^3_{\rm P}}\ +\ {\rm h.c.},\nonumber\\
K^{(2)}_3 \!&\equiv&\! \frac{\widehat{S}^{\ast 3} (\widehat{H}^T_1 i\tau_2
  \widehat{H}_2)}{M^3_{\rm P}}\ +\ {\rm h.c.},\qquad\
K^{(3)}_3\ \equiv\ \frac{\widehat{S}^\ast (\widehat{H}^T_1 i\tau_2
  \widehat{H}_2)^2}{M^3_{\rm P}}\ +\ {\rm h.c.},\nonumber\\
K_5 \!&\equiv&\! \frac{\widehat{S} (\widehat{H}^T_1 i\tau_2
  \widehat{H}_2)^3}{M^5_{\rm P}}\ +\ {\rm h.c.},\qquad\quad\
K_6\ \equiv\ \frac{\widehat{S}^4 (\widehat{H}^T_1 i\tau_2
  \widehat{H}_2)^2}{M^6_{\rm P}}\ +\ {\rm h.c.},\nonumber\\
K_7 \!&\equiv&\! \frac{\widehat{S}^7 (\widehat{H}^T_1 i\tau_2
  \widehat{H}_2)}{M^7_{\rm P}}\ +\ {\rm h.c.},
\end{eqnarray}
where $\delta  (\bar{\theta}   )$    is the  usual    Grassmann-valued
$\delta$-function.  Notice that the terms $K^{(1)}_0$, $K^{(2)}_0$ and
$K^{(3)}_0$ represent  the usual Higgs-superfield propagators and have
no direct effect on our power counting  rules.  These terms are merely
needed  to contract the superfields in  propagator lines and so form a
loop  supergraph. Furthermore, from  Eqs.\ (\ref{W}) and (\ref{K}), we
observe that ${\cal Z}^R_5$  forbids the appearance  of superpotential
operators of $d=2$ ($W_2$)  and of K\"ahler-potential terms of $d=1,4$
($K_1$,$K_4$).

In the following, we shall systematically analyze all possible sets of
vertices compatible  with the conditions  in Eq.\  (\ref{Dcond}) up to
five loops.  At the one-loop level ($n=1$), with  $\sum_d V_d = 0$, we
readily find  from Eq.\ (\ref{D1}) that $\sum_d  d U_d = 1$, entailing
the absence   of  contributing   operators.   The   situation  becomes
increasingly more    involved for  $n =    2,\ 3,\ 4$    and  5.  More
explicitly, our systematic  search  for   the existence of    possible
harmful tadpoles may be summarized as follows:
\begin{eqnarray}
  \label{n2}
{\bf I.}\quad n = 2:\qquad && \nonumber\\
{\rm a)} && D\ =\ 2,\qquad \sum_d V_d\ =\ 2,\qquad 
\sum_d d\,V_d\: +\: \sum_d d\,U_d\ =\ 1:\nonumber\\[-0.1cm]
&& \{ W_0,\ W_1 \};\nonumber\\[0.2cm]
{\rm b)} && D\ =\ 4,\qquad \sum_d V_d\ =\ 0,\qquad 
\sum_d d\,U_d\ =\ 3:\nonumber\\[-0.1cm]
&& \{ K^{(i)}_3 \}.\\
  \label{n3}
{\bf II.}\quad n = 3:\qquad && \nonumber\\
{\rm a)} && D\ =\ 2,\qquad \sum_d V_d\ =\ 4,\qquad 
\sum_d d\,V_d\: +\: \sum_d d\,U_d\ =\ 1:\nonumber\\[-0.1cm]
&& \{ 3W_0,\ W_1 \};\nonumber\\[0.2cm]
{\rm b)} && D\ =\ 4,\qquad \sum_d V_d\ =\ 2,\qquad 
\sum_d d\,V_d\: +\: \sum_d d\,U_d\ =\ 3:\nonumber\\[-0.1cm]
&& \{ W_0,\ W_3\},\quad \{ W_0,\ W_1,\ K_2\},\quad 
\{ 2W_0,\  K^{(i)}_3 \}; \nonumber\\
{\rm c)} && D\ =\ 6,\qquad \sum_d V_d\ =\ 0,\qquad 
\sum_d d\,U_d\ =\ 5:\nonumber\\[-0.1cm]
&& \{ K_5 \},\quad \{ K_2,\  K^{(i)}_3 \}.\\
  \label{n4}
{\bf III.}\quad n = 4:\qquad && \nonumber\\
{\rm a)} && D\ =\ 2,\qquad \sum_d V_d\ =\ 6,\qquad 
\sum_d d\,V_d\: +\: \sum_d d\,U_d\ =\ 1:\nonumber\\[-0.1cm]
&& \{ 5W_0,\ W_1 \};\nonumber\\[0.2cm]
{\rm b)} && D\ =\ 4,\qquad \sum_d V_d\ =\ 4,\qquad 
\sum_d d\,V_d\: +\: \sum_d d\,U_d\ =\ 3:\nonumber\\[-0.1cm]
&& \{ 3W_0,\ W_3\},\quad \{W_0,\ 3W_1\},\quad 
\{ 3W_0,\ W_1,\ K_2\},\nonumber\\ 
&&\{ 4W_0,\  K^{(i)}_3 \};\nonumber\\[0.2cm]
{\rm c)} && D\ =\ 6,\qquad \sum_d V_d\ =\ 2,\qquad 
\sum_d d\,V_d\: +\: \sum_d d\,U_d\ =\ 5:\nonumber\\[-0.1cm]
&& \{ W_0,\ W_5\},\quad \{ W_1,\ W_4\},\quad \{ W_0,\ W_3,\
K_2\},\quad \{ W_0,\ W_1,\ 2K_2\},\nonumber\\
&&\{ 2W_1,\  K^{(i)}_3 \},\quad \{ 2W_0,\ K_5 \},\quad 
\{ 2W_0,\ K_2,\  K^{(i)}_3 \}; \nonumber\\
{\rm d)} && D\ =\ 8,\qquad \sum_d V_d\ =\ 0,\qquad 
\sum_d d\,U_d\ =\ 7:\nonumber\\[-0.1cm]
&& \{ 2K_2,\  K^{(i)}_3 \},\quad \{ K_2,\ K_5 \},\quad \{ K_7 \}.\\
  \label{n5}
{\bf IV.}\quad n = 5:\qquad && \nonumber\\
{\rm a)} && D\ =\ 2,\qquad \sum_d V_d\ =\ 8,\qquad 
\sum_d d\,V_d\: +\: \sum_d d\,U_d\ =\ 1:\nonumber\\[-0.1cm]
&& \{ 7W_0,\ W_1 \};\nonumber\\[0.2cm]
{\rm b)} && D\ =\ 4,\qquad \sum_d V_d\ =\ 6,\qquad 
\sum_d d\,V_d\: +\: \sum_d d\,U_d\ =\ 3:\nonumber\\[-0.1cm]
&& \{ 3W_0,\ 3W_1\},\quad \{ 5W_0,\ W_3\},\quad 
\{5W_0,\ W_1,\ K_2\},\quad \{ 6W_0,\ K^{(i)}_3\};\nonumber\\[0.2cm] 
{\rm c)} && D\ =\ 6,\qquad \sum_d V_d\ =\ 4,\qquad 
\sum_d d\,V_d\: +\: \sum_d d\,U_d\ =\ 5:\nonumber\\[-0.1cm]
&& \{ 3W_0,\ W_5\},\quad \{ 2W_0,\ W_1,\ W_4\},\quad 
\{ W_0,\ 2W_1,\ W_3\},\nonumber\\
&& \{ 3W_0,\ W_3,\ K_2\},\quad 
\{ W_0,\ 3W_1,\ K_2\},\quad \{ 2W_0,\ 2W_1,\ K^{(i)}_3 \},\nonumber\\
&& \{ 3W_0,\ W_1,\ 2K_2\},\quad 
\{ 4W_0,\ K_5\},\quad \{ 4W_0,\ K_2,\ K^{(i)}_3 \};\nonumber\\[0.2cm]
{\rm d)} && D\ =\ 8,\qquad \sum_d V_d\ =\ 2,\qquad 
\sum_d d\,V_d\: +\: \sum_d d\,U_d\ =\ 7:\nonumber\\[-0.1cm]
&& \{ W_0,\ W_7\},\quad \{ W_1,\ W_6\},\quad \{ W_3,\ W_4\},\quad 
\{ W_0,\ W_5,\ K_2\},\nonumber\\
&& \{ W_1,\ W_4,\ K_2\},\quad \{ W_0,\ W_4,\ K^{(i)}_3 \},\quad 
\{ W_1,\ W_3,\ K^{(i)}_3 \},\nonumber\\
&&\{ W_0,\ W_3,\ 2K_2 \},\quad 
\{ 2W_1,\ K_5\},\quad \{ 2W_1,\ K_2,\ K^{(i)}_3 \},\nonumber\\
&& \{ W_0,\ W_1,\ K_6\},\quad \{ W_0,\ W_1,\ 3K_2\},\quad 
\{ W_0,\ W_1,\ K^{(i)}_3,\ K^{(j)}_3 \},\nonumber\\
&&\{ 2W_0,\ K_7\},\quad \{ 2W_0,\ K_2,\ K_5\},\quad 
\{ 2W_0,\ 2K_2,\ K^{(i)}_3\};\nonumber\\[0.2cm]
{\rm e)} && D\ =\ 10,\qquad \sum_d V_d\ =\ 0,\qquad 
\sum_d d\,U_d\ =\ 9:\nonumber\\[-0.1cm]
&& \{ K_2,\ K_7 \},\quad \{ 2K_2,\ K_5 \},\quad 
\{ 3K_2,\ K^{(i)}_3 \},\quad \{ K_6,\ K^{(i)}_3 \},\nonumber\\
&& \{ K^{(i)}_3,\ K^{(j)}_3,\ K^{(k)}_3 \}.
\end{eqnarray}
Here $i,j,k=1,2,3$.   The remaining task is to  show  that the sets of
vertices listed above do not produce tadpole graphs.  This can be best
verified  case by case algebraically in  the following manner.  First,
we    multiply  all the  vertices  belonging  to   a set  and formally
substitute        $\widehat{H}^\dagger_1               \widehat{H}_1,\ 
\widehat{H}^\dagger_2     \widehat{H}_2$    and      $\widehat{S}^\ast
\widehat{S}$ with  1 into the product of  vertices.  Then,  we examine
whether terms linear in $\widehat{S}$ or $\widehat{S}^\ast$ survive in
the resulting expression.  In this way, we have carefully checked that
there are no such  terms linear in $\widehat{S}$ or $\widehat{S}^\ast$
for all sets of vertices listed  in Eqs.\ (\ref{n2})--(\ref{n5}), thus
implying the absence of harmful tadpole graphs up to five loops.

At a higher loop level, we can  construct tadpole supergraphs by making
free use   of the renormalizable superpotential   vertex $W_0$ in Eq.\ 
(\ref{W}) together  with some of the above vertices e.g., 
the    higher-dimensional  K\"ahler-potential
vertices $K_2$  and $K_5$ defined in  Eq.\ (\ref{K}). Specifically, we
find that the set of vertices
\begin{equation}
  \label{n6}
\{ 4W_0,\ K_2,\ K_5\}
\end{equation}
leads  to the  typical  six-loop    tadpole graph depicted  in   Fig.\ 
\ref{f1}(a).  Also, it is not difficult to see that the above graph is
actually   a harmful  divergent  one since   the   set of vertices  in
(\ref{n6}) satisfies the global constraint of Eq.\ (\ref{Dcond}), with
$n=6$, $D=8=$ even, $\sum_d V_d = 4$, and $\sum_d d V_d + \sum_d d U_d
= \sum_d d\,U_d = 7$.

\subsection{The $Z^R_7$ case}

In this section we shall show that  the symmetry $Z^R_7$ prohibits the
presence of all possible harmful tadpole divergences up  to six loops. 
Following a line of arguments similar to the $Z^R_5$ case, we conclude
that only superpotential and K\"ahler-potential  operators with $d \le
9$ are of  interest  in this  case.   Therefore, we list all  possible
vertices  of $d\le  9$,  respecting the  discrete  $R$-symmetry ${\cal
  Z}^R_7$ (cf.\ Eq.\ (\ref{Z7R})):
\begin{eqnarray}
  \label{W7}
W:\quad 
W_0 \!&\equiv&\! \widehat{S} (\widehat{H}^T_1 i\tau_2 \widehat{H}_2)\,
\delta (\bar{\theta} )\
     +\ {\rm h.c.},\qquad\  
W_2\ \equiv\ \frac{\widehat{S}^5}{M^2_{\rm P}}\,
\delta (\bar{\theta} )\ +\ {\rm h.c.},\nonumber\\
W_3 \!&\equiv&\!
\frac{(\widehat{H}^T_1 i\tau_2 \widehat{H}_2)^3}{M^3_{\rm P}}
\, \delta (\bar{\theta} )\ 
+\ {\rm h.c.},\qquad\ \
W_5\ \equiv\ 
\frac{\widehat{S}^4 (\widehat{H}^T_1 i\tau_2 \widehat{H}_2)^2}{M^5_{\rm P}}\,
 \delta (\bar{\theta} )\ +\ {\rm h.c.},\nonumber\\
W_7 \!&\equiv&\!
\frac{\widehat{S}^8 (\widehat{H}^T_1 i\tau_2 \widehat{H}_2)}{M^7_{\rm
    P}}\, \delta (\bar{\theta} )\ +\ {\rm h.c.},\qquad
W_8\ \equiv\ 
\frac{\widehat{S}^3 (\widehat{H}^T_1 i\tau_2
  \widehat{H}_2)^4}{M^8_{\rm P}}\, \delta (\bar{\theta} )\ 
+\ {\rm h.c.},\nonumber\\
W_9 \!&\equiv&\! 
\frac{\widehat{S}^{12}}{M^9_{\rm P}}\, \delta (\bar{\theta} )\ +\ 
{\rm h.c.}\\[0.3cm]
  \label{K7}
K:\quad
K^{(1)}_0 \!&\equiv&\! \widehat{H}^\dagger_1 \widehat{H}_1\,,\qquad
K^{(2)}_0\ \equiv\ \widehat{H}^\dagger_2 \widehat{H}_2\,,\qquad
K^{(3)}_0\ \equiv\ \widehat{S}^\ast \widehat{S}\,,\nonumber\\[0.2cm]
K^{(1)}_3 \!&\equiv&\! \frac{\widehat{S}^3 (\widehat{H}^T_1 i\tau_2
  \widehat{H}_2)}{M^3_{\rm P}}\ +\ {\rm h.c.},\qquad\ \
K^{(2)}_3\ \equiv\ \frac{\widehat{S}^\ast (\widehat{H}^T_1 i\tau_2
  \widehat{H}_2)^2}{M^3_{\rm P}}\ +\ {\rm h.c.},\nonumber\\
K_4 \!&\equiv&\! \frac{\widehat{S}^{\ast 4} (\widehat{H}^T_1 i\tau_2
  \widehat{H}_2)}{M^4_{\rm P}}\ +\ {\rm h.c.},\qquad\quad
K_5\ \equiv\ \frac{\widehat{S}^7}{M^5_{\rm P}}\ +\ {\rm h.c.},\nonumber\\
K_6 \!&\equiv&\! \frac{\widehat{S}^2 (\widehat{H}^T_1 i\tau_2
  \widehat{H}_2)^3}{M^6_{\rm P}}\ +\ {\rm h.c.},\qquad\
K^{(1)}_8\ \equiv\ \frac{\widehat{S}^6 (\widehat{H}^T_1 i\tau_2
  \widehat{H}_2)^2}{M^8_{\rm P}}\ +\ {\rm h.c.},\nonumber\\
K^{(2)}_8  \!&\equiv&\! \frac{\widehat{S}^{\ast 2} 
(\widehat{H}^T_1 i\tau_2 \widehat{H}_2)^4}{M^8_{\rm P}}\ +\ {\rm
h.c.},\qquad
K^{(1)}_9\ \equiv\ \frac{\widehat{S} (\widehat{H}^T_1 i\tau_2 
\widehat{H}_2)^5}{M^9_{\rm P}}\ +\ {\rm h.c.},\nonumber\\
K^{(2)}_9  \!&\equiv&\! \frac{\widehat{S}^{\ast 5} 
(\widehat{H}^T_1 i\tau_2 \widehat{H}_2)^3}{M^9_{\rm P}}\ +\ 
{\rm h.c.}\,.
\end{eqnarray}
Note that  the   symmetry ${\cal  Z}^R_7$  forbids  the  occurrence of
superpotential operators of $d=1,4,6$ ($W_1,W_4,W_6$)   as well as  of
K\"ahler-potential terms of $d=1,2,7$ ($K_1,K_2,K_7$).

As we did for the $Z^R_5$ case, we shall determine all possible sets of
vertices compatible with the  conditions  in Eq.\ (\ref{Dcond}) up  to
six loops. Again,  it  is not  difficult to  see that at  the one-loop
level ($n=1$), with $\sum_d  V_d = 0$ and  $\sum_d d U_d  = 1$, one is
unable   to find contributing operators.   Furthermore, the absence of
$d=1$ operators leads to the constraint
\begin{equation}
\sum\limits_d d\,V_d\: +\: \sum\limits_d d\,U_d\ =\ D\: -\: 1\ >\ 1\,,
\end{equation}
i.e.\ $D > 2$.  This observation simplifies further the search for the
existence of possible harmful tadpoles. Thus, for $n  = 2,\ 3,\ 4,\ 5$
and 6, we find the set of vertices:
\begin{eqnarray}
  \label{nn2}
{\bf I.}\quad n = 2:\qquad && \nonumber\\
{\rm a)} && D\ =\ 4,\qquad \sum_d V_d\ =\ 0,\qquad 
\sum_d d\,U_d\ =\ 3:\nonumber\\[-0.1cm]
&& \{ K^{(i)}_3 \}.\\
  \label{nn3}
{\bf II.}\quad n = 3:\qquad && \nonumber\\
{\rm a)} && D\ =\ 4,\qquad \sum_d V_d\ =\ 2,\qquad 
\sum_d d\,V_d\: +\: \sum_d d\,U_d\ =\ 3:\nonumber\\[-0.1cm]
&& \{ W_0,\ W_3\},\quad \{ 2W_0,\  K^{(i)}_3 \}; \nonumber\\
{\rm b)} && D\ =\ 6,\qquad \sum_d V_d\ =\ 0,\qquad 
\sum_d d\,U_d\ =\ 5:\nonumber\\[-0.1cm]
&& \{ K_5 \}\, .\\
  \label{nn4}
{\bf III.}\quad n = 4:\qquad && \nonumber\\
{\rm a)} && D\ =\ 4,\qquad \sum_d V_d\ =\ 4,\qquad 
\sum_d d\,V_d\: +\: \sum_d d\,U_d\ =\ 3:\nonumber\\[-0.1cm]
&& \{ 3W_0,\ W_3\},\quad \{ 4W_0,\ K^{(i)}_3\};\nonumber\\[0.2cm] 
{\rm b)} && D\ =\ 6,\qquad \sum_d V_d\ =\ 2,\qquad 
\sum_d d\,V_d\: +\: \sum_d d\,U_d\ =\ 5:\nonumber\\[-0.1cm]
&& \{ W_0,\ W_5\},\quad \{ W_2,\ W_3\},\quad \{ W_0,\ W_2,\
K^{(i)}_3 \},\quad \{ 2W_0,\ K_5\}; \nonumber\\
{\rm c)} && D\ =\ 8,\qquad \sum_d V_d\ =\ 0,\qquad 
\sum_d d\,U_d\ =\ 7:\nonumber\\[-0.1cm]
&& \{ K^{(i)}_3,\ K_4 \}.\\
  \label{nn5}
{\bf IV.}\quad n = 5:\qquad && \nonumber\\
{\rm a)} && D\ =\ 4,\qquad \sum_d V_d\ =\ 6,\qquad 
\sum_d d\,V_d\: +\: \sum_d d\,U_d\ =\ 3:\nonumber\\[-0.1cm]
&& \{ 5W_0,\ W_3\},\quad \{ 6W_0,\ K^{(i)}_3\};\nonumber\\[0.2cm] 
{\rm b)} && D\ =\ 6,\qquad \sum_d V_d\ =\ 4,\qquad 
\sum_d d\,V_d\: +\: \sum_d d\,U_d\ =\ 5:\nonumber\\[-0.1cm]
&& \{ 3W_0,\ W_5\},\quad \{ 2W_0,\ W_2,\ W_3\},\quad 
\{ 3W_0,\ W_2,\ K^{(i)}_3 \},\quad \{ 4W_0,\ K_5\};\nonumber\\[0.2cm]
{\rm c)} && D\ =\ 8,\qquad \sum_d V_d\ =\ 2,\qquad 
\sum_d d\,V_d\: +\: \sum_d d\,U_d\ =\ 7:\nonumber\\[-0.1cm]
&& \{ W_0,\ W_7\},\quad \{ W_2,\ W_5\},\quad \{ W_0,\ W_2,\ K_5\},\nonumber\\
&& \{ W_0,\ W_3,\ K_4\},\quad \{ 2W_2,\ K^{(i)}_3 \},\quad 
\{ 2W_0,\ K^{(i)}_3,\ K_4 \};\nonumber\\[0.2cm]
{\rm d)} && D\ =\ 10,\qquad \sum_d V_d\ =\ 0,\qquad 
\sum_d d\,U_d\ =\ 9:\nonumber\\[-0.1cm]
&& \{ K^{i}_9 \},\quad \{ K_4,\ K_5 \},\quad 
\{ K^{(i)}_3,\ K_6 \},\quad \{ K^{(i)}_3,\ K^{(j)}_3,\ K^{(k)}_3
\};\nonumber\\[0.2cm]
  \label{nn6}
{\bf V.}\quad n = 6:\qquad && \nonumber\\
{\rm a)} && D\ =\ 4,\qquad \sum_d V_d\ =\ 8,\qquad 
\sum_d d\,V_d\: +\: \sum_d d\,U_d\ =\ 3:\nonumber\\[-0.1cm]
&& \{ 7W_0,\ W_3\},\quad \{ 8W_0,\ K^{(i)}_3\};\nonumber\\[0.2cm] 
{\rm b)} && D\ =\ 6,\qquad \sum_d V_d\ =\ 6,\qquad 
\sum_d d\,V_d\: +\: \sum_d d\,U_d\ =\ 5:\nonumber\\[-0.1cm]
&& \{ 5W_0,\ W_5\},\quad \{ 4W_0,\ W_2,\ W_3\},\quad 
\{ 5W_0,\ W_2,\ K^{(i)}_3 \},\quad \{ 6W_0,\ K_5\};\nonumber\\[0.2cm]
{\rm c)} && D\ =\ 8,\qquad \sum_d V_d\ =\ 4,\qquad 
\sum_d d\,V_d\: +\: \sum_d d\,U_d\ =\ 7:\nonumber\\[-0.1cm]
&& \{ 3W_0,\ W_7\},\quad \{ 2W_0,\ W_2,\ W_5\},\quad 
\{ W_0,\ 2W_2,\ W_3\},\nonumber\\
&& \{ 2W_0,\ 2W_2,\ K^{(i)}_3 \},\quad 
\{ 3W_0,\ W_3,\  K_4 \},\quad \{ 3W_0,\ W_2,\  K_5 \},\nonumber\\
&& \{ 4W_0,\ K^{(i)}_3,\  K_4 \};\nonumber\\[0.2cm]
{\rm d)} && D\ =\ 10,\qquad \sum_d V_d\ =\ 2,\qquad 
\sum_d d\,V_d\: +\: \sum_d d\,U_d\ =\ 9:\nonumber\\[-0.1cm]
&& \{ W_0,\ W_9\},\quad \{ W_2,\ W_7\},\quad \{ 2W_0,\ K^{(i)}_9\},\nonumber\\
&& \{ W_0,\ W_3,\ K_6\},\quad \{ 2W_2,\ K_5 \},\quad 
\{ W_0,\ W_5, K_4 \},\nonumber\\
&& \{ W_2,\ W_3, K_4 \},\quad \{ 2W_3,\ K^{(i)}_3 \},\quad
\{ W_0,\ W_2,\ K^{(i)}_3,\ K_4 \},\nonumber\\
&& \{ 2W_0,\ K^{(i)}_3,\ K_6 \},\quad  
\{ 2W_0,\ K_4,\ K_5 \},\quad 
\{ W_0,\ W_3,\ K^{(i)}_3,\ K^{(j)}_3 \},\nonumber\\
&&\{ 2W_0,\ K^{(i)}_3,\ K^{(j)}_3,\ K^{(k)}_3 \};\nonumber\\[0.2cm]
{\rm e)} && D\ =\ 12,\qquad \sum_d V_d\ =\ 0,\qquad 
\sum_d d\,U_d\ =\ 11:\nonumber\\[-0.1cm]
&& \{ K^{(i)}_3,\ K^{(j)}_8 \},\quad \{ K_5,\ K_6 \},\quad 
\{ K^{(i)}_3,\ 2K_4 \}, \nonumber\\
&& \{ K^{(i)}_3, K^{(j)}_3,\ K_5 \}\, .
\end{eqnarray}
The   indices  $i,j,k$   take admissible  values    according  to Eq.\ 
(\ref{K7}).   Again, we  have carefully  checked  that terms linear in
$\widehat{S}$ or $\widehat{S}^\ast$ do not survive when the product of
all vertices within each set  listed in Eqs.\ (\ref{nn2})--(\ref{nn6})
is formed by  formally replacing  the bilinears $\widehat{H}^\dagger_1
\widehat{H}_1,\     \widehat{H}^\dagger_2     \widehat{H}_2$       and
$\widehat{S}^\ast \widehat{S}$ with 1.

Nevertheless, at the seven-loop level, we  can still construct tadpole
supergraphs  by   combining  the renormalizable  superpotential vertex
$W_0$  four  times   with the  higher-dimensional   K\"ahler-potential
vertices $K^{(1)}_3$ and $K_6$ in Eq.\ (\ref{K7}). In other words, the
set of vertices
\begin{equation}
  \label{nn7}
\{ 4W_0,\ K^{(1)}_3,\ K_6\}
\end{equation}
gives  rise   to  the  typical  seven-loop   tadpole  graph   of Fig.\ 
\ref{f1}(b).  Finally, we can check that the global constraint of Eq.\ 
(\ref{Dcond}) is satisfied, with  $n=7$,  $D=10=$ even, $\sum_d V_d  =
4$, and $\sum_d d V_d + \sum_d d U_d = \sum_d d\,U_d = 9$.

\setcounter{equation}{0}
\section{The Peccei--Quinn-symmetric limit}

In  this appendix  we shall  derive the  analytic expressions  for the
Higgs-boson masses  and couplings  pertinent to the  two gauge-singlet
SUSY  extensions  of  the   MSSM  in  the  PQ-symmetric  limit,  i.e.\ 
$\kappa/\lambda,\  t_S,\  m^2_{12} \to  0$.   Of  course, a  kinematic
situation  close  to the  PQ-symmetric  limit  can  more naturally  be
realized in the NMSSM rather than in the MNSSM where $\lambda t_S/\mu$
is   expected  to  be   unsuppressed  of   order  $M^2_{\rm   SUSY}$.  
Additionally, we  shall assume that $M_{H^+}\gg  M_W$.  For notational
simplicity, we have everywhere dropped the superscript (0), e.g.\ from
$M^{2(0)}_{H^+}$,  $M^{2(0)}_a$ etc.,  as all  quantities  involved in
this appendix are evaluated at the tree level.

In the limit of a heavy $H^+$, the quantity $\delta = \sqrt{ \lambda^2
  v^2/(2M^2_a)}$ defined in Eq.~(\ref{delta})  is much less than~1 and
therefore serves  as an expansion  parameter in our  calculations.  In
this limit, it is a reasonable approximation to set $\mu$ to its value
in  the   middle  of   the  allowed  $\mu^2$-interval   determined  by
Eq.~(\ref{muint}), i.e.~$\mu^2 = \mu^2_{\rm mid} = s^2_\beta c^2_\beta
M^2_a$, at  which $M_{H_1}$ is  expected to approximately  acquire its
maximum.  Then, the tree-level CP-even Higgs-boson mass matrix $M^2_S$
may  be  cast,  up  to  terms  of order  $\delta^2  M^2_a$,  into  the
approximately diagonal form:
\begin{equation}
  \label{MSPQ}
(O^H)^T M^2_S O^H = \left(\! \begin{array}{ccc}
\frac{1}{2}\,\lambda^2 v^2 s^2_{2\beta} & 0 & 0 \\
0 & M^2_Z c^2_{2\beta} + \frac{1}{2}\,\lambda^2 v^2 s^2_{2\beta}
& \Big( \frac{1}{2}\, \lambda^2 v^2 - M^2_Z \Big) s_{2\beta}
c_{2\beta} \\
0 & \Big( \frac{1}{2}\, \lambda^2 v^2 - M^2_Z \Big) s_{2\beta}
c_{2\beta} & M^2_a + \frac{1}{2}\,\lambda^2 v^2 c^2_{2\beta}
- \Big( \frac{1}{2}\, \lambda^2 v^2 - M^2_Z \Big) s^2_{2\beta}
\end{array} \!\right)\,,
\end{equation}
by virtue of the orthogonal matrix $O^H$
\begin{equation}
  \label{OHPQ}
O^H \ =\ \left(\! \begin{array}{ccc}
{\rm sign}\,(\lambda\mu)\, \delta s_\beta c_{2\beta} & c_\beta & -s_\beta\,
(1 - \frac{1}{2}\,\delta^2 c^2_{2\beta} ) \\
-{\rm sign}\,(\lambda\mu)\, \delta c_\beta c_{2\beta} & s_\beta & 
c_\beta\, (1 - \frac{1}{2}\,\delta^2 c^2_{2\beta} )\\
1 - \frac{1}{2}\,\delta^2 c^2_{2\beta} & 0 & {\rm sign}\,(\lambda\mu)\,
\delta c_{2\beta} 
\end{array} \!\right)\ + \ {\cal O}(\delta^3 )\, ,
\end{equation}
where we have used the  short-hand notation $s_{2\beta} = \sin 2\beta$
and $c_{2\beta} = \cos 2\beta$.

Likewise, the  orthogonal matrix $O^A$, which  diagonalizes the CP-odd
Higgs-boson mass  matrix $M^2_P$ in the PQ-symmetric  limit, is easily
found to be
\begin{equation}
  \label{OPPQ}  
O^A\ =\ \frac{1}{\sqrt{1+\delta^2}}\ \left(\! \begin{array}{cc}
-{\rm sign}\,(\lambda\mu)\, \delta & 1 \\
-1 & -{\rm sign}\,(\lambda\mu)\, \delta \end{array} \!\right)\ .
\end{equation}
In  the PQ-symmetric  limit, the  CP-odd mass  matrix $M^2_P$  has one
massless eigenstate $A_1$ and one massive one $A_2$, with $M^2_{A_2} =
M^2_a + \frac{1}{2}\,\lambda^2 v^2$.

Substituting    Eqs.~(\ref{OHPQ})   and   (\ref{OPPQ})    into   Eqs.\ 
(\ref{gHHZ}) and (\ref{gHAZ}), we  obtain, up to order $\delta^2$, all
the couplings $H_iZZ$ and $H_iA_jZ$:
\begin{eqnarray}
  \label{couplPQ}
g_{H_1ZZ} \!& =&\! 0\,,\quad 
g_{H_2ZZ} \ =\ 1,\quad  g_{H_3ZZ}\ =\ 0\,,\nonumber\\
g_{H_1A_1Z} \!& =&\! -\,\delta^2\, c_{2\beta}\,,\quad 
g_{H_2A_1Z}\ =\ 0\,,\quad g_{H_3A_1Z}\ =\ 
                        -{\rm sign}\, (\lambda\mu )\, \delta\,,\nonumber\\
g_{H_1A_2Z} \!& =&\! {\rm sign}\, (\lambda\mu )\, \delta\, c_{2\beta}\,,\quad 
g_{H_2A_2Z}\ =\ 0\,,\quad g_{H_3A_2 Z}\ =\ 
           1 - \frac{\delta^2}{2}\, (1 + c^2_{2\beta})\, .\quad
\end{eqnarray}

\end{appendix}

%
%%%
%

\begin{figure}[p]
   \leavevmode
 \begin{center}
   \epsfxsize=16.5cm
    \epsffile[0 0 539 652]{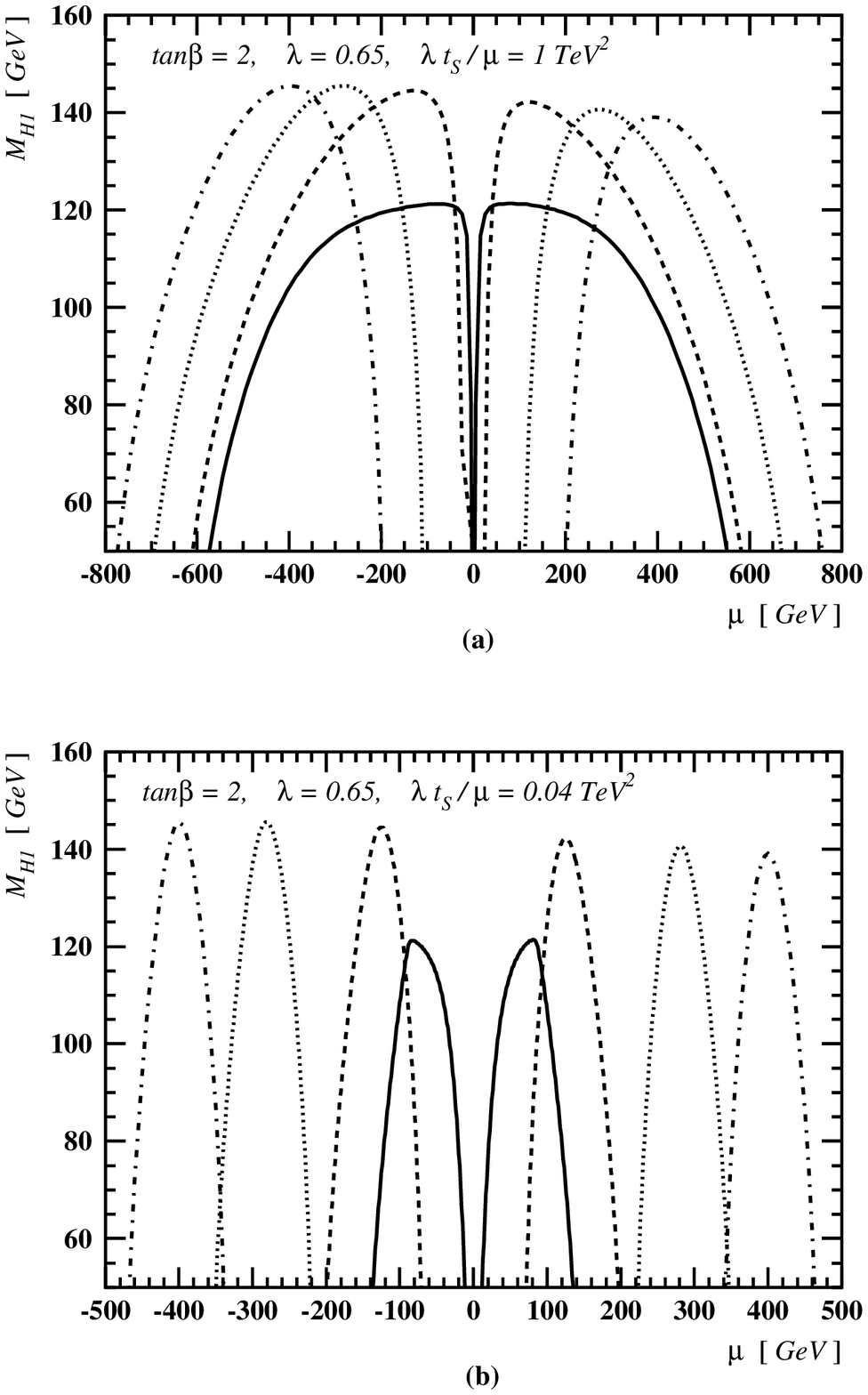}
 \end{center}
 \vspace{-0.5cm} 
\caption{Numerical predictions for $M_{H_1}$ as a function of $\mu$ in
  the MNSSM with $m^2_{12} = 0$, for $M_{H^+} = 0.1$ (solid line), 0.3
  (dashed line), 0.7 (dotted line), 1 (dash-dotted line)
  TeV.}\label{fig:nmssm1}
\end{figure}

%
%%%
%

\begin{figure}[p]
   \leavevmode
 \begin{center}
   \epsfxsize=16.5cm
    \epsffile[0 0 539 652]{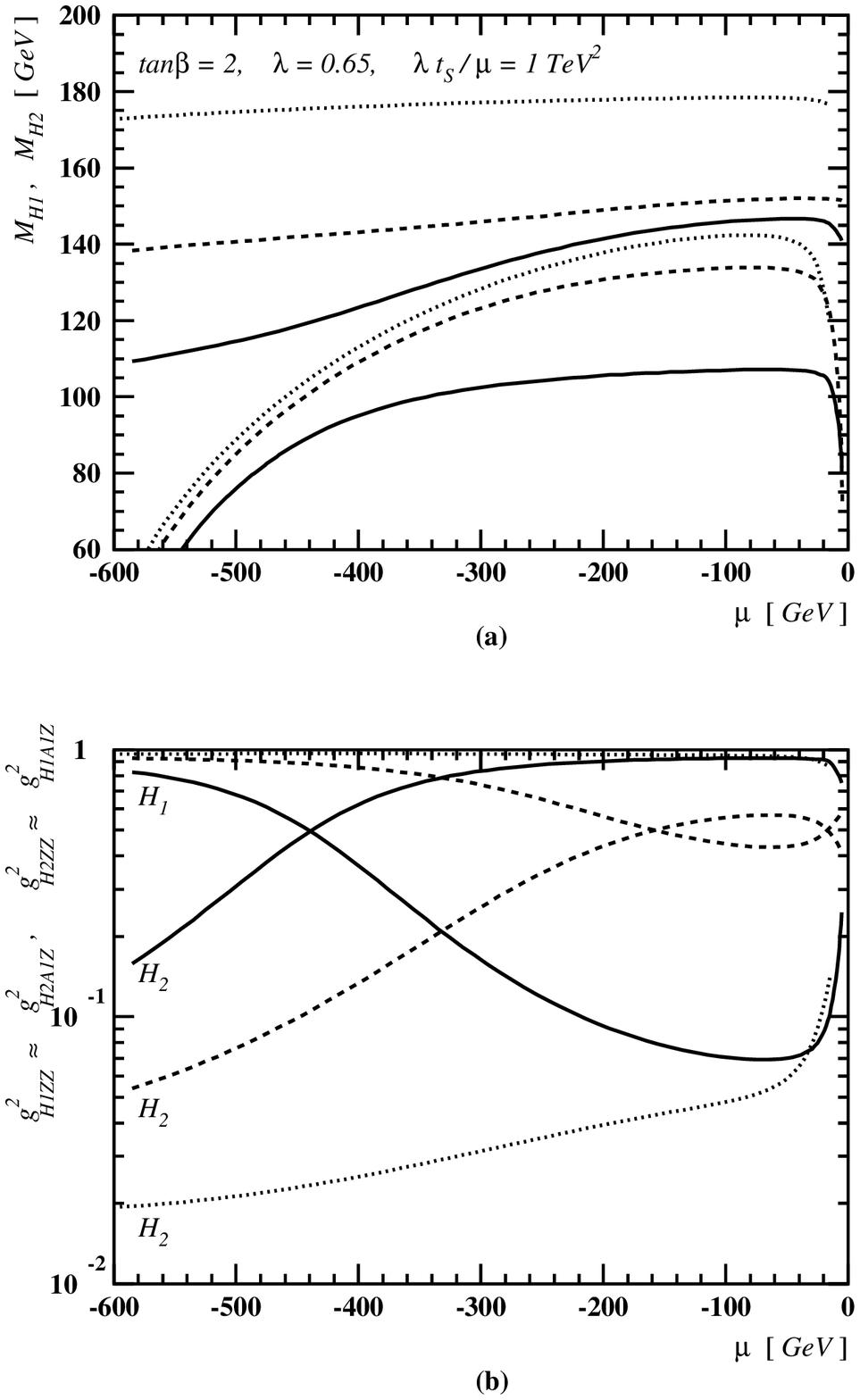}
 \end{center}
 \vspace{-1.2cm} 
\caption{Numerical estimates of (a) $M_{H_1}$ and $M_{H_2}$ and of 
(b) $g^2_{H_1ZZ}$ and $g^2_{H_2ZZ}$, as functions of $\mu$ in the 
MNSSM with $m^2_{12} = 0$, for $M_{H^+} = 80$ (solid line), 120
  (dashed line) and 160 (dotted line) GeV.}\label{fig:nmssm2}
\end{figure}

%
%%%
%

\begin{figure}[p]
   \leavevmode
 \begin{center}
   \epsfxsize=16.5cm
    \epsffile[0 0 539 652]{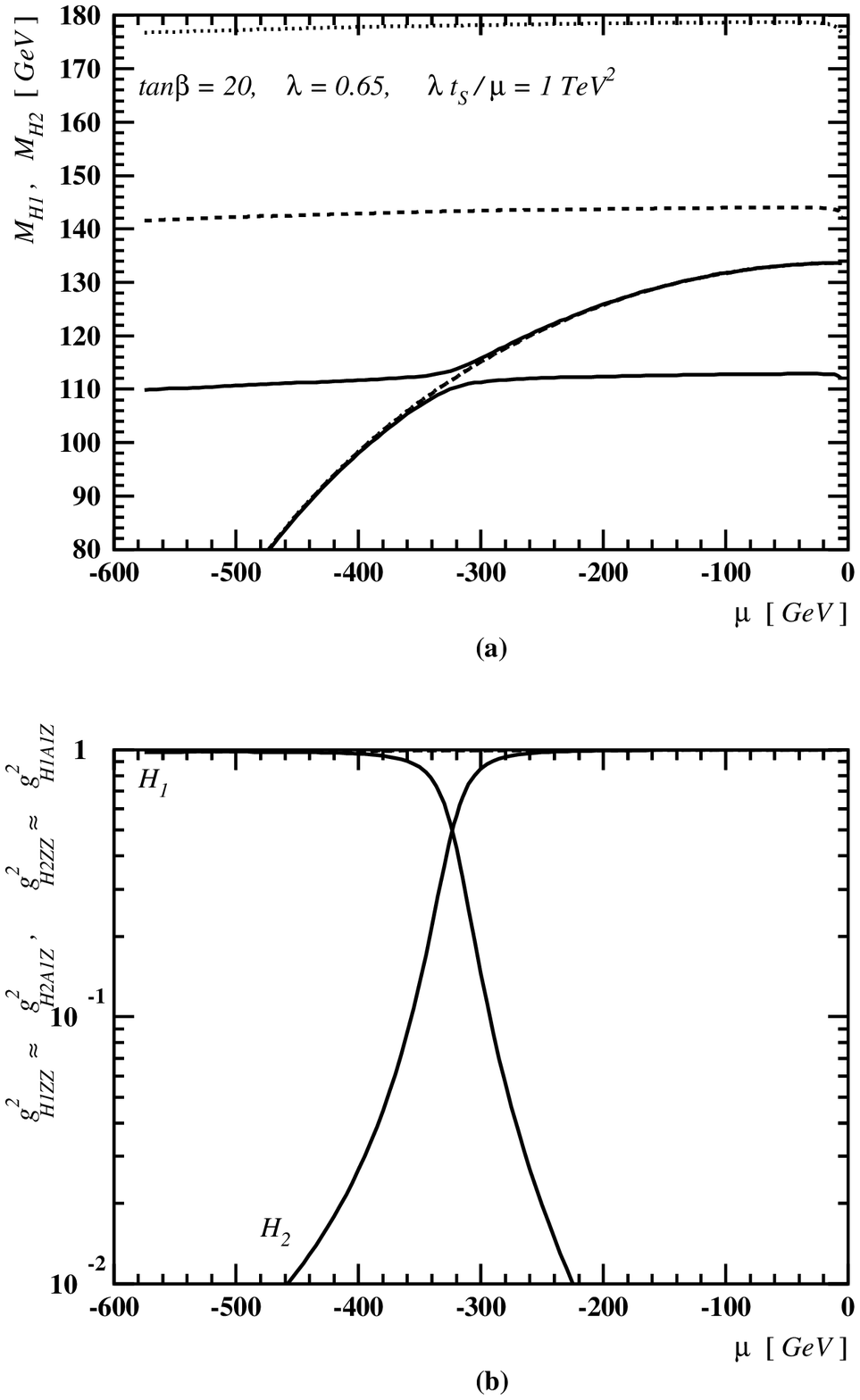}
 \end{center}
 \vspace{-0.8cm} 
\caption{The same as in Fig.\ \ref{fig:nmssm2}, but with $\tan\beta =
  20$.}\label{fig:nmssm11}
\end{figure}

%
%%%
%

\begin{figure}[p]
   \leavevmode
 \begin{center}
   \epsfxsize=16.5cm
    \epsffile[0 0 539 652]{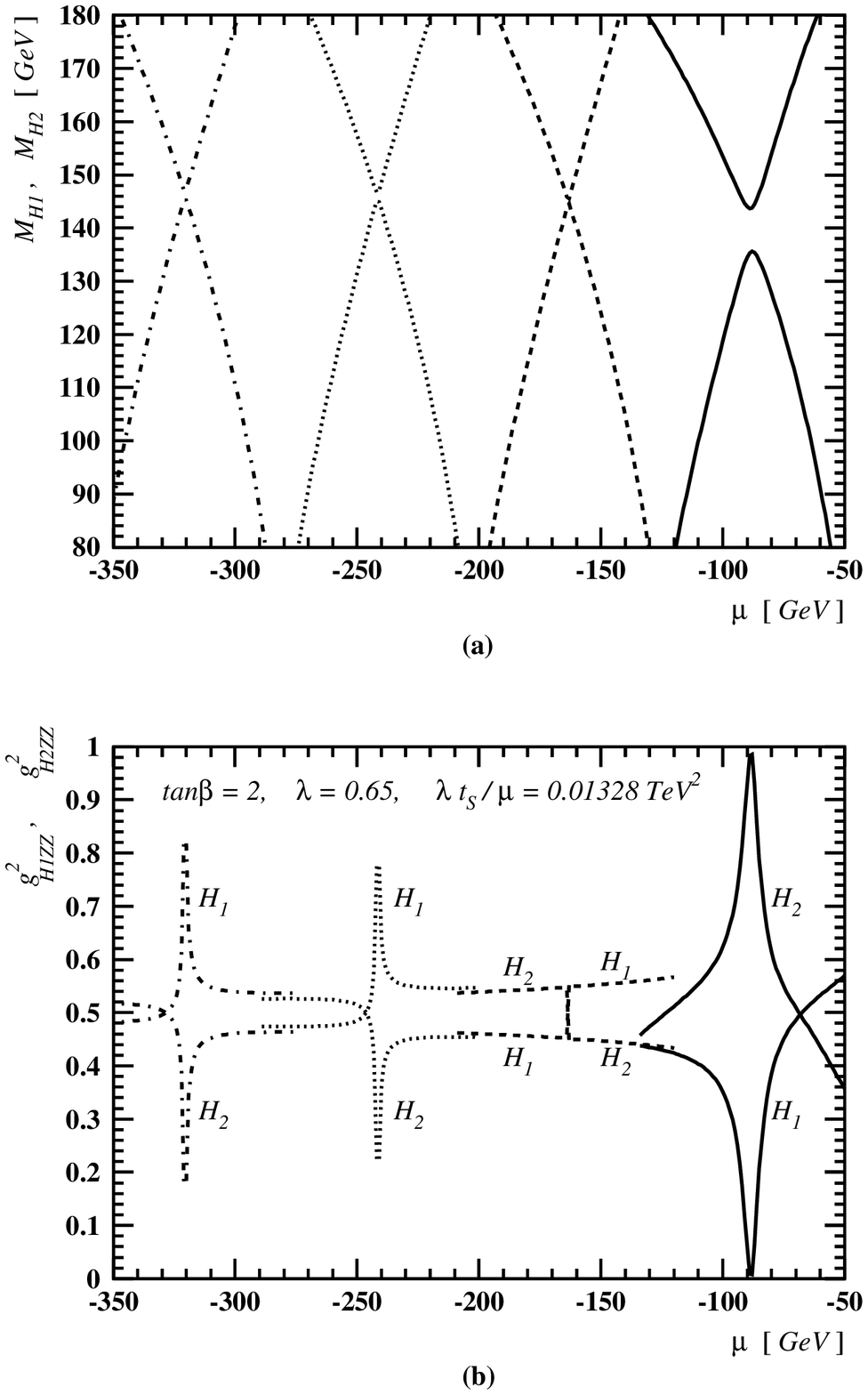}
 \end{center}
 \vspace{-1.2cm} 
\caption{Numerical estimates of (a) $M_{H_1}$ and $M_{H_2}$ and of 
  (b) $g^2_{H_1ZZ}$ and $g^2_{H_2ZZ}$, as functions of $\mu$ in the
  MNSSM with $m^2_{12}=0$, for $M_{H^+} = 0.2$ (solid line), 0.4
  (dashed line), 0.6 (dotted line) and 0.8 (dash-dotted line)
  TeV.}\label{fig:nmssm7}
\end{figure}

%
%%%
%
\begin{figure}[p]
   \leavevmode
 \begin{center}
   \epsfxsize=16.5cm
    \epsffile[0 0 539 652]{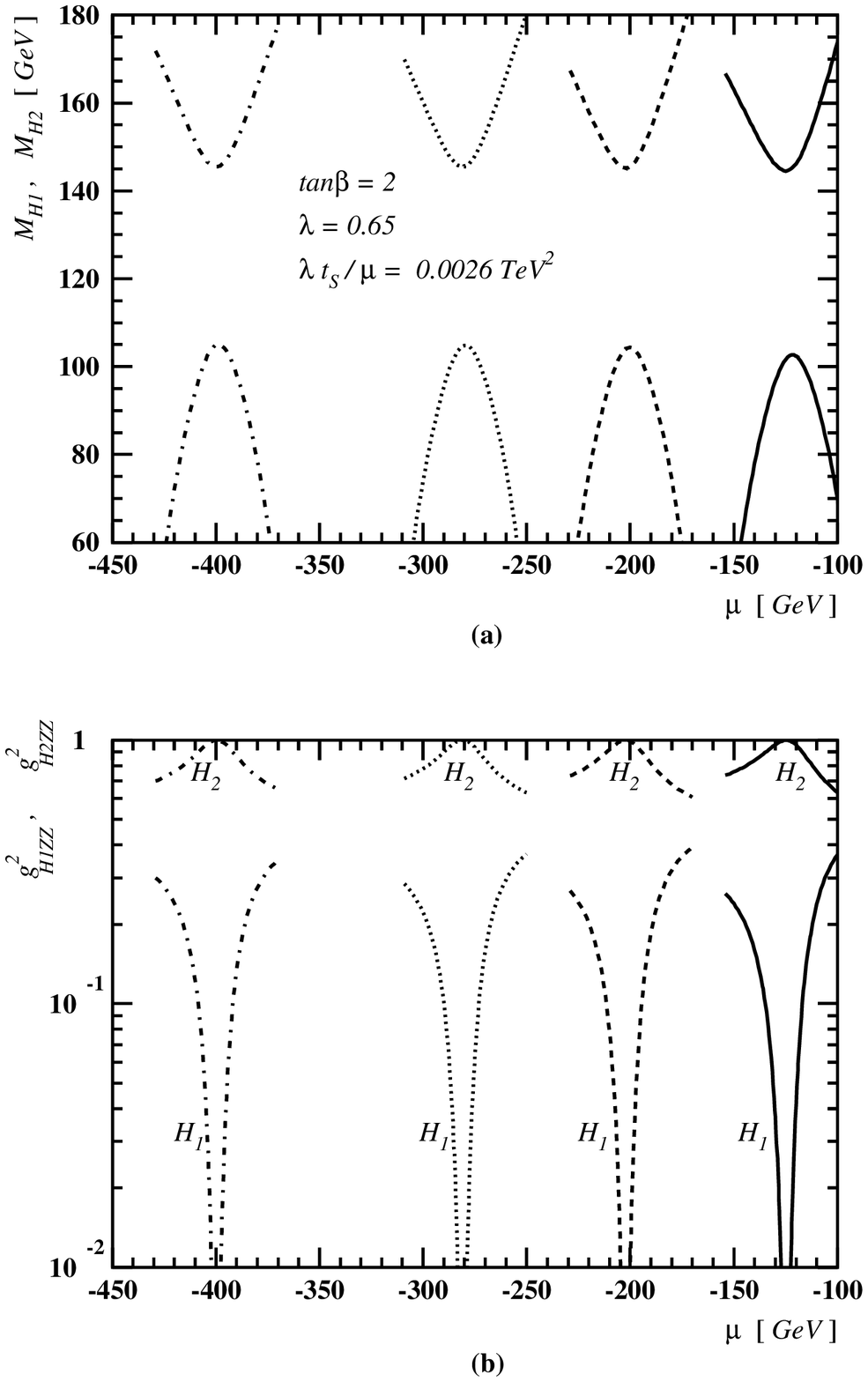}
 \end{center}
 \vspace{-1.2cm} 
\caption{Numerical predictions for (a) $M_{H_1}$ and $M_{H_2}$ and for
  (b) $g^2_{H_1ZZ}$ and $g^2_{H_2ZZ}$, as functions of $\mu$ in the
  MNSSM with $m^2_{12}=0$, for $M_{H^+} = 0.3$ (solid line), 0.5
  (dashed line), 0.7 (dotted line) and 1 (dash-dotted line)
  TeV.}\label{fig:nmssm8}
\end{figure}

%
%%%
%

\begin{figure}[p]
   \leavevmode
 \begin{center}
   \epsfxsize=16.5cm
    \epsffile[0 0 539 652]{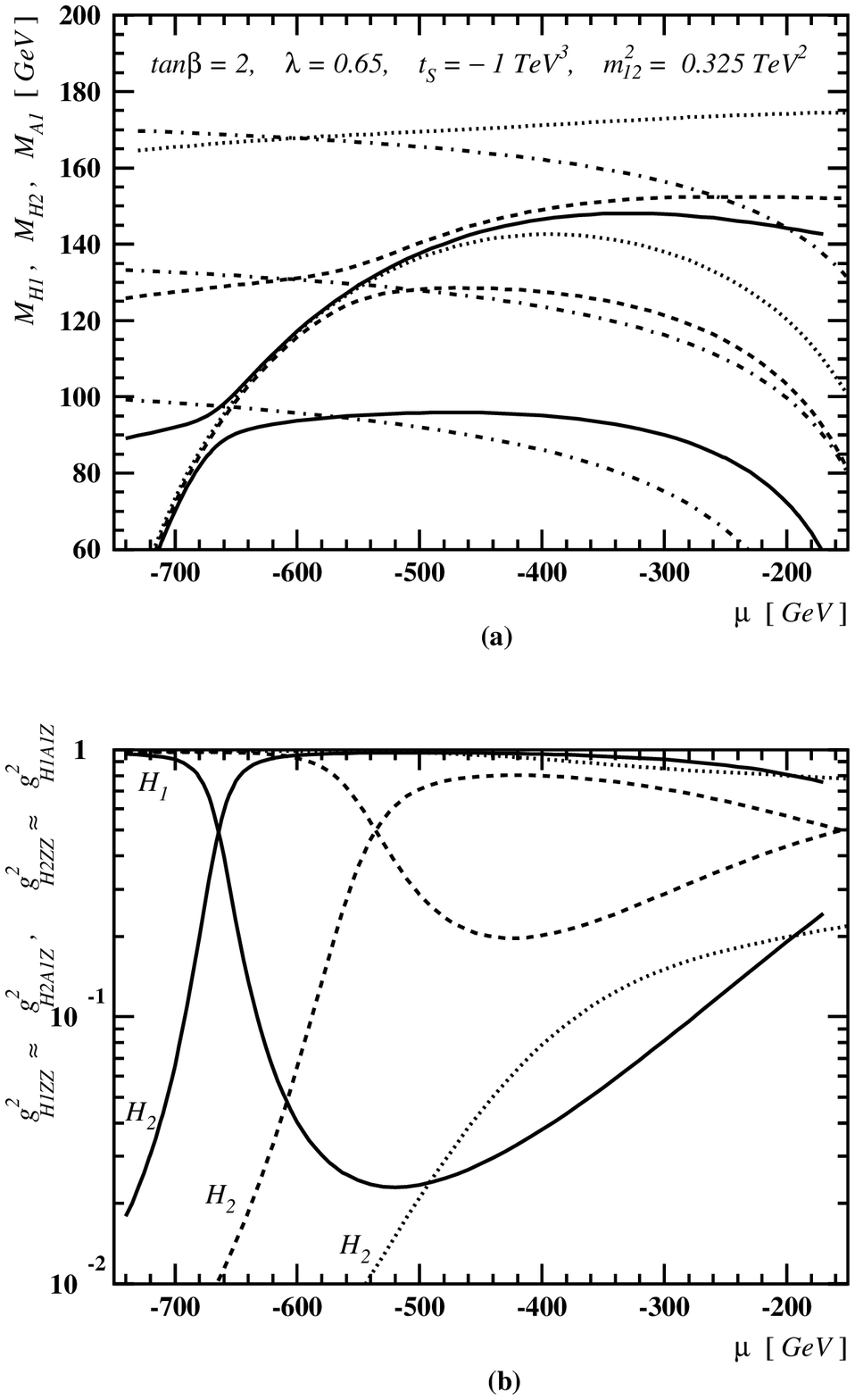}
 \end{center}
 \vspace{-1.2cm} 
\caption{Numerical predictions for (a) $M_{H_1}$, $M_{H_2}$ and
  $M_{A_1}$, and for (b) $g^2_{H_1ZZ}$ and $g^2_{H_2ZZ}$, as functions
  of $\mu$ in the MNSSM, for $M_{H^+} = 80$ (solid line), 120 (dashed
  line) and 160 (dotted line) GeV.  Numerical estimates of $M_{A_1}$
  are indicated by dash-dotted lines.}\label{fig:nmssm9}
\end{figure}

%
%%%
%

\begin{figure}[p]
   \leavevmode
 \begin{center}
   \epsfxsize=18cm
    \epsffile[0 0 567 454]{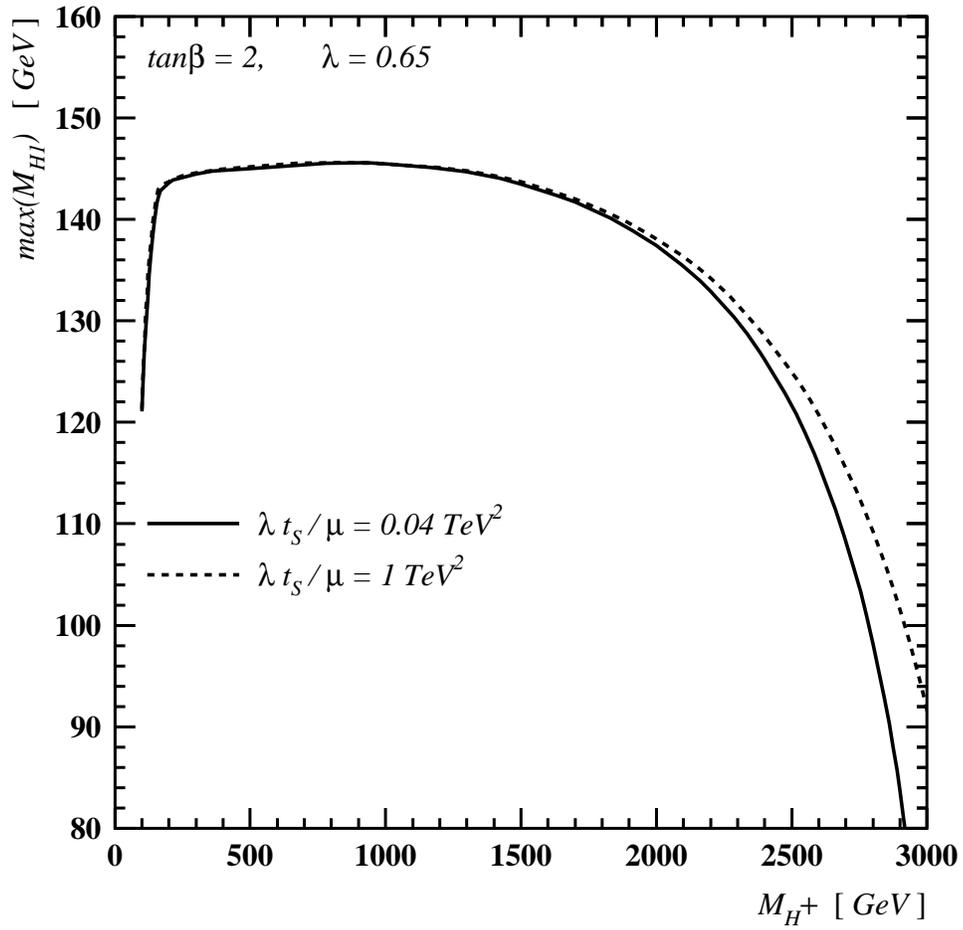}
 \end{center}
 \vspace{-0.8cm} 
\caption{The maximal predicted value of $M_{H_1}$ 
  as a function of the charged Higgs-boson mass $M_{H^+}$ in the MNSSM
  with $m^2_{12} =0$.}\label{fig:nmssm3}
\end{figure}

%
%%%
%

\begin{figure}[p]
   \leavevmode
 \begin{center}
   \epsfxsize=16.5cm
    \epsffile[0 0 539 652]{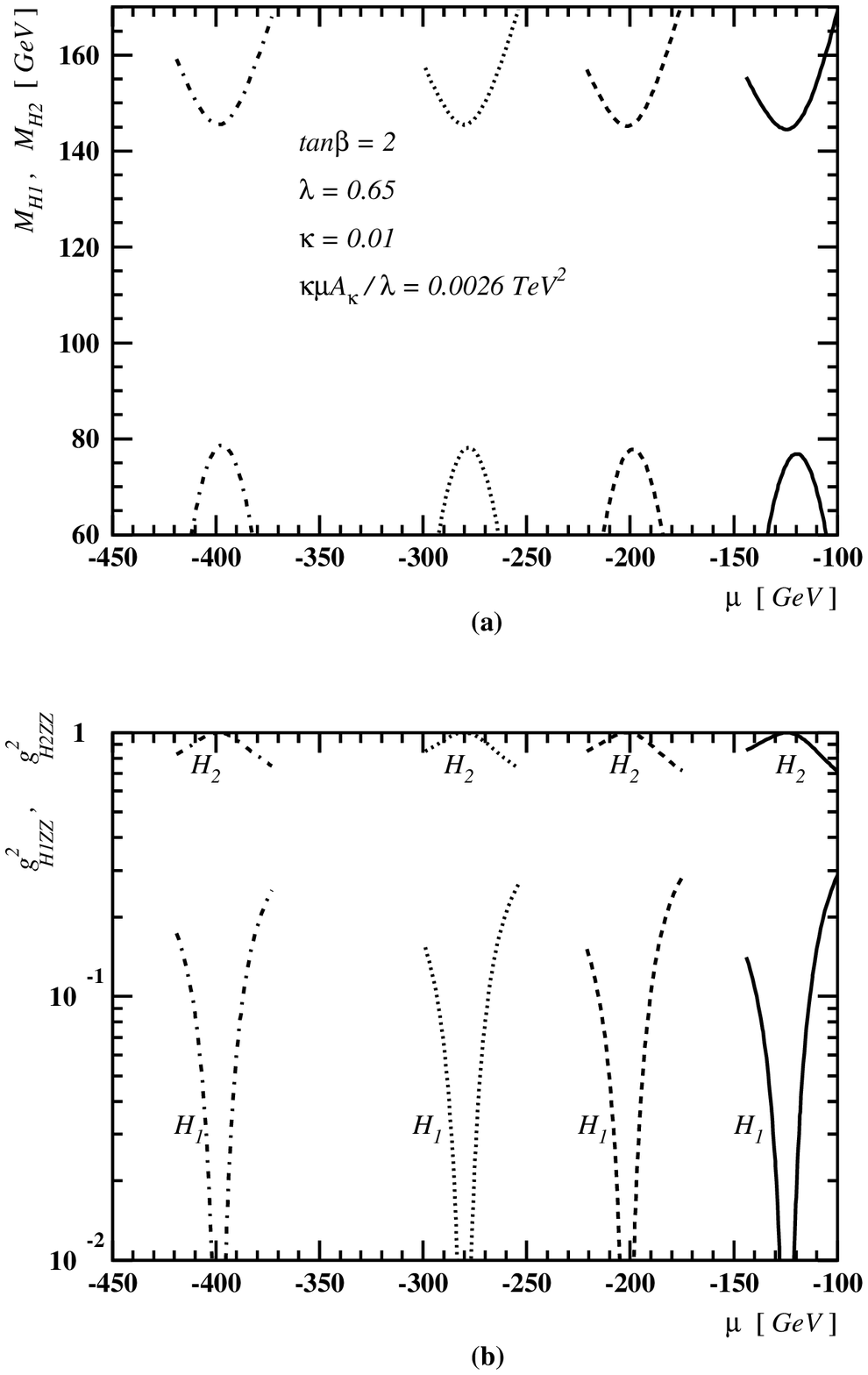}
 \end{center}
 \vspace{-1.2cm} 
\caption{Numerical estimates of (a) $M_{H_1}$ and $M_{H_2}$ and of 
  (b) $g^2_{H_1ZZ}$  and $g^2_{H_2ZZ}$ as functions  of $\mu$ in the
  NMSSM, for  $M_{H^+} =  0.3$ (solid  line), 0.5 (dashed  line),  0.7
  (dotted line) and 1 (dash-dotted line) TeV.}\label{fig:nmssm5}
\end{figure}

%
%%%
%

\begin{figure}[p]
   \leavevmode
 \begin{center}
   \epsfxsize=16.5cm
    \epsffile[0 0 539 652]{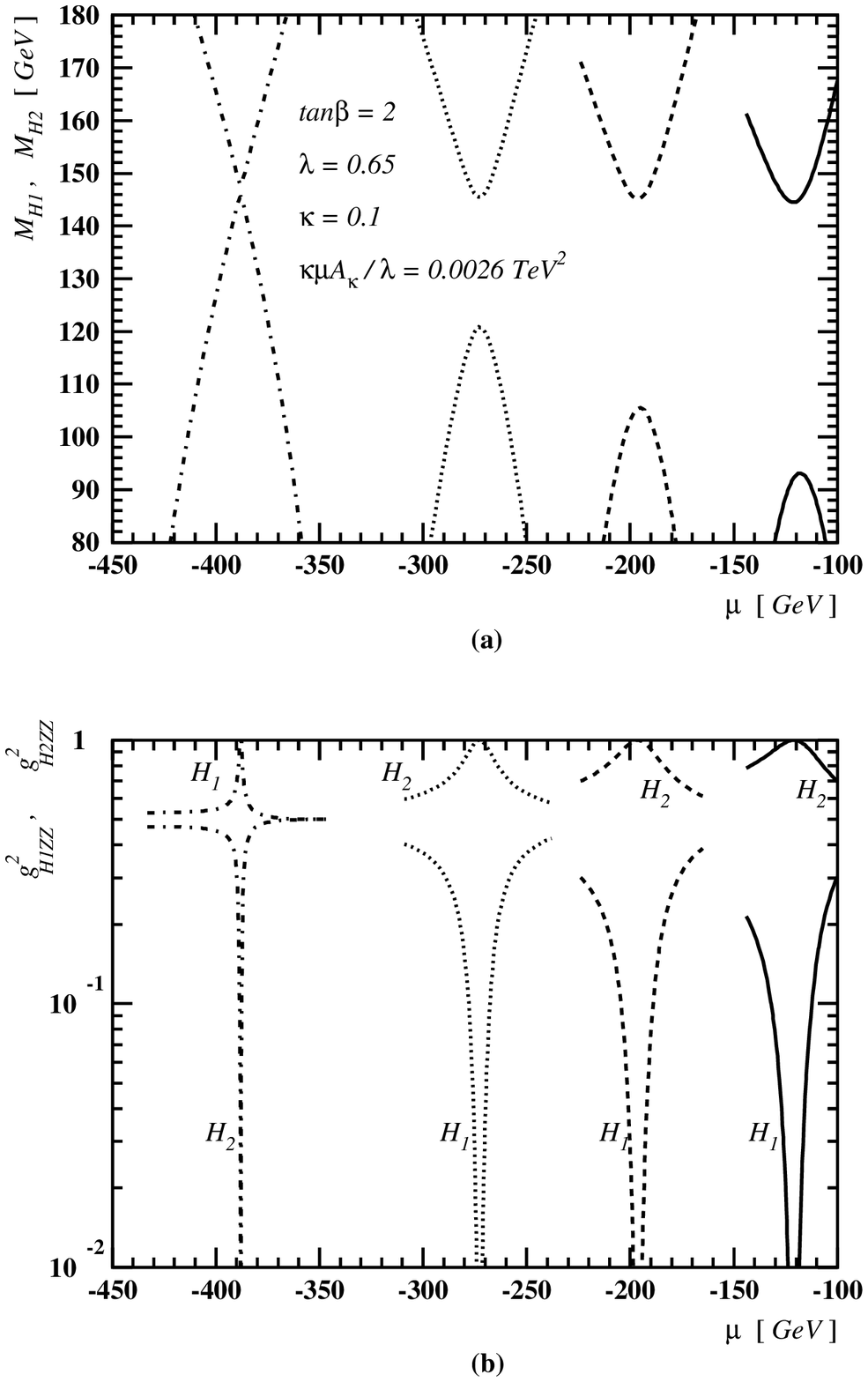}
 \end{center}
 \vspace{-1.cm} 
\caption{The same as in Fig.\ \ref{fig:nmssm5}, but with $\kappa
  =0.1$.}\label{fig:nmssm4}
\end{figure}

%
%%%
%

\begin{figure}[p]
   \leavevmode
 \begin{center}
   \epsfxsize=16.5cm
    \epsffile[0 0 539 652]{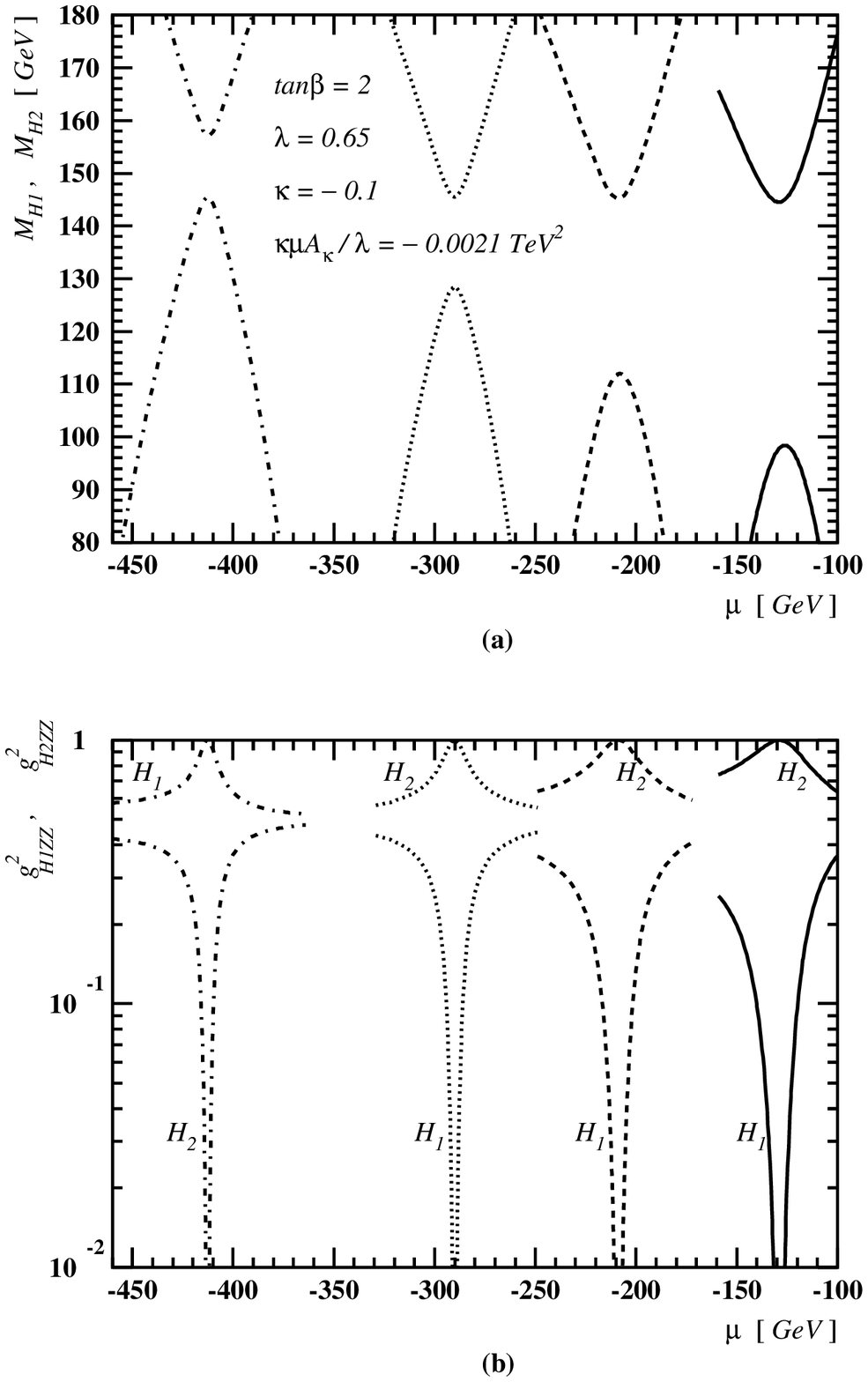}
 \end{center}
 \vspace{-1.cm} 
\caption{The same as in Fig.\ \ref{fig:nmssm5}, but with $\kappa
  =-0.1$ and $\kappa\mu A_\kappa/\lambda =
  -0.0021$~TeV$^2$.}\label{fig:nmssm10}
\end{figure}

%
%%%
%

\begin{figure}[p]
   \leavevmode
 \begin{center}
   \epsfxsize=16.5cm
    \epsffile[0 0 539 652]{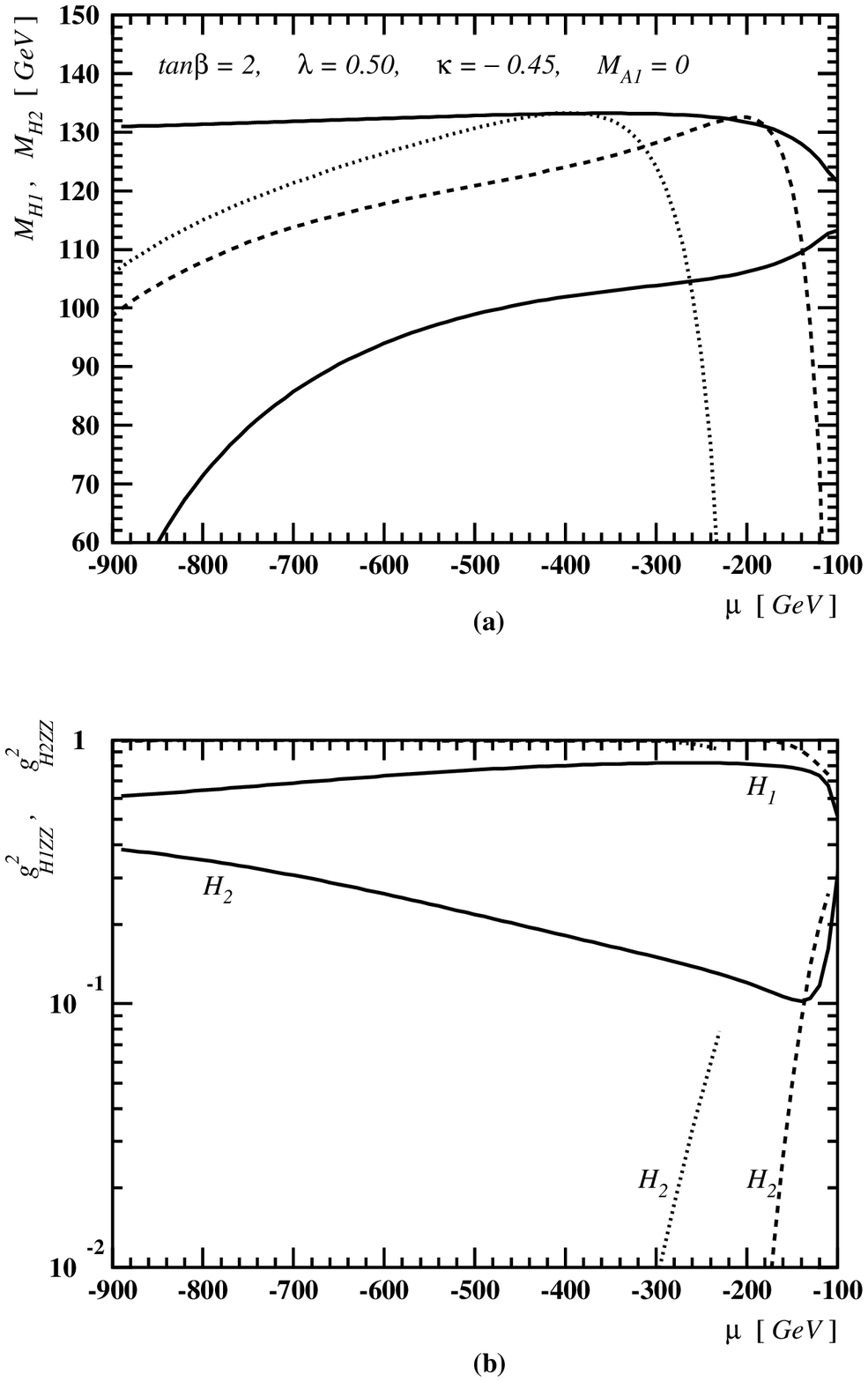}
 \end{center}
 \vspace{-1.2cm} 
\caption{Numerical estimates of (a) $M_{H_1}$ and $M_{H_2}$ and of 
  (b) $g^2_{H_1ZZ}$ and $g^2_{H_2ZZ}$ as functions of $\mu$ in the
  NMSSM, with the constraint $M_{A_1} = 0$, for $M_{H^+} = 120$ (solid
  line), 400 (dashed line) and 800 (dotted line)
  GeV.}\label{fig:nmssm6}
\end{figure}

\end{document}